\documentclass[aps,preprint,onecolumn,secnumarabic,nobalancelastpage,amsmath,amssymb,
nofootinbib]{revtex4}

\RequirePackage{fix-cm}
\usepackage{enumerate}

\usepackage{pst-plot}
\usepackage{pstricks,pstricks-add}
\usepackage[scriptsize,nooneline,hang]{caption}
\usepackage[hang,nooneline,scriptsize]{subfigure}

\usepackage{graphicx}      
\usepackage{url}           
\usepackage{dcolumn}
\usepackage{bm}            
\usepackage{mathrsfs}
\usepackage{epstopdf}
\usepackage{color}

\begin{document}

\preprint{APS/123-QED}

\title{Detailed study of geodesics in the Kerr-Newman-(A)dS spactime and the rotating charged black hole spacetime in $f(R)$ gravity}

\author{Saheb Soroushfar${}^1$}
\author{Reza Saffari${}^1$}%
\email{rsk@guilan.ac.ir}
\author{Sobhan Kazempour${}^1$}
\author{Saskia Grunau${}^2$}
\author{Jutta Kunz${}^2$}

 \affiliation{${}^1$Department of Physics, University of Guilan, 41335-1914, Rasht, Iran.\\
${}^2$Institut f\"ur Physik, Universit\"at Oldenburg, Postfach 2503 D-26111 Oldenburg, Germany.
}%

\date{\today}

\begin{abstract}
We perform a detailed study of the geodesic equations in the spacetime of
the static and rotating charged black hole 
corresponding to the Kerr-Newman-(A)dS spacetime. We derive the
equations of motion for test particles and light rays and present their solutions 
in terms of the Weierstrass $\wp$, $\zeta$ and $\sigma$ functions as well
as the Kleinian $\sigma$ function. With the help of parametric diagrams and effective
potentials we analyze the geodesic motion and classify the possible orbit types.
This spacetime is also a solution of $f(R)$ gravity with a constant
curvature scalar.
\end{abstract}

\maketitle

\section{INTRODUCTION}
A large body of observational evidence has been gathered
in support of an accelerated expansion of the universe at the present time.
This includes, in particular, measurements of the luminosity distance of type Ia supernovae
\cite{Riess:1998cb},
the anisotropy of the cosmic microwave background
\cite{Spergel:2006hy,Ade:2013sjv},
weak lensing \cite{Jain:2003tba},
baryon acoustic oscillations \cite{Eisenstein:2005su},
and the large scale structure of the universe \cite{Tegmark:2003ud}.

To explain the current accelerated expansion of the Universe
is one of the most challenging problems of modern cosmology. 
In the current cosmological standard model, the $\Lambda$CDM model,
a small positive cosmological constant is included in the
Einstein field equations to model this acceleration.
It is therefore clearly of considerable interest, to study the influence
of a cosmological constant on further solutions of the Einstein equations,
such as black hole spacetimes, in particular.

For a deeper understanding of the gravitational field of massive objects 
and in order to accurately predict observational 
effects (such as light deflection, gravitational time delay, perihelion
shift and Lense-Thirring effect), 
it is mandatory to have very good knowledge of the motion
of test particles and light rays in the spacetimes of interest.
But only analytical methods allow for arbitrarily high accuracy
of the prediction of this motion and the associated observables.

In the Schwarzschild spacetime
the equations of motion of test particles and light 
rays were solved analytically in terms of elliptic functions by Hagihara in
1931 \cite{Y. Hagihara}.
The geodesic equations in the 
Reissner-Nordstr\"om, Kerr, and Kerr-Newman space-times have the same
mathematical structure \cite{Chandrasekhar:1985kt} and can be solved analogously.
This analytical method was recently further advanced and applied to the
hyperelliptical case, where the analytical solution of the equations
of motion in the 4-dimensional Schwarzschild-(A)dS,
Reissner-Nordstr\"om-(A)dS and Kerr-(A)dS spacetimes was
presented \cite{Hackmann:2008zza,Hackmann:2008tu,Hackmann:2009nh,Hackmann:2010zz,Grunau:2010gd,Enolski:2010if}.

These mathematical tools were also
applied to the geodesic motion in Taub-NUT
and wormhole spacetimes \cite{Kagramanova:2010bk,Kagramanova:2013mwv},
and to higher dimensional spacetimes, including static black hole spacetimes
and Myers-Perry spacetimes 
\cite{Hackmann:2008tu,Kagramanova:2012hw,Diemer:2014lba,Diemer:2013fza},
while in five dimensional black ring spacetimes, the
equations of motion could be solved analytically in special cases
\cite{Grunau:2012ai,Grunau:2012ri}.
Moreover, the motion of test particles was studied in various black
string spacetimes including field theoretical cosmic string
spacetimes and black holes pierced by a black string
\cite{Aliev:1988wv,Galtsov:1989ct,Chakraborty:1991mb,Ozdemir:2003km,Ozdemir:2004ne,Grunau:2013oca,Hackmann:2009rp,Hackmann:2010ir}. 
In addition, the geodesic equations were solved analytically in a static black hole
spacetime of $f(R)$ gravity \cite{Soroushfar:2015wqa},
for Ho\v{r}ava-Lifshitz black holes \cite{Enolski:2011id},
and for BTZ and GMGHS black holes 
\cite{Soroushfar:2015dfz,Soroushfar:2016yea}. 

On the other hand, the large body of current cosmological data
could also be taken to indicate that 
General Relativity itself should be extended.
The latter would also be supported by
the necessity for dark matter as revealed from astrophysical and
cosmological observations
(see e.g.~\cite{Bertone:2004pz})
and moreover by theoretical arguments
at the ultraviolet scale (e.g., quantum gravity, initial singularities).

Consequently, numerous theoretical attempts 
to model the evolution of the Universe
are based on the modification of gravity
(see e.g.~the reviews 
\cite{Capozziello:2011et,Clifton:2011jh,Joyce:2014kja,Jain:2010ka,Koyama:2015vza}).
Popular suggestions to modify gravity include theories with
higher powers of the Riemann and Ricci tensors as well as the curvature scalar $R$.
Lovelock theory \cite{Lovelock:1971yv}
and $f(R)$ gravity \cite{Sotiriou:2008rp,DeFelice:2010aj,Capozziello:2011et}
are such examples, where the Einstein-Hilbert action is 
generalized accordingly.

A change of the action has influence on the dynamics 
of the Universe, but it may also affect the dynamics at the galactic or solar system scales.
Clearly, any modifications of the action must retain the well tested sector of
General Relativity, like its description of the solar system. 
However, modified theories may yield different answers from General Relativity in the
strong field regime.
It is therefore essential to inquire about the existence of black holes and about
their properties in modified theories of gravity
(see e.g.~\cite{Berti:2015itd}).
In general, 
the study of black holes in these theories may reveal interesting features
not present in General Relativity.

Focussing on black holes in $f(R)$ theories
\cite{Brevik:2004sd,Cognola:2005de,Saffari:2007zt,delaCruzDombriz:2009et, Capozziello:2009jg, Sebastiani:2010kv, Moon:2011hq,Larranaga:2011fv, Cembranos:2011sr, delaCruzDombriz:2012xy, Hendi:2014mba},
we note that a particular class of solutions 
is obtained when the curvature scalar is constant, $R=R_0$.
Taking the trace of the field equations then specifies this constant in terms of the
function $f(R)$ and its derivative at $R_0$. A comparison with General Relativity and its
black hole solutions reveals, that the finite curvature scalar
acts basically like a cosmological constant. In vacuum therefore the 
Schwarzschild-(A)dS
and Kerr-(A)dS solutions are recovered, when certain rescalings are performed.
When adding charge to the solutions by including an electromagnetic field,
the Reissner-Nordstr\"om-(A)dS and the Kerr-Newman--(A)dS solutions can be
recovered after certain rescalings,
because the trace of the energy momentum tensor vanishes.

In this paper we study the geodesic motion in the spacetime of the static and 
rotating charged black hole (Kerr-Newman-(A)dS spacetime). 
Since after certain rescalings the Kerr-Newman-(A)dS black hole of 
General Relativity also describes a rotating charged black hole 
in $f(R)$ gravity, the current analysis can also be applied 
to this black hole solution in $f(R)$ gravity.
Let us mention, however, that the stability of this $f(R)$ black hole
can only be established, after the function $f(R)$ is specified.

We here analyze the possible
orbit types using effective potential techniques and parametric
diagrams. Furthermore, we present the analytical solutions of the
equations of motion for test particles and light. The equations of
motion are of elliptic and hyperelliptic type and the solutions are
given in terms of the Weierstrass $\wp$, $\zeta$ and $\sigma$
functions as well as the Kleinian $\sigma$ functions.
Complete integrability of the geodesic equations is guaranteed
by the presence of four integrals of motion of the Kerr-Newman-(A)dS
spacetime, with the forth one being the Carter constant.

Similar analyses of geodesic motion and solutions of the equations 
of motion were presented in the Kerr-(A)dS spacetime \cite{Hackmann:2010zz} 
and in the Kerr-Newman spacetime \cite{Hackmann:2013pva}, 
but in the Kerr-Newman-(A)dS spacetime the geodesic motion 
has not been analyzed analytically before in great detail, although some aspects were studied in \cite{Heisnam:2014,Stuchlik:1997gk}. 
The analytical solution of the geodesic equation of light and a study of the gravitational lensing and frame dragging of light were presented in \cite{Kraniotis:2014paa}. However, neither an analysis of all possible orbits for particles and light was done nor the full 
set of analytic solutions to the equations of motions was found.

Our paper is organized as follows: In Sec.~(\ref{field}) 
we give a brief review of the field equations 
and the metric of the rotating black hole in $f(R)$ gravity 
and its connection to General Relativity. 
In Sec.~(\ref{rotatinggeodesic}) we present the equations of motion 
in the Kerr-Newman-(A)dS spacetime. We analyze the geodesic motion 
in Sec.~(\ref{sec:analysis}) and give a list of all possible orbit types. 
The analysis is given separately for the static case 
(Reissner-Nordstr\"om-(A)dS) and the rotating case (Kerr-Newman-(A)dS). 
In Sec.~(\ref{analytical solutions}) we present the full set of 
analytical solutions of the geodesic equations in the general rotating case 
of the Kerr-Newman-(A)dS black hole. Some example orbits in the static 
and the rotating case are shown in Sec.~(\ref{sec:orbits}). 
We conclude in Sec.~(\ref{conclusions}).

\boldmath
\section{FIELD EQUATIONS IN $f(R)$ MODIFIED GRAVITY AND RESCALINGS}\label{field}
\unboldmath

In this section we give a brief review of the field equations and the metric of the rotating black hole in $f(R)$, which represents the Kerr-Newman-(A)dS spacetime after certain rescalings.
In four dimensions the action of $f(R)$ gravity with a Maxwell field is given by
\begin{align}\label{action}
S=S_{g}+S_{M},
\end{align}
where $S_{g}$ and $S_{M}$ are the gravitational and the electromagnetic actions
\begin{equation}
S_{g}=\dfrac{1}{16 \pi G} \int d^{4} x \sqrt{ -g  }(R+f(R)),
\end{equation}
\begin{equation}\label{action2}
S_{M}=\dfrac{-1}{16 \pi} \int d^{4} x \sqrt{-g }[F_{\mu\nu}F^{\mu\nu}],
\end{equation}
where $G$ is the gravitational constant, which we will set to one,
$g$ is the determinant of the metric, $R$ is the curvature scalar, and $R + f(R)$ is the
function defining the modified gravity theory under consideration.
From the above action, the Maxwell
equations take the form
\begin{equation}
\nabla_{\mu}F^{\mu\nu}=0 ,
\end{equation}
while the field equations in the metric formalism are
\begin{align}\label{rmiyo}
R_{\mu\nu}\big(1+ f'(R)\big)-\frac{1}{2}\big(R+f(R)\big)g_{\mu\nu}
+\big(g_{\mu\nu}\nabla^{2}-\nabla_{\mu}\nabla_{\nu}\big)f'(R)=2T_{\mu\nu},
\end{align}
where $R_{\mu\nu}$ is the Ricci tensor, $\nabla$ denotes the usual
covariant derivative, and the stress-energy tensor of the
electromagnetic field is given by
\begin{equation}
T_{\mu\nu}=F_{\mu\rho}F_{\nu}^{\rho}-\dfrac{g_{\mu\nu}}{4}F_{\rho\sigma}F^{\rho\sigma},
\end{equation}
and has vanishing trace
\begin{equation}
T^{\mu} _{\,\,\mu}=0 .
\label{tracezero}
\end{equation}

Taking the trace of Eq.~(\ref{rmiyo}) under the assumption, that 
the curvature scalar is constant, $R=R_{0}$, leads to
\begin{equation}
R_{0}\big(1+f'(R_{0})\big)-2\big(R_{0}+f(R_{0})\big)=0 .
\label{R_0}
\end{equation}
This is the same equation as in the vacuum case, because 
the matter field has vanishing trace, Eq.~(\ref{tracezero}),
and it determines the constant value of the curvature scalar
\begin{equation}
R_{0}=\dfrac{2f(R_{0})}{f'(R_{0})-1} .
\end{equation}
Using this relation in Eq.~(\ref{rmiyo}) gives the field equations
\begin{align}
R_{\mu\nu} - \dfrac{1}{2} \dfrac{f(R_{0})}{f'(R_{0})-1}  g_{\mu\nu}
=\dfrac{2}{1+f'(R_{0})}T_{\mu\nu}.
\end{align}
Comparison with the Einstein equations in the presence of a cosmological constant
$\Lambda$ then indicates an equivalence of the two sets of equations
for $R_0=4 \Lambda$, when we further rescale the left hand side 
of the equations 
adequately.

Consequently, up to rescalings, the Kerr-Newman-(A)dS solution of
General Relativity is also a solution of the field equations in $f(R)$ gravity
\cite{Larranaga:2011fv, Cembranos:2011sr, delaCruzDombriz:2012xy}.
Thus, the stationary black hole solution can be obtained in Boyer-Lindquist like
coordinates $(t,r,\theta,\varphi)$ as follows \cite{Larranaga:2011fv}
\begin{align}\label{rotatingmetric}
ds^{2}=-\dfrac{\Delta_{r}}{\rho^{2}} \left[dt-\dfrac{a \sin^{2}\theta
d\varphi}{\Xi}\right]^{2}
+\dfrac{\rho^{2}}{\Delta_{r}}dr^{2}+\dfrac{\rho^{2}}{\Delta_{\theta}}d\theta^{2}
+\dfrac{\Delta_{\theta} \sin^{2}\theta}{\rho^{2}}\left[adt-\dfrac{r^{2}+a^{2}}{\Xi}d\varphi
\right]^{2} ,
\end{align}
with
\begin{align}
\Delta_{r}=(r^{2}+a^{2}) \left(1-\dfrac{R_{0}}{12}r^{2} \right)-2Mr+\dfrac{Q^{2}}{(1+f'(R_{0}))},
\end{align}
\begin{align}
\Xi=1+\dfrac{R_{0}}{12}a^{2}, \qquad
\rho^{2}=r^{2}+a^{2} \cos^{2}\theta, \qquad
\Delta_{\theta}=1+\dfrac{R_{0}}{12}a^{2} \cos^{2}\theta ,
\end{align}
where $ Q $ is the electric charge, $ a $ is the angular momentum per mass of the black hole,
and $R_{0}$ enters like a cosmological constant ($R_{0}=4\Lambda $),
yielding a non-asymptotically flat de Sitter or anti-de Sitter spacetime, when $R_0$ is finite.
Note, that the electric charge enters with a scaling factor in the metric.
As in the Kerr spacetime there is a ringlike singularity defined by $\rho^2=0$, 
and the horizons are given by $\Delta_r=0$.

\section{THE GEODESIC EQUATIONS}\label{rotatinggeodesic}

In this section we derive the equations of motion for a rotating charged black hole Eq.(\ref{rotatingmetric}), 
using the Hamilton-Jacobi formalism, and later introduce effective potentials for the $r$- and $\theta$-motion. 

The Hamilton-Jacobi equation
\begin{equation}\label{Hamilton}
\dfrac{\partial S}{\partial\tau} +\frac{1}{2} \ g^{ij}\dfrac{\partial S}{\partial x^{i}}\dfrac{\partial S}{\partial x^{j}}=0 
\end{equation}
can be solved with an ansatz for the action
\begin{equation}
\label{S}
S=\frac{1}{2}\varepsilon \tau - Et+L_{z}\phi +S_{\theta}(\theta) + S_{r} (r).
\end{equation}
The constants of motion are the energy $E$ and the angular momentum $L$  which are given
by the generalized momenta $P_{t}$ and $P_{\phi}$
\begin{equation}\label{constants of motion}
P_{t}=g_{tt}\dot{t}+g_{t \varphi}\dot{\varphi}=-E,  \qquad P_{\phi}=g_{\varphi \varphi}\dot\varphi +g_{t \varphi}\dot{t} =L.
\end{equation}
Using Eqs.(\ref{Hamilton})--(\ref{constants of motion}) we get
\begin{align}\label{ds/dr.ds/dtheta}
\Delta_{\theta}\left(\dfrac{\partial S}{\partial\theta}\right)^{2}+\varepsilon a^{2} \cos^{2}\theta -\dfrac{2aEL\Xi - E^{2}a^{2}\sin^{2}\theta}{\Delta_{\theta}}+\dfrac{L^{2}\Xi^{2}}{
\Delta_{\theta}\sin^{2}\theta}=-\Delta_{r}(\dfrac{\partial S}{\partial r})^{2}-\nonumber\\ \varepsilon r^{2}+\dfrac{(a^{2}+r^{2})^{2}E^{2}+a^{2}L^{2} \Xi^{2}-2aEL\Xi (r^{2}+a^{2})}{\Delta_{r}} ,
\end{align}
where each side depends on $r$ or $\theta$ only. 
With the separation ansatz Eq.(\ref{S}) and with the help of the Carter constant \cite{Carter:1968rr},
we derive the equations of motion:
\begin{align}\label{Joda}
\rho^{4}\left(\dfrac{dr}{d\tau}\right)^{2}=-\Delta_{r}(K+\varepsilon r^{2})+\big[(a^{2}+r^{2})E-aL\Xi \big]^{2}=R(r),
\end{align}
\begin{align}\label{thetatho}
\rho^{4}\left(\dfrac{d\theta}{d\tau}\right)^{2}=\Delta_{\theta}(K-\varepsilon a^{2}\cos^{2}\theta)-\dfrac{1}{\sin^{2}\theta}\big(aE \sin^{2}\theta -L\Xi \big)^{2} =\Theta(\theta),
\end{align}
\begin{align}
\rho^{2}\left(\dfrac{d\varphi}{d\tau}\right)=\dfrac{aE\Xi (a^{2}+r^{2})-a^{2}\Xi^{2}L}{\Delta_{r}}-\dfrac{1}{\Delta_{\theta}\sin^{2}\theta}(a\Xi E \sin^{2}\theta -\Xi^{2}L),
\end{align}
\begin{align}\label{ttho}
\rho^{2}\left(\dfrac{dt}{d\tau}\right)=\dfrac{E(r^{2}+a^{2})^{2}-aL\Xi(r^{2}+a^{2})}{\Delta_{r}}-\dfrac{\sin^{2}\theta}{\Delta_{\theta}}\left(E a^{2}-\dfrac{L\Xi a}{\sin^{2}\theta}\right).
\end{align}
In the following we will explicitly solve these equations. 
Eq.(\ref{Joda}) suggests the introduction of an effective potential $V_{{\rm eff},r}$ such that $V_{{\rm eff},r}=E$ 
corresponds to $\left(\dfrac{dr}{d\tau}\right)^{2}=0$ 
\begin{equation}
V_{{\rm eff},r}=\dfrac{L\Xi a\pm \sqrt{\Delta_{r}(K+\varepsilon r^{2})}}{a^{2}+r^{2}},
\end{equation} 
where $\left(\dfrac{dr}{d\tau}\right)^{2}\geq 0$ for $E\leq V_{{\rm eff},r}^{-}$ and $E\geq V_{{\rm eff},r}^{+}$. 
In the same way an effective potential
corresponding to Eq.(\ref{thetatho}) can be introduced
\begin{align}
V_{{\rm eff},\theta}=\dfrac{L\Xi \pm \sqrt{\Delta_{\theta}\sin^{2}\theta (K-\varepsilon a^{2}\cos^{2}\theta)}}{a \sin^{2}\theta}.
\end{align}
but here $\left(\dfrac{d\theta}{d\tau}\right)^{2}\geq 0$ for $V_{\rm eff,\theta}^{-}\leq E \leq V_{\rm eff,\theta}^{+}$.
\\
Introducing the Mino time $\lambda$ \cite{Mino:2003yg} connected to the proper time 
$\tau$ by $\dfrac{d\tau}{d\lambda}=\rho^{2}$, the equations of motions read
\begin{align}\label{d}
\left(\dfrac{dr}{d\lambda}\right)^{2}=-\Delta_{r}(K+\varepsilon r^{2})+\big[(a^{2}+r^{2})E-aL\Xi \big]^{2}=R(r),
\end{align}
\begin{align}
\left(\dfrac{d\theta}{d\lambda}\right)^{2}=\Delta_{\theta}(K-\varepsilon a^{2}\cos^{2}\theta)-\dfrac{1}{\sin^{2}\theta}\big(aE \sin^{2}\theta -L\Xi \big)^{2} =\Theta(\theta),
\end{align}
\begin{align}
\left(\dfrac{d\varphi}{d\lambda}\right)=\dfrac{aE\Xi (a^{2}+r^{2})-a^{2}\Xi^{2}L}{\Delta_{r}}-\dfrac{1}{\Delta_{\theta}\sin^{2}\theta}(a\Xi E \sin^{2}\theta -\Xi^{2}L),
\end{align}
\begin{align}\label{t}
\left(\dfrac{dt}{d\lambda}\right)=\dfrac{E(r^{2}+a^{2})^{2}-aL\Xi(r^{2}+a^{2})}{\Delta_{r}}-\dfrac{\sin^{2}\theta}{\Delta_{\theta}}(E a^{2}-\dfrac{L\Xi a}{\sin^{2}\theta}).
\end{align}
we introduce dimensionless quantities to rescale the parameters
\begin{align}
\tilde{r}=\dfrac{r}{M} , \qquad \tilde{a}=\dfrac{a}{M} , \qquad \tilde{t}=\dfrac{t}{M} , \qquad \tilde{L}=\dfrac{L}{M} , \qquad \tilde{K}=\dfrac{K}{M^{2}} , \nonumber\\ \tilde{R_{0}}=R_{0}M^{2}, \qquad  \tilde{Q}=\dfrac{Q}{M},\qquad  \gamma = M\lambda.
\end{align}
Then the equations (\ref{d}) - (\ref{t}) can be rewritten as
\begin{align}\label{drd}
\left(\dfrac{d\tilde{r}}{d\gamma}\right)^{2}=-\Delta_{\tilde{r}}(\tilde{K}+\varepsilon
\tilde{r}^{2})+ \left[ (\tilde{a}^{2}+\tilde{r}^{2})E
-\tilde{a}\tilde{L}\Xi \right]^{2}=\tilde{R}(\tilde{r}),
\end{align}
\begin{align}\label{dthetad}
\left(\dfrac{d\theta}{d\gamma}\right)^{2}=\Delta_{\theta}(\tilde{K}
-\varepsilon \tilde{a}^{2}\cos^{2}\theta)
-\dfrac{1}{\sin^{2}\theta}\left(\tilde{a}E \sin^{2}\theta
-\tilde{L}\Xi \right)^{2}=\tilde{\Theta}(\theta),
\end{align}
\begin{align}\label{dphi}
\left(\dfrac{d\varphi}{d\gamma}\right)=\dfrac{ \tilde{a}E \Xi (\tilde{a}^{2}+\tilde{r}^{2})-\tilde{a}^{2}\Xi^{2}\tilde{L}}{\Delta_{\tilde{r}}}
-\dfrac{1}{\Delta_{\theta}\sin^{2}\theta} \big(\tilde{a}\Xi E
\sin^{2}\theta -\Xi^{2}\tilde{L}  \big),
\end{align}
\begin{align}\label{dtd}
\left(\dfrac{d\tilde{t}}{d\gamma}\right)=\dfrac{E(\tilde{r}^{2}+\tilde{a}^{2})^{2} - \tilde{a}\tilde{L}\Xi
(\tilde{r}^{2}+\tilde{a}^{2})}
{\Delta_{\tilde{r}}}-\dfrac{\sin^{2}\theta}{\Delta_{\theta}} \left(E
\tilde{a}^{2}-\dfrac{\tilde{L}\Xi \tilde{a}}{\sin^{2}\theta} \right).
\end{align}

\section{ANALYSIS OF THE GEODESIC EQUATIONS}\label{sec:analysis}
In this section we will analyze the geodesic equations and give a list of all possible orbits. First we will study the special case of a static charged black hole (Reissner-Nordstr\"om-(A)dS) and then we will give a full analysis of the general rotating charged black hole solution (Kerr-Newman-(A)dS).

\subsection{The static case}

\begin{figure}[h]
	\centering
	\subfigure[$\varepsilon=1$, $\tilde{R}_0=\frac{1}{3}\cdot 10^{-5}$, $\tilde{q}=0.25$]{
		\includegraphics[width=0.36\textwidth]{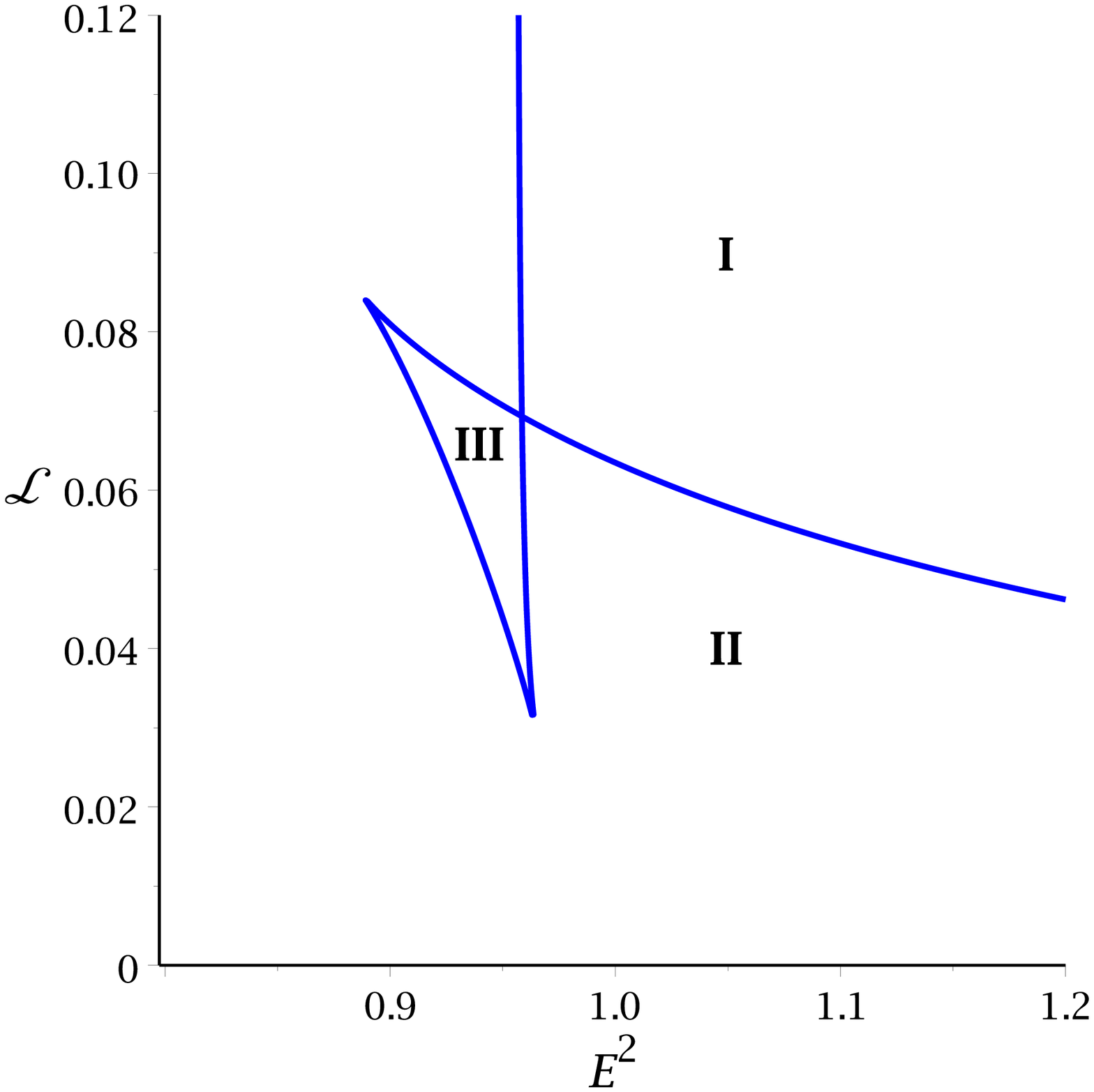}
	}
	\subfigure[$\varepsilon=0$, $\tilde{R}_0=\frac{1}{3}\cdot 10^{-5}$, $\tilde{q}=0.25$]{
		\includegraphics[width=0.36\textwidth]{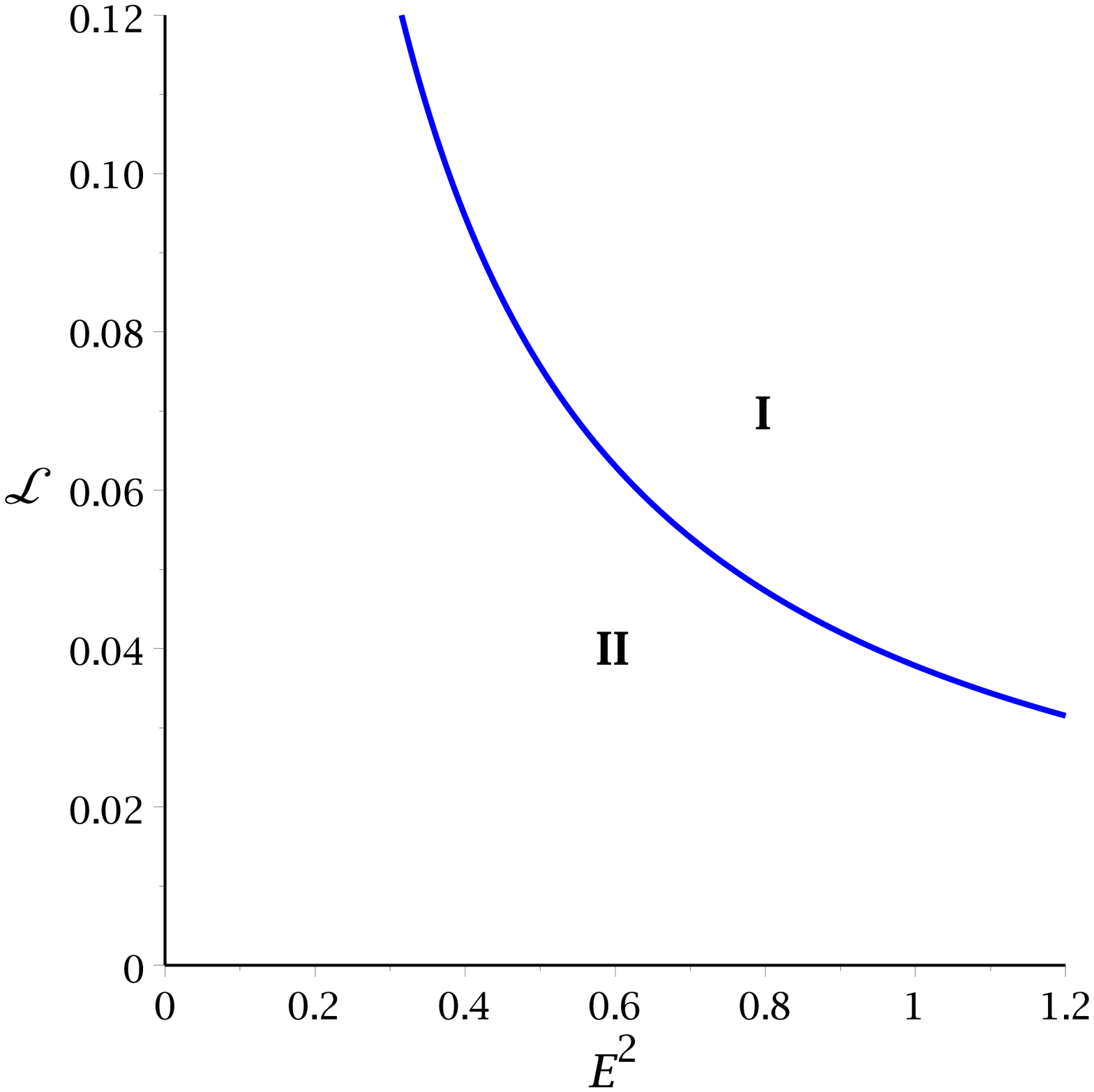}
	}
	\caption{Parametric $\mathcal{L}$-$E^2$-diagrams of the $\tilde{r}$-motion. 
For $\tilde{R}_0>0$ the polynomial  $R(\tilde{r})$ has a single positive zero in region I, three positive zeros in region II 
and five positive zeros in region III. If $\varepsilon =0$ then region III vanishes, implying that 
stable bound orbits for light do not exist outside the horizons. If $\tilde{R}_0<0$ then there are two positive zeros in the regions I,II, and four positive zeros in region III.}
 \label{pic:staticparametric-diagrams}
\end{figure}

\begin{figure}[h]
	\centering
	\subfigure[Effective potential in the range $\tilde{r}\in {[0,250 ]} $ ]{
		\includegraphics[width=0.38\textwidth]{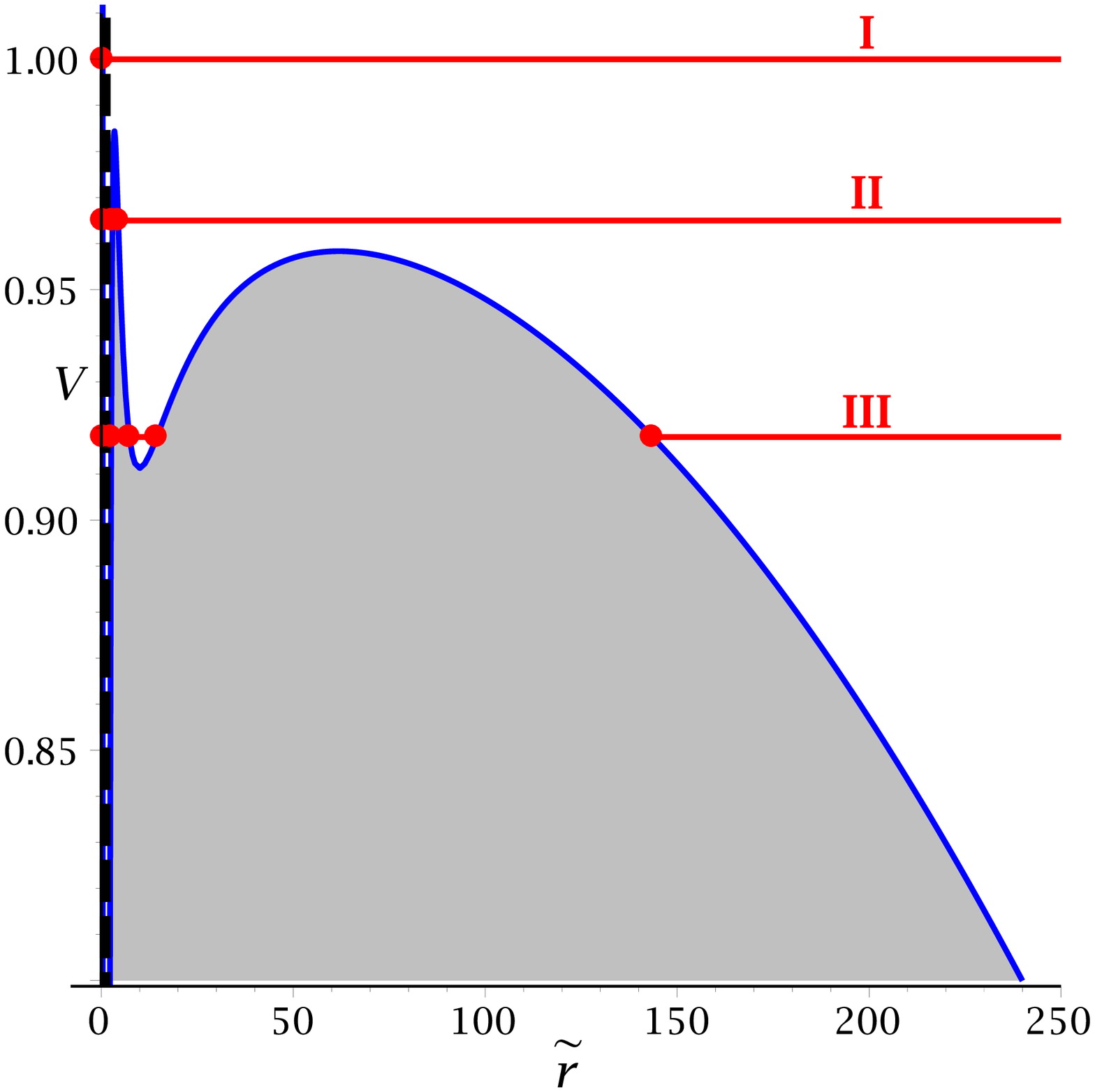}
	}
	\subfigure[Closeup of figure (a)]{
		\includegraphics[width=0.38\textwidth]{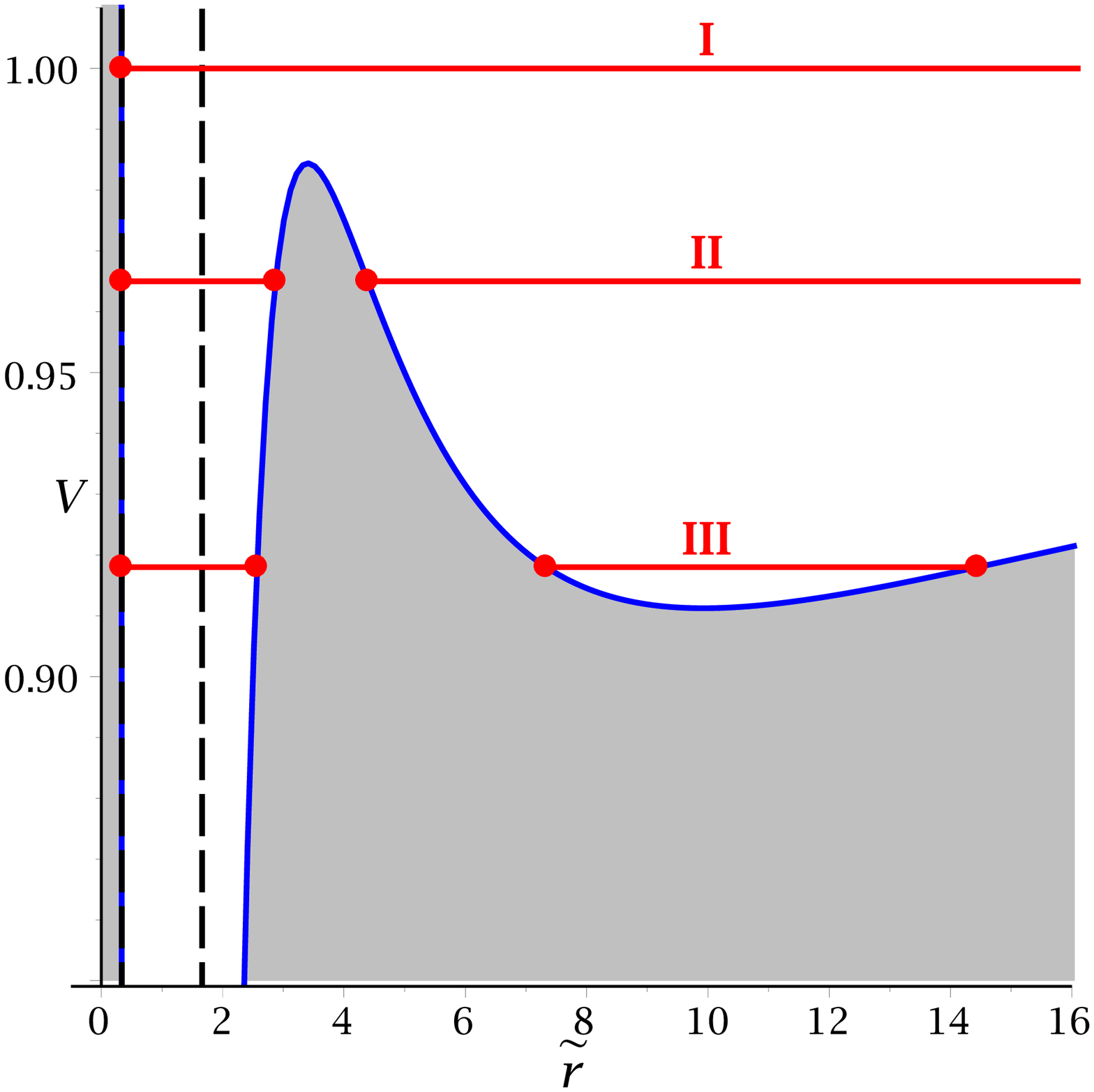}
	}
	\caption{Effective potential (blue) with parameters $\varepsilon=1$, $\tilde{R}_0=\frac{1}{3}\cdot 10^{-5}$, $\tilde{q}=0.75$ and $\mathcal{L}=0.076$. The horizons are depicted as vertical dashed lines. Example energies of region I, II and III (compare figure \ref{pic:parametric-diagrams} and table \ref{tab:orbit-types}) are given as red horizontal lines.}
 \label{pic:potentials}
\end{figure}

In this section, we investigate the possible orbit types in the static case with the help of the analytical solutions which are described in previous sections, parameter diagrams (see Fig.\ref{pic:staticparametric-diagrams}) and the effective potential (see Fig.\ref{pic:potentials}).

In the static case $a=0$ the motion is confined to a plane and therefore the geodesic equations reduce to
\begin{equation}\label{dr/dphi}
(\frac{dr}{d\varphi})^2=\frac{r^4}{L^2}(E^2-(1-\frac{2M}{r}+\frac{q^{2}}{r^{2}}-\frac{1}{12}R_{0}{r}^{2})(\varepsilon+\frac{L^2}{r^2}))=:R(r),
\end{equation}
\begin{equation}\label{dr/dt}
(\frac{dr}{dt})^2=\frac{1}{E^2}(1-\frac{2M}{r}+\frac{q^{2}}{r^{2}}-\frac{1}{12}R_{0}{r}^{2})^2(E^2-(1-\frac{2M}{r}+\frac{q^{2}}{r^{2}}-\frac{1}{12}R_{0}{r}^{2})(\varepsilon+\frac{L^2}{r^2})),
\end{equation}
where we introduced $ q^{2}=\frac{Q^{2}}{(1+f^{'}(R_{o}))} $. An effective potential can be defined as
\begin{equation}
V_{eff}=(1-\frac{2m}{r}+\frac{q^{2}}{r^{2}}-\frac{1}{12}R_{0}{r}^{2})(\varepsilon+\frac{L^2}{r^2}).
\label{eqn:veff}
\end{equation}
The shape of an orbit depends on the energy $E$ and the angular momentum $L$
of the test particle or light ray, as well as the charge $ q $ and the cosmological constant $ \Lambda $.
The mass can be absorbed through a rescaling of the radial coordinate and the parameters
\begin{equation}
\tilde{r}=\frac{r}{M},\qquad
\tilde{q}=\frac{q}{M},\qquad \mathcal{L}=\frac{M^{2}}{L^{2}},\qquad \tilde{R_{0}}=\frac{1}{12}R_{0} M^{2}.
\end{equation}
Thus, Eq.~(\ref{dr/dphi}) can be written as
\begin{equation}\label{dr/dphin}
(\frac{d\tilde{r}}{d\varphi})^2=\varepsilon\tilde{R_{0}}\mathcal{L}\tilde{r}^{6}+((E^{2}-\varepsilon)\mathcal{L}+\tilde{R_{0}})\tilde{r}^{4}+(2\varepsilon\mathcal{L})\tilde{r}^{3}-(1+\varepsilon\mathcal{L}\tilde{q}^{2} )\tilde{r}^{2}+2\tilde{r}-\tilde{q}^{2}=R(\tilde{r}).
\end{equation}

In following we give a list of the possible orbits types. Let $\tilde{r}_-$ be the inner horizon and $\tilde{r}_+$ be the outer event horizon. 
\begin{enumerate}
	\item \textit{Escape orbit} (EO) with range $\tilde{r} \in [r_1, \infty)$ where $r_1>\tilde{r}_+$.
	\item \textit{Two-world escape orbit} (TEO) with range $[r_1, \infty)$ where $0<r_1 < r_-$.
	\item \textit{Bound orbit} (BO) with range $\tilde{r} \in [r_1, r_2]$ with $r_1, r_2  > r_+$.
	\item \textit{Many-world bound orbit} (MBO) with range $\tilde{r} \in [r_1, r_2]$ where $0<r_1 \leq r_-$ and $r_2 \geq r_+$.
	\item \textit{Terminating orbit} (TO) with ranges 
	\begin{enumerate}
	 \item either $\tilde{r} \in [0, \infty)$ (\textit{Terminating escape orbit} -- TEO) 
	 \item or $\tilde{r} \in [0, r_1]$ with $r_1\geq \tilde{r}_+$ (\textit{Terminating bound orbit} -- TBO). 
	\end{enumerate}
	TOs only occur for $q\neq 0$, otherwise the charge will provide a potential barrier preventing the geodesic from reaching the singularity at $\tilde{r}=0$.
\end{enumerate}

These five regular types of geodesic motion correspond to different arrangements
of the real and positive zeros of $ R(r)$ defining the borders of $ R(r) \geq 0 $ or, equivalently,
$ E^2 \geq V_{eff} $.

Eq.~(\ref{dr/dphin}) implies that $ R(\tilde{r}) \geq 0 $ is a necessary condition for the existence of a
geodesic and, thus, that the positive zeros of $ R(\tilde{r}) $ are the turning points of the orbits.
If for a given set of parameters
$ \tilde{R_{0}}, \tilde{q}, \varepsilon, E^{2}, \mathcal{L} $
the polynomial $ R(\tilde{r}) $ has $n$ positive
zeros, then for varying $ E^{2} $ and $L$
this number can only change if two zeros merge to one.
Solving $ R(\tilde{r})=0, \frac{d R(\tilde{r}) }{d\tilde{r}}=0  $ for $ E^{2} $  and $ \mathcal{L} $, for $ \varepsilon=1 $, yields
\begin{equation}\label{LE^2T}
E^{2}=\frac{(\tilde{r}(\tilde{r}-2)+\tilde{q}^{2}-\tilde{R_{0}}\tilde{r}^{4})^{2}}{\tilde{r}^{2}(\tilde{r}^{2}-3\tilde{r}+2\tilde{q}^{2})},\qquad\quad
\mathcal{L}=-\frac{\tilde{r}^{2}-3\tilde{r}+2\tilde{q}^{2}}{\tilde{r}^{2}(\tilde{R_{0}}\tilde{r}^{4}+\tilde{q}^{2}-\tilde{r})}
\end{equation}
and for $ \varepsilon=0 $, yields
\begin{equation}\label{LE^2N}
\mathcal{L}=\frac{1}{E^{2}}
(\frac{2(1+\sqrt{9-8\tilde{q}^{2}})}{(3+\sqrt{9-8\tilde{q}^{2}})^{3}}-
\tilde{R_{0}}).
\end{equation}
In Fig.~\ref{pic:staticparametric-diagrams},
the results of this analysis are shown for both test particles ($ \varepsilon=1 $)
and light rays ($ \varepsilon=0 $).

In the Parametric $\mathcal{L}$-$E^2$-diagrams three regions of geodesic motion with different numbers of zeros can be identified (in the following $r_{i}<r_{i+1}$ is assumed):
\begin{enumerate}
	\item Region I: 
	\begin{enumerate}
		\item $\tilde{R}_0>0$: $R (\tilde{r})$ has a single positive real zero $r_1$ and $ R(\tilde{r})\geq 0 $ for $\tilde{r} \geq r_1$. The only possible orbit type is EO.
		\item $\tilde{R}_0<0$:  $R (\tilde{r})$ has two positive zeros $r_1, r_2$ and $ R(\tilde{r})\geq 0 $ for $ {r_{1}} \leq \tilde{r} \leq {r_{2}} $. The only possible orbit type is MBO.
	\end{enumerate}
	\item Region II: 
	\begin{enumerate}
		\item $\tilde{R}_0>0$: $R (\tilde{r})$ has three positive zeros $r_1, r_2, r_3$ with $ R(\tilde{r})\geq 0 $ for   $ {r_{1}} \leq \tilde{r} \leq {r_{2}} $ and  $ r_{3} \leq\tilde{r} $. Possible orbit types are MBO and EO.
		\item $\tilde{R}_0<0$: $R (\tilde{r})$ has two positive zeros $r_1, r_2$ and $ R(\tilde{r})\geq 0 $ for $ {r_{1}} \leq \tilde{r} \leq {r_{2}} $. The only possible orbit type is MBO.
	\end{enumerate}
	\item Region III: 
	\begin{enumerate}
		\item $\tilde{R}_0>0$: $R (\tilde{r})$ has five positive zeros $r_1, r_2, r_3, r_4, r_5$ with  $ R(\tilde{r})\geq 0 $ for  $ {r_{1}} \leq \tilde{r} \leq {r_{2}} $,
$ {r_{3}} \leq \tilde{r} \leq {r_{4}} $ and $ r_{5} \leq\tilde{r} $. Possible orbit types are MBO, BO and EO.
		\item $\tilde{R}_0<0$: $R (\tilde{r})$ has four positive zeros $r_1, r_2, r_3, r_4$ with $ R(\tilde{r})\geq 0 $ for $ {r_{1}} \leq \tilde{r} \leq {r_{2}} $ and $ {r_{3}} \leq \tilde{r} \leq {r_{4}} $. Possible orbit types are MBO and BO.
	\end{enumerate}
\end{enumerate}

Terminating orbits are possible in all three regions iff the black hole is uncharged $\tilde{q}=0$. For light rays only regions I and II appear and therefore stable bound orbits do not exist for $\varepsilon =0$. A summary of possible orbit types for $\tilde{R}_0>0$ and $\tilde{R}_0<0$
can be found in Tables~\ref{tab:staticorbit-types} and~\ref{tab:staticorbit-types2} respectively. 

\begin{table}[h]
\begin{center}
\begin{tabular}{|cccc|}\hline
zeros & region  & range of $\tilde{r}$ & orbit \\
\hline\hline
1 & I &
\begin{pspicture}(-2.5,-0.2)(4,0.2)
\psline[linewidth=0.5pt]{->}(-2.5,0)(4,0)
\psline[linewidth=0.5pt](-2.5,-0.2)(-2.5,0.2)
\psline[linewidth=0.5pt,doubleline=true](-1.5,-0.2)(-1.5,0.2)
\psline[linewidth=0.5pt,doubleline=true](-0.5,-0.2)(-0.5,0.2)
\psline[linewidth=1.2pt]{*-}(-2,0)(4,0)
\end{pspicture}
  & TEO
\\  \hline
3 & II &
\begin{pspicture}(-2.5,-0.2)(4,0.2)
\psline[linewidth=0.5pt]{->}(-2.5,0)(4,0)
\psline[linewidth=0.5pt](-2.5,-0.2)(-2.5,0.2)
\psline[linewidth=0.5pt,doubleline=true](-1.5,-0.2)(-1.5,0.2)
\psline[linewidth=0.5pt,doubleline=true](-0.5,-0.2)(-0.5,0.2)
\psline[linewidth=1.2pt]{*-*}(-2,0)(0,0)
\psline[linewidth=1.2pt]{*-}(1,0)(4,0)
\end{pspicture}
  & MBO, EO
\\  
 &  &
\begin{pspicture}(-2.5,-0.2)(4,0.2)
\psline[linewidth=0.5pt]{->}(-2.5,0)(4,0)
\psline[linewidth=0.5pt](-2.5,-0.2)(-2.5,0.2)
\psline[linewidth=0.5pt,doubleline=true](-1.5,-0.2)(-1.5,0.2)
\psline[linewidth=0.5pt,doubleline=true](-0.5,-0.2)(-0.5,0.2)
\psline[linewidth=1.2pt]{*-*}(-1.5,0)(-0.5,0)
\psline[linewidth=1.2pt]{*-}(1,0)(4,0)
\end{pspicture}
  & MBO, EO
\\  \hline
5 & III &
\begin{pspicture}(-2.5,-0.2)(4,0.2)
\psline[linewidth=0.5pt]{->}(-2.5,0)(4,0)
\psline[linewidth=0.5pt](-2.5,-0.2)(-2.5,0.2)
\psline[linewidth=0.5pt,doubleline=true](-1.5,-0.2)(-1.5,0.2)
\psline[linewidth=0.5pt,doubleline=true](-0.5,-0.2)(-0.5,0.2)
\psline[linewidth=1.2pt]{*-*}(-2,0)(0,0)
\psline[linewidth=1.2pt]{*-*}(1,0)(2,0)
\psline[linewidth=1.2pt]{*-}(3,0)(4,0)
\end{pspicture}
  & MBO, BO, EO
\\ \hline\hline
\end{tabular}
\caption{Types of orbits in the spacetime of a static charged black hole for $\tilde{q}\neq 0$ and a positive cosmological constant $\tilde{R}_0>0$. The range of the orbits is represented by thick lines. The dots show the turning points of the orbits. The positions of the horizons are marked by vertical double lines. The single vertical line indicates $\tilde{r}=0$. Terminating orbits exist in all three regions only if $\tilde{q}= 0$.}
\label{tab:staticorbit-types}
\end{center}
\end{table}

\begin{table}[h]
\begin{center}
\begin{tabular}{|cccc|}\hline
zeros & region  & range of $\tilde{r}$ & orbit \\
\hline\hline
2 & I, II &
\begin{pspicture}(-2.5,-0.2)(4,0.2)
\psline[linewidth=0.5pt]{->}(-2.5,0)(4,0)
\psline[linewidth=0.5pt](-2.5,-0.2)(-2.5,0.2)
\psline[linewidth=0.5pt,doubleline=true](-1.5,-0.2)(-1.5,0.2)
\psline[linewidth=0.5pt,doubleline=true](-0.5,-0.2)(-0.5,0.2)
\psline[linewidth=1.2pt]{*-*}(-2,0)(0,0)
\end{pspicture}
  & MBO
\\  
 & II &
\begin{pspicture}(-2.5,-0.2)(4,0.2)
\psline[linewidth=0.5pt]{->}(-2.5,0)(4,0)
\psline[linewidth=0.5pt](-2.5,-0.2)(-2.5,0.2)
\psline[linewidth=0.5pt,doubleline=true](-1.5,-0.2)(-1.5,0.2)
\psline[linewidth=0.5pt,doubleline=true](-0.5,-0.2)(-0.5,0.2)
\psline[linewidth=1.2pt]{*-*}(-1.5,0)(-0.5,0)
\end{pspicture}
  & MBO
\\  \hline
4 & III &
\begin{pspicture}(-2.5,-0.2)(4,0.2)
\psline[linewidth=0.5pt]{->}(-2.5,0)(4,0)
\psline[linewidth=0.5pt](-2.5,-0.2)(-2.5,0.2)
\psline[linewidth=0.5pt,doubleline=true](-1.5,-0.2)(-1.5,0.2)
\psline[linewidth=0.5pt,doubleline=true](-0.5,-0.2)(-0.5,0.2)
\psline[linewidth=1.2pt]{*-*}(-2,0)(0,0)
\psline[linewidth=1.2pt]{*-*}(1,0)(2,0)
\end{pspicture}
  & MBO, BO
\\ \hline\hline
\end{tabular}
\caption{Types of orbits in the spacetime of a static charged black hole for $\tilde{q}\neq 0$ and a negative cosmological constant $\tilde{R}_0<0$. The range of the orbits is represented by thick lines. The dots show the turning points of the orbits. The positions of the horizons are marked by vertical double lines. The single vertical line indicates $\tilde{r}=0$. Terminating orbits exist in all three regions only if $\tilde{q}= 0$.}
\label{tab:staticorbit-types2}
\end{center}
\end{table}

For $\tilde{R}_0>0$ a plot of the effective potential introduced in Eq.\eqref{eqn:veff} with example energies corresponding to the different regions is shown in Fig~\ref{pic:potentials}. Here the possible orbit types can be identified.

\subsection{The rotating case}

In this section we analyse the equations of motion in the rotating case (Kerr-Newman-(A)dS spacetime) and investigate the possible orbit types.

\subsubsection{Types of latitudinal motion}
 
In this subsection and the next subsection, we use the function $ \tilde{\Theta}(\theta) $ in Eq.~(\ref{dthetad}) and the polynomial $\tilde{R}(\tilde{r}) $ in Eq.~(\ref{drd}), to determine the possible orbits of light and test particles.

First we substitute $\upsilon = \cos^{2}\theta$ with $ \theta \in [0, 1]$ in the function $\tilde{\Theta}(\theta)$:
 \begin{align}
 \tilde{\Theta}(\upsilon)=(1+\dfrac{\tilde{R}_{0}}{12}\tilde{a}^{2}\upsilon)(\tilde{K}-\varepsilon \tilde{a}^{2} \upsilon)- \left(\tilde{a}^{2}E^{2}(1-\upsilon)-2\tilde{L}\Xi \tilde{a}E+\dfrac{\tilde{L}^{2}\Xi^{2}}{(1-\upsilon)} \right) ,
 \end{align} 
Geodesic motion is possible if $\tilde{\Theta}(\theta)\geqslant 0$, then real values of the coordinate $ \theta $ are obtained. This condition also implies that $\tilde{K}>0$ for all geodesics with $\tilde{R}_0>-\frac{12}{\tilde{a}^2}$, or $\Lambda >-\frac{3}{\tilde{a}^2}$. From the observational side $\Lambda >-\frac{3}{\tilde{a}^2}$ is always true, since the cosmological constant acquires a very small positive value.

\begin{figure}[h]
 \centering
 \includegraphics[width=0.38\textwidth]{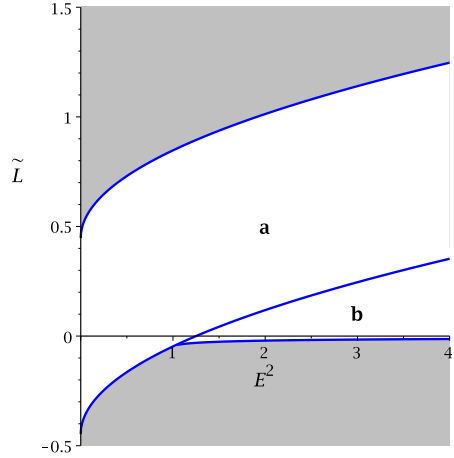}
 \caption{ \label{fig:parameterplot-theta} $\varepsilon =1$, $\tilde{a}=0.4$,  $\tilde{K}=0.2$, $\tilde{R_0}=4\cdot 10^{-5}$: Parametric $\tilde{L}$-$E^2$-diagram for the function  $\tilde{\Theta}$. $\tilde{\Theta}$ possesses one zero in region a and two zeros in region b. In the grey areas geodesic motion is not possible.}
\end{figure}

The number of zeros of $ \tilde{\Theta}(\theta) $, which are the turning points of the latitudinal motion, only changes if a zero crosses 
 $0$ or $1$, or if a double zero occurs. $\upsilon=0$ is a zero of $\tilde{\Theta}$ if
 \begin{align}
 \tilde{\Theta}(\upsilon =0)=\tilde{K}- \big(\tilde{a}^{2}E^{2}-2\tilde{L}\Xi \tilde{a}E+\tilde{L}^{2}\Xi^{2} \big) ,
 \end{align}
and therefore
 \begin{align}
 \tilde{L}=\dfrac{E\tilde{a}\pm \sqrt{\tilde{K}}}{\Xi}.
 \end{align}
 Since $\upsilon=1$ is a pole of $\tilde{\Theta}(\upsilon)$ for $\tilde{L}\neq 0$, it is only possible that $\upsilon=1$ is a zero of $\tilde{\Theta}(\upsilon)$ if $\tilde{L}=0$,
 \begin{align}
 \tilde{\Theta}(\upsilon =1, \tilde{L}=0)=\left(1+\dfrac{\tilde{R}_{0}}{12}\tilde{a}^{2}\right)(\tilde{K}-\varepsilon \tilde{a}^{2}).
 \end{align}
To remove the pole of $\tilde{\Theta}(\upsilon)$ at $\upsilon = 1$ we consider
 \begin{align}\label{thetanoo}
 \tilde{\Theta}^{'}({\upsilon})=(1-\upsilon)\left(1+\dfrac{\tilde{R}_{0}}{12}\tilde{a}^{2}\upsilon\right)(\tilde{K}-\varepsilon \tilde{a}^{2}\upsilon)-\big( \tilde{a}E(1-\upsilon)-\tilde{L}\Xi \big)^{2},
 \end{align}
where $ \tilde{\Theta}({\upsilon})  =\frac{1}{1-\upsilon}\tilde{\Theta}^{'}({\upsilon}) $. Then double zeros fulfill the conditions
 \begin{align}\label{x}
  \tilde{\Theta}^{'}({\upsilon})=0 \qquad  \text{and}  \qquad \frac{ d\tilde{\Theta}^{'}({\upsilon})}{d\upsilon}=0,
 \end{align}
which yields
 \begin{align}
 \tilde{L}=\dfrac{\left( 6E \pm \sqrt{36 E^{2}+3 \tilde{K}\tilde{R}_{0}} \right) (\tilde{R}_{0}\tilde{a}^{2}+12)}{12\tilde{R}_{0}\tilde{a}\Xi}.
 \end{align}
From the condition of $\upsilon=0$ being a zero and the condition of double zeros, we can plot parametric $\tilde{L}$-$E^2$-diagrams, see figure \ref{fig:parameterplot-theta}. These reveal two regions in which geodesic motion is possible. The function $\tilde{\Theta}$ has a single zero $\upsilon_0$ in region a, therefore the geodesics will cross the equatorial plane ($\tilde{K}>(E\tilde{a}-\tilde{L}\Xi)^2$). In region b, the function $\tilde{\Theta}$ has two zeros $\upsilon_1$, $\upsilon_2$, which corresponds to motion above or below the equatorial plane ($\tilde{K}<(E\tilde{a}-\tilde{L}\Xi)^2$). If $\tilde{K}=(E\tilde{a}-\tilde{L}\Xi)^2$, the geodesics will reside in the equatorial plane.

\subsubsection{Types of radial motion}

The zeros of the polynomial $\tilde{R}$ are the turning points of orbits of
light and test particles, and therefore $\tilde{R}$ determines the possible
types of orbits
\begin{align}
\tilde{R}(\tilde{r})=-\Delta_{\tilde{r}}(\varepsilon \tilde{r}^{2}
+\tilde{K})+ \big[(\tilde{a}^{2}
+\tilde{r}^{2})E-\tilde{a}\tilde{L}\Xi \big]^{2},
\end{align}
with
\begin{align}
\Delta_{\tilde{r}}=(\tilde{r}^{2} +\tilde{a}^{2})\left(1+\dfrac{\tilde{R_{0}}}{12}\tilde{r}^{2}\right)
-2\tilde{r}+q^2
\end{align}
where we introduced
\begin{align}
q^2=\frac{\tilde{Q}^{2}}{(1+f'(R_{0}))} \ .
\end{align}

There are \emph{bound orbits}, where test particles move back and forth between two turning points, and \emph{escape orbits}, where the black hole is approached, but the test particles turn around at a certain point to escape towards infinity. \emph{Terminating orbits} end in the singularity, if they reach simultaneously $\tilde{r}=0$ and $\vartheta=\frac{\pi}{2}$, such that $\rho^2=0$.
If a test particle crosses the black hole horizons twice or even multiple times, 
it can enter another universe. These orbits are  called \emph{two-world orbits} or \emph{many-world orbits}. Due to the ring singularity it is possible that a geodesic crosses $\tilde{r}=0$, 
which is then called \emph{transit orbit} or \emph{crossover orbit} \cite{Hackmann:2010zz}. 
Below we give a list of all possible orbits. Let $\tilde{r}_+$ be the outer event horizon and $\tilde{r}_-$ be the inner horizon of the black hole:
\begin{enumerate}
	\item \textit{Transit orbit} (TrO) with range $\tilde{r} \in (-\infty, \infty)$.
	\item \textit{Escape orbit} (EO) with range $\tilde{r} \in [r_1, \infty)$ with $r_1>\tilde{r}_+$, or with range $\tilde{r} \in (-\infty, r_1]$ with  $r_1<0$.
	\item \textit{Two-world escape orbit} (TEO) with range $[r_1, \infty)$ where $0<r_1 < r_-$.
	\item \textit{Crossover two-world escape orbit} (CTEO) with range $[r_1, \infty)$ where $r_1 < 0$.
	\item \textit{Bound orbit} (BO) with range $\tilde{r} \in [r_1, r_2]$ with
	\begin{enumerate}
		\item $r_1, r_2  > r_+$ or 
		\item $ 0 < r_1, r_2 < r_-$
	\end{enumerate}
	\item \textit{Many-world bound orbit} (MBO) with range $\tilde{r} \in [r_1, r_2]$ where $0<r_1 \leq r_-$ and $r_2 \geq r_+$.
	\item \textit{Terminating orbit} (TO) with ranges either $\tilde{r} \in [0, \infty)$ or $\tilde{r} \in [0, r_1]$ with
	\begin{enumerate}
		\item $r_1\geq \tilde{r}_+$ or 
		\item $0<r_1<\tilde{r}_-$
	\end{enumerate}
\end{enumerate}

\begin{figure}[h]
	\centering
	\subfigure[$\varepsilon=1$, $\tilde{a}=0.7$, $\tilde{K}=12$, $R_0=4\cdot 10^{-5}$, $q=0.7$]{
		\includegraphics[width=0.3\textwidth]{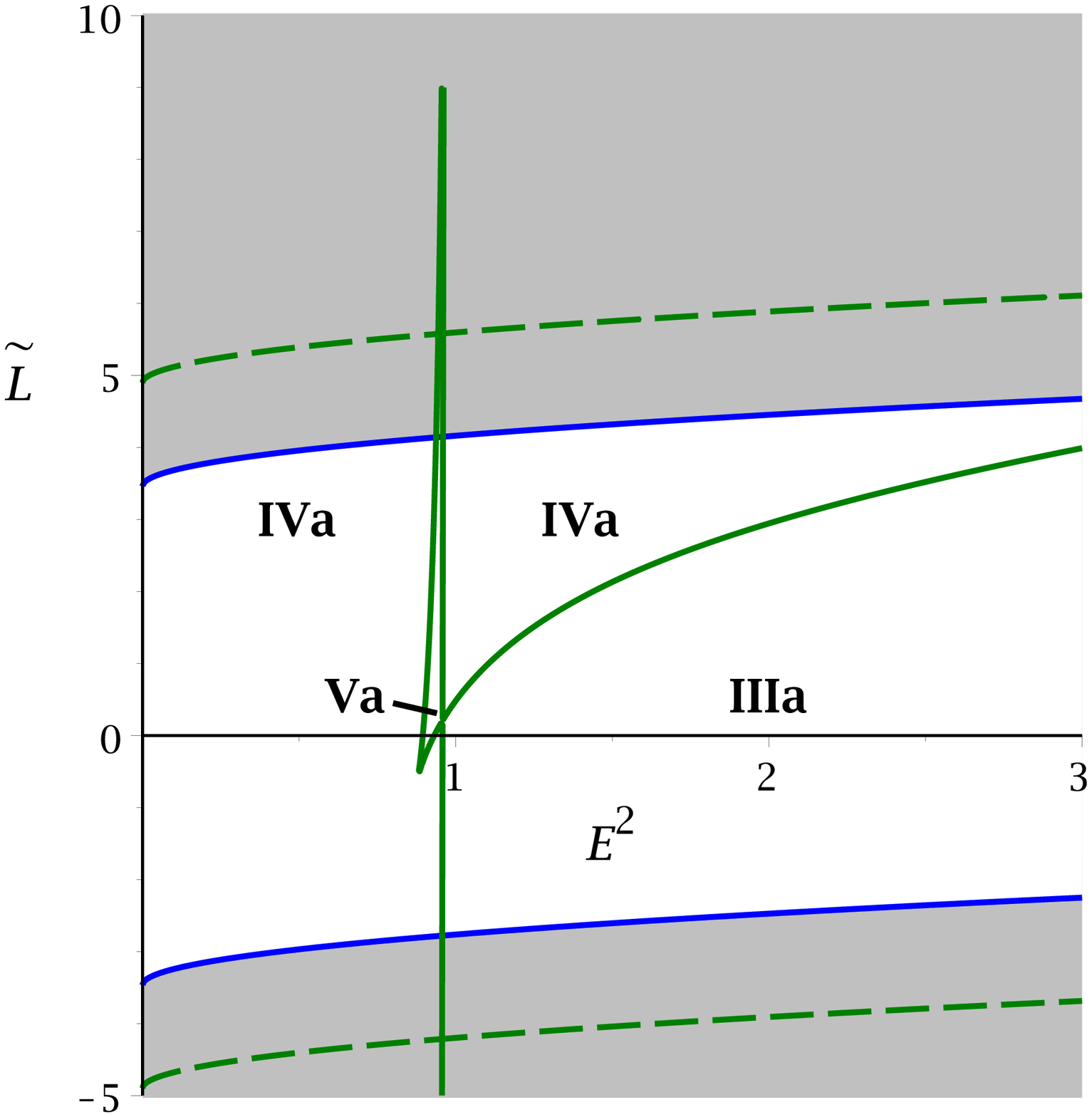}
	}
	\subfigure[$\varepsilon=1$, $\tilde{a}=0.7$, $\tilde{K}=12$, $R_0=4\cdot 10^{-5}$, $q=0.7$]{
		\includegraphics[width=0.3\textwidth]{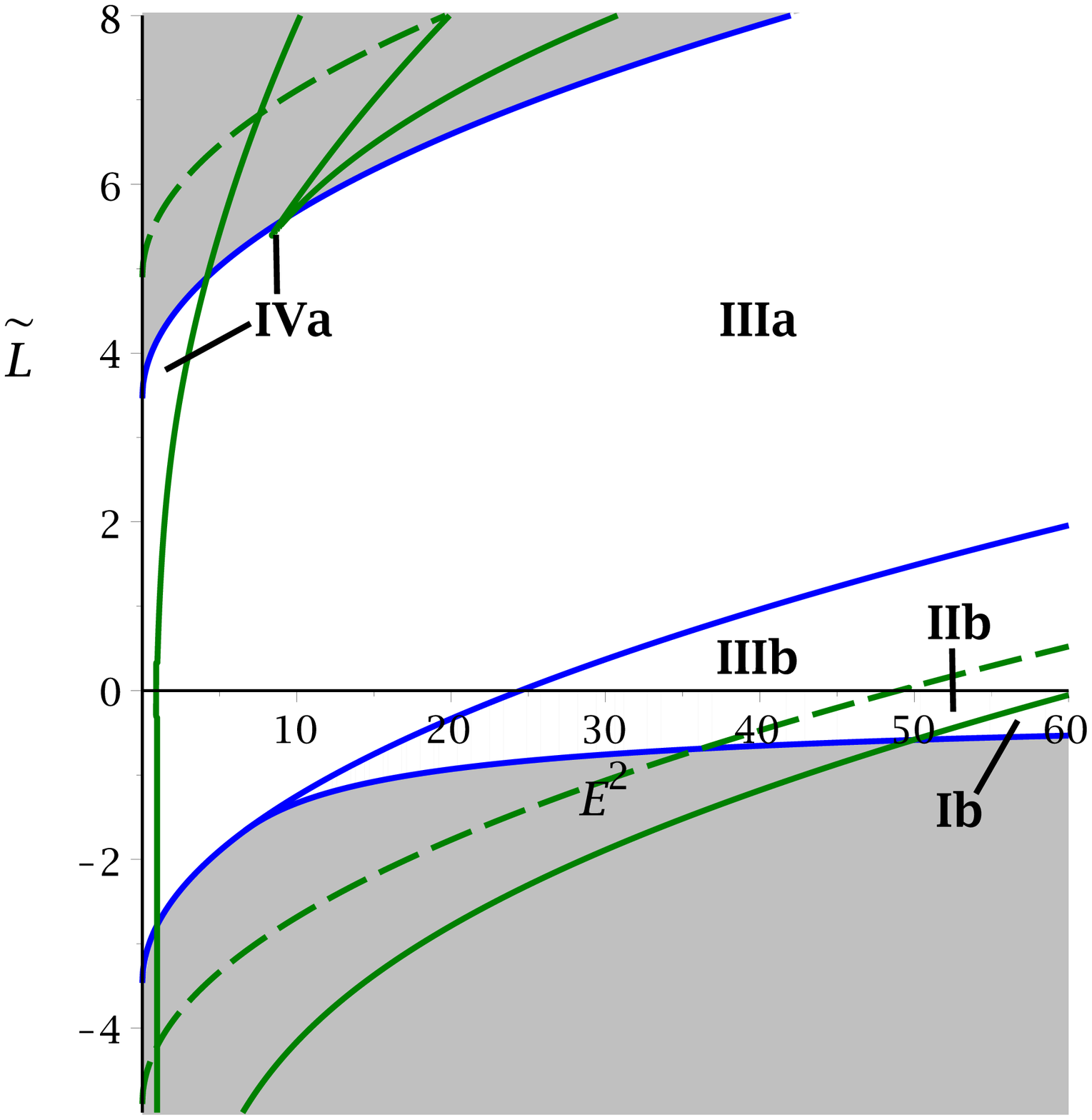}
	}
	\subfigure[$\varepsilon=1$, $\tilde{a}=0.9$, $\tilde{K}=0.3$, $R_0=4\cdot 10^{-5}$, $q=0.2$]{
		\includegraphics[width=0.3\textwidth]{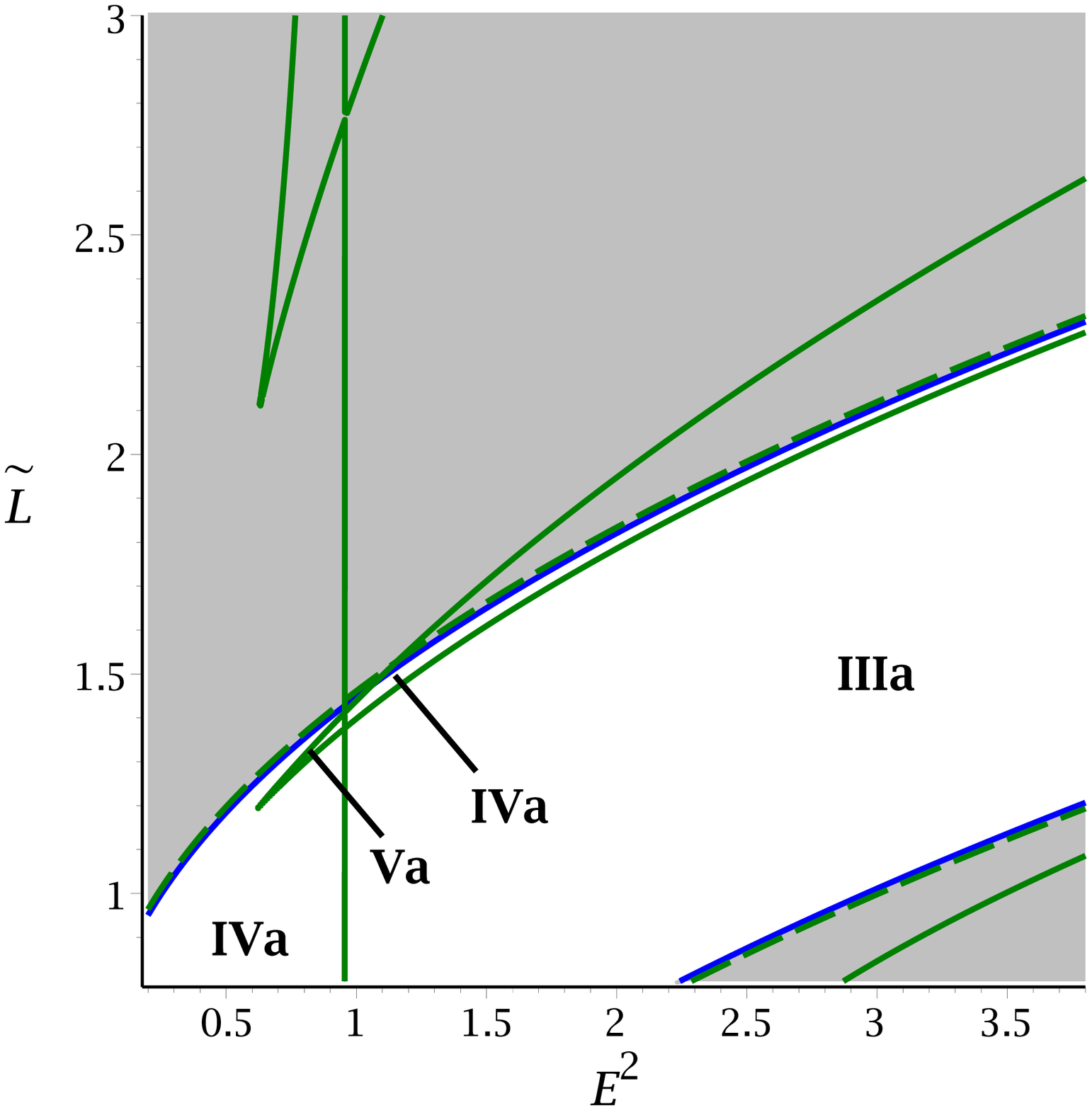}
	}
	\caption{\label{pic:parametric-diagrams} Combined $\tilde{L}$-$E^2$-diagrams of the $\tilde{r}$-motion (green lines) and $\theta$-motion (blue lines). The dashed green lines show, where  $\tilde{R}(\tilde{r}=0)=0$. The polynomial $\tilde{R}$ has no zero in region I, 2 negative zeros in region II, 1 negative and 1 positive zeros in region III, 3 positive and 1 negative zeros in region IV, 5 positive and 1 negative zeros in region V. Inside the grey areas the $\theta$-equation does not allow geodesic motion. In regions marked with the letter a, the orbits cross $\theta=\frac{\pi}{2}$, but not $\tilde{r}=0$. Whereas in regions marked with the letter b, $\tilde{r}=0$ can be crossed but $\theta=\frac{\pi}{2}$ is never crossed.}
\end{figure}

The type of an orbit is determined by the number of real zeros of the polynomial $\tilde{R}$. This number changes if a double zero occurs
\begin{equation}
	\tilde{R}(\tilde{r})=0 \qquad \text{and} \qquad \frac{d \tilde{R}(\tilde{r})}{d\tilde{r}}=0 \, .
	\label{eqn:doublezero}
\end{equation}
Additionally the distinction between positive ($r$) and negative ($r$) zeros is interesting, 
since the geodesics can cross $\tilde{r}=0$. The number of positive and negative zeros changes if $\tilde{R}(\tilde{r}=0)=0$. Taking both these conditions into account, we can plot parametric $\tilde{L}$-$E^2$-diagrams, which show five regions with different numbers of zeros. In region I $\tilde{R}$ has no zeros. In region II there are two negative zeros. A negative and a positive zero are possible in region III. Region IV has three positive and a single negative zero. Five positive zeros  and a negative zero appear in region V.  In figure \ref{pic:parametric-diagrams} examples of the  parametric $\tilde{L}$-$E^2$-diagrams of the $\tilde{r}$-motion can be seen. We combined them with the parametric diagrams of the $\theta$-motion.

The regions I and II intersect only with region b so here the orbits will not cross the equatorial plane. The regions IV and V only intersect with region a, therefore the orbits will cross the equatorial plane. Region III intersects both with the regions a and b.
In the regions I and II the geodesics can cross $\tilde{r}=0$, but in the regions III, IV and V $\tilde{r}=0$ cannot be crossed. 
 
We conclude that the only way for a geodesic to reach the singularity (Terminating Orbit) is $\tilde{R}(\tilde{r}=0)=0$ and $\tilde{\Theta}(\theta=\frac{\pi}{2})=0$. This is the case if $\tilde{K}=(E\tilde{a}-\tilde{L}\Xi)^2$ and additionally $q=0$.\\

\begin{figure}[h]
	\centering
	\subfigure[$\varepsilon=1$, $\tilde{a}=0.7$, $\tilde{K}=12$, $R_0=4\cdot 10^{-5}$, $q=0.7$, $\tilde{L}=0.5$]{
		\includegraphics[width=0.3\textwidth]{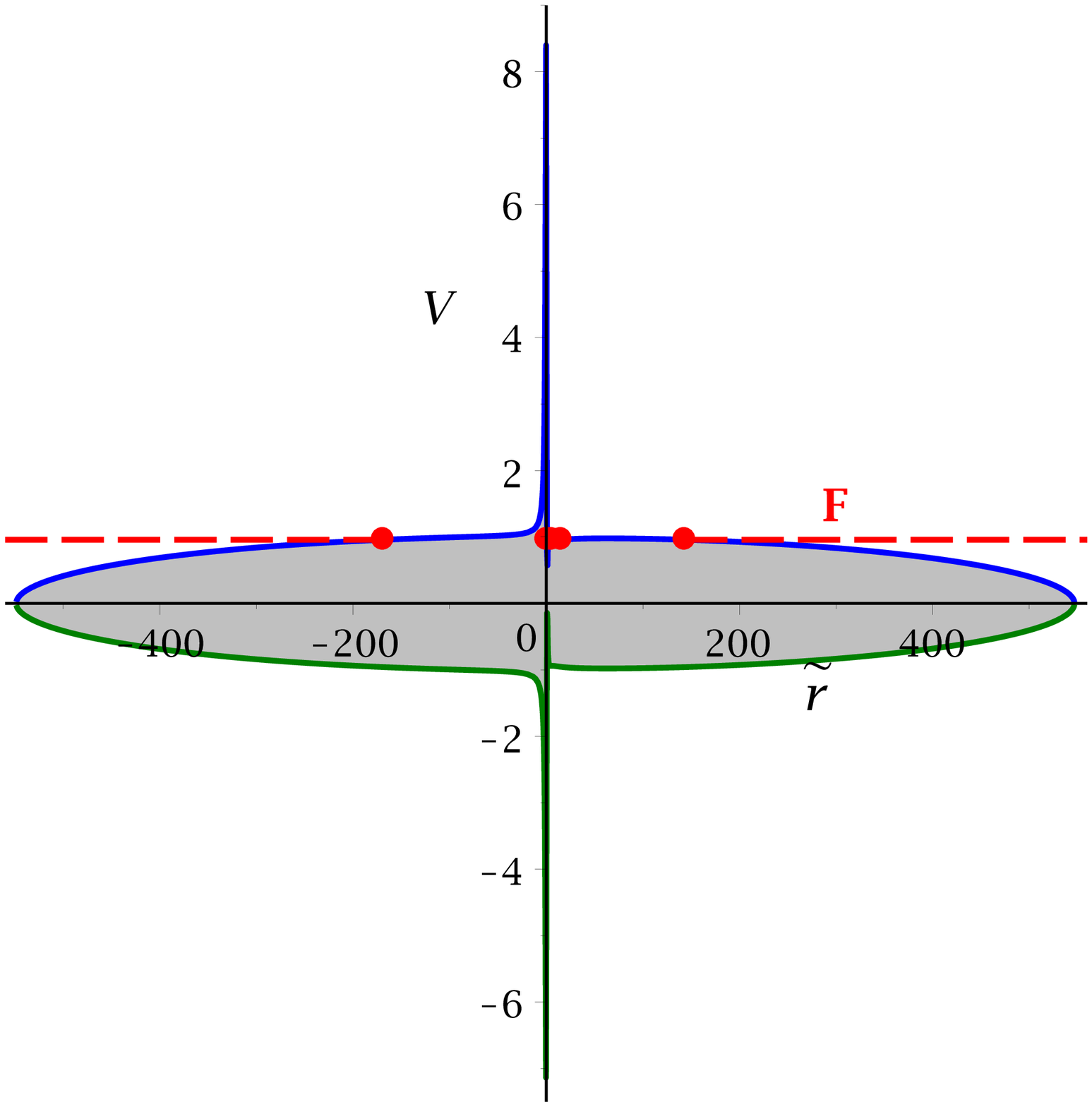}
	}
	\subfigure[closeup of figure (a)]{
		\includegraphics[width=0.3\textwidth]{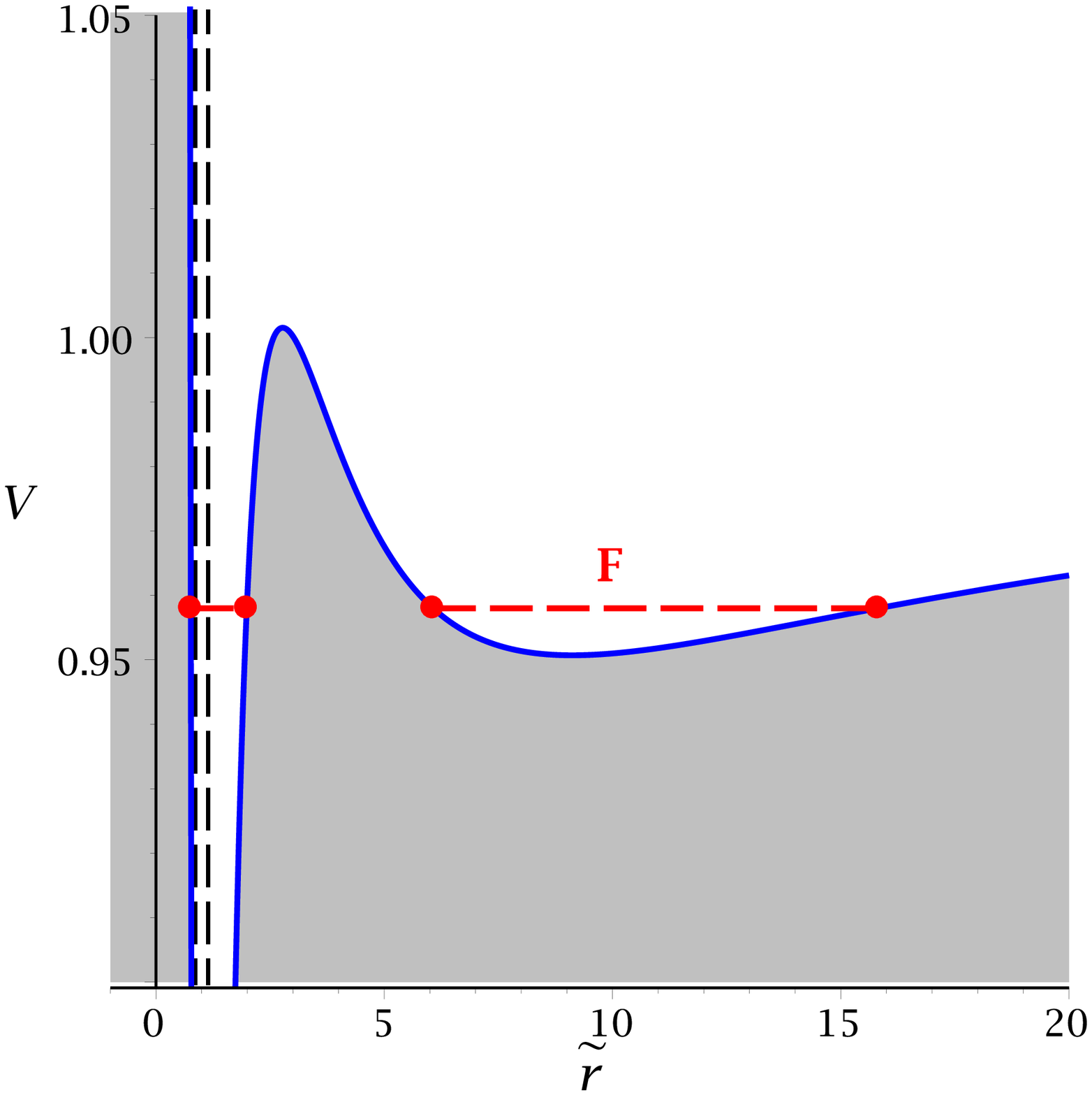}
	}
	\subfigure[$\varepsilon=1$, $\tilde{a}=0.7$, $\tilde{K}=1$, $R_0=4\cdot 10^{-5}$, $q=0.7$, $\tilde{L}=0.5$]{
		\includegraphics[width=0.3\textwidth]{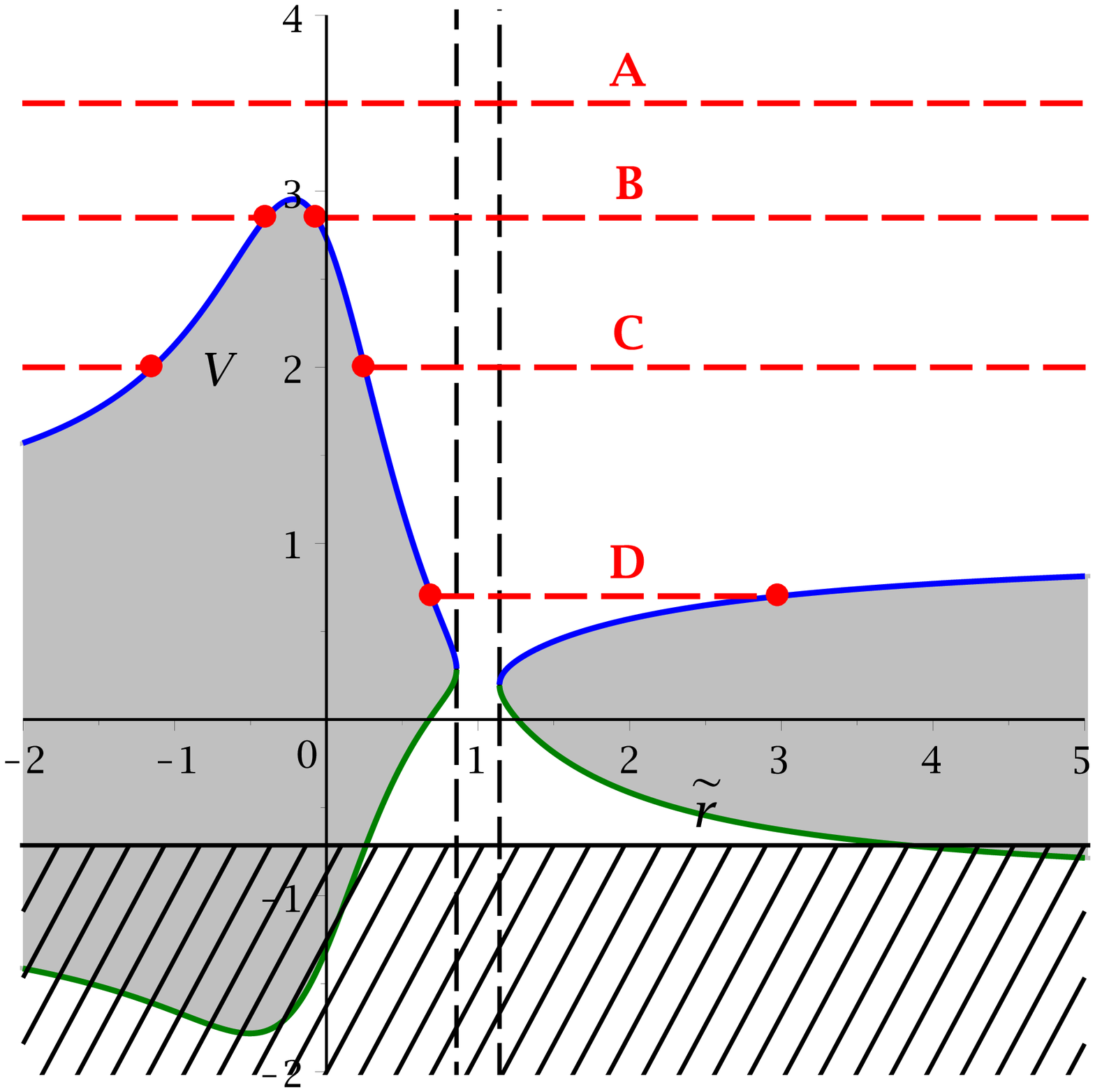}
	}
	\subfigure[$\varepsilon=1$, $\tilde{a}=0.9$, $\tilde{K}=0.3$, $R_0=4\cdot 10^{-5}$, $q=0.3$, $\tilde{L}=1.25$]{
		\includegraphics[width=0.3\textwidth]{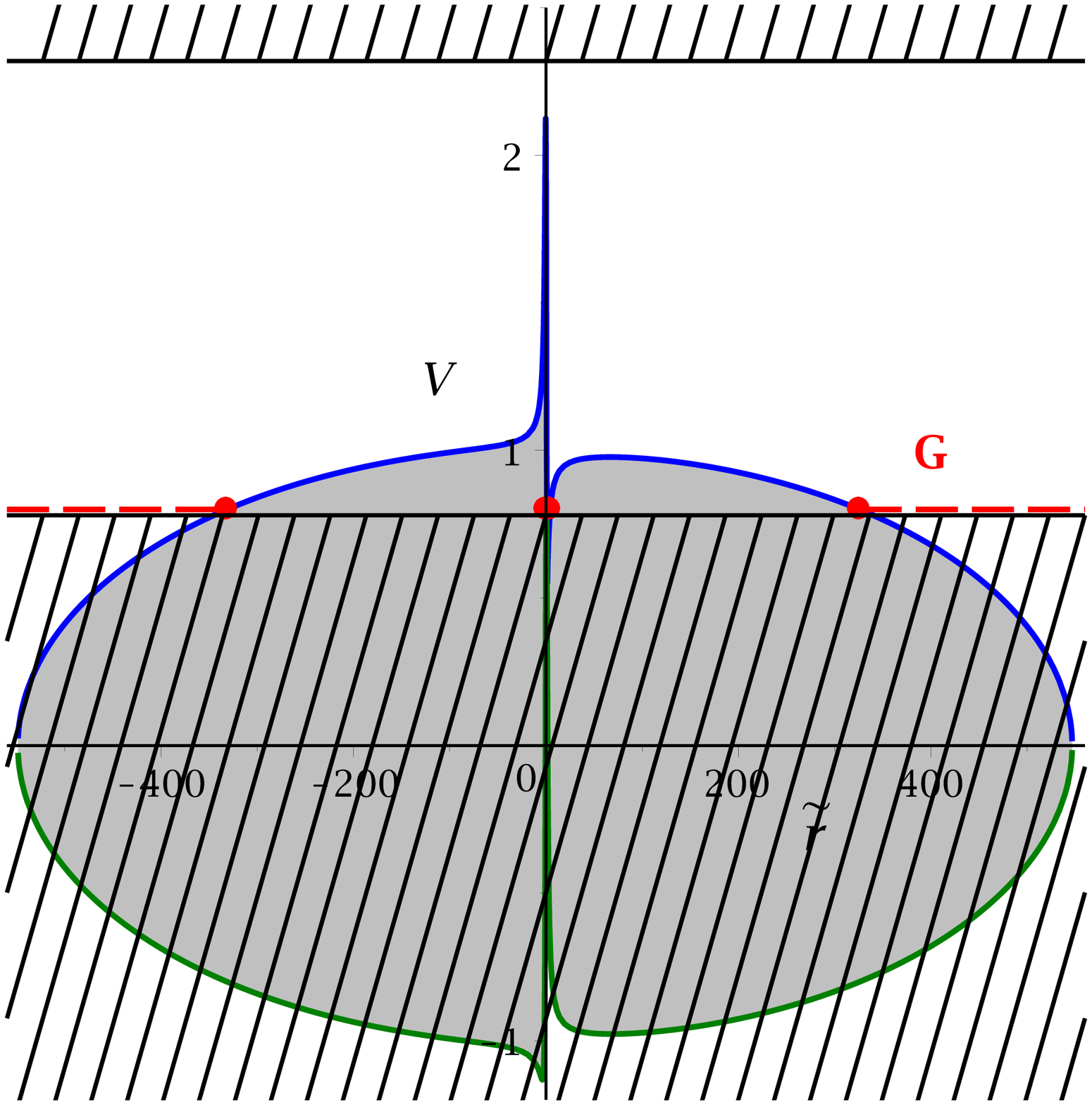}
	}
	\subfigure[closeup of figure (d)]{
		\includegraphics[width=0.3\textwidth]{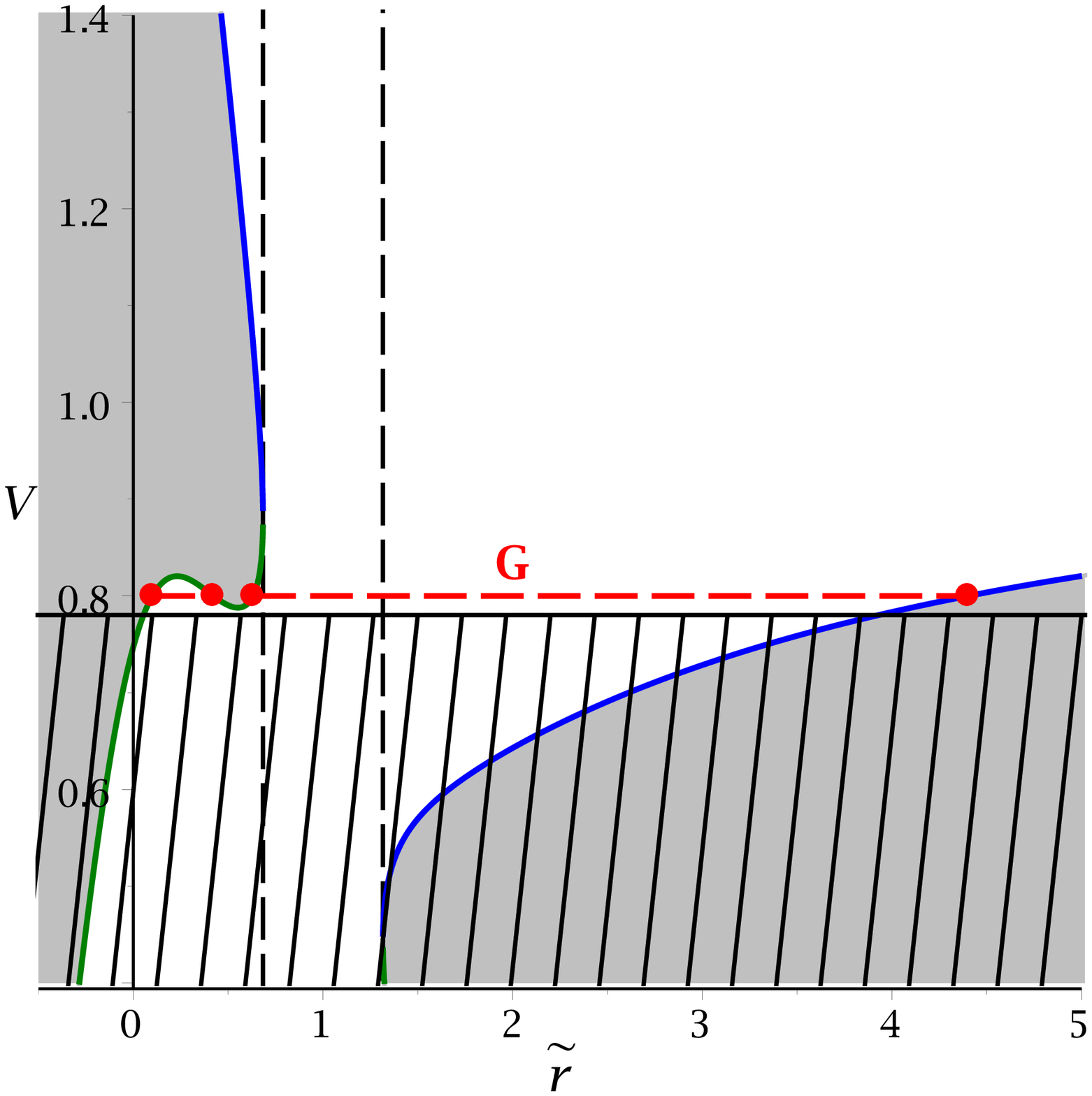}
	}
	\subfigure[$\varepsilon=1$, $\tilde{a}=0.9$, $\tilde{K}=0.3$, $R_0=4\cdot 10^{-5}$, $q=0.2$, $\tilde{L}=1.45$]{
		\includegraphics[width=0.3\textwidth]{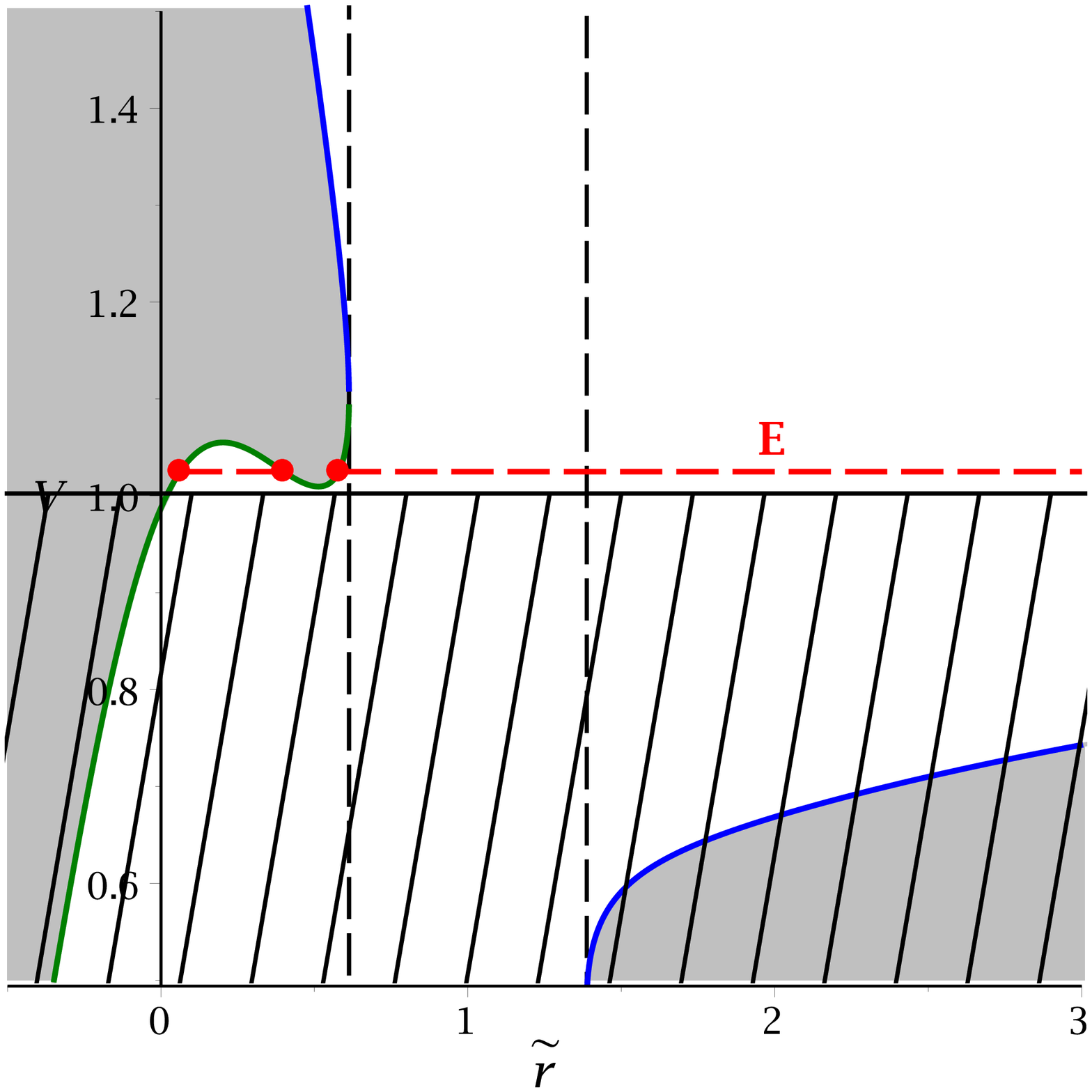}
	}
	\caption{ \label{pic:potential} Plots of the effective potential together with examples of energies for the different orbit types of table \ref{tab:orbit-types}. The blue and green curves represent the two branches of the effective potential. In the grey area the red dashed lines correspond to energies. The red dots mark the zeros of the polynomial $R$, which are the turning points of the orbits. In the grey area no motion is possible since $\tilde{R} <0$. In the dashed area the $\theta$-equation does not allow geodesic motion ($\tilde{\Theta} <0$). The vertical black dashed lines show the position of the horizons.}
\end{figure}

We use the parametric $\tilde{L}$-$E^2$-diagrams and the effective potential of the $\tilde{r}$-equation (see figure \ref{pic:potential}) to determine all possible orbit types. Table \ref{tab:orbit-types} shows an overview. If null geodesics ($\varepsilon =0$) are considered, the region V vanishes from the parametric $\tilde{L}$-$E^2$-diagrams, therefore orbits of type F and G (see table \ref{tab:orbit-types}) are not possible for light. This implies that bound orbits outside the horizons ($\tilde{r}>\tilde{r}_+$) are only possible for particles but not for light.\\

The preceding analysis was done for $\tilde{R}_0 >0$ which implies a positive cosmological constant. Since the motion of test particles and light in this spacetime is similar to the Kerr-(A)dS spacetime we refer to \cite{Hackmann:2010zz} for an analysis concerning a negative cosmogical constant. In comparison with the parametric diagrams of \cite{Hackmann:2010zz} it is obvious that the overall behaviour and the possible orbit types are the same, still, there are some differences to the Kerr-(A)dS spacetime caused by the parameter $q$. In the spacetime of a rotating black hole, there are orbits that do not cross $\tilde{r}=0$ and also do not cross the equatorial plane ($\theta=\frac{\pi}{2}$). This occurs in the region IIIb (see figure \ref{pic:parametric-diagrams} and table \ref{tab:orbit-types}), which is not present for the Kerr-(A)dS case $q=0$. In the Kerr-(A)dS spacetime the green dashed line in figure \ref{pic:parametric-diagrams} will coincide with the blue lines from the $\theta$ parametric plot, so that region III is not splitted into an a and a b part. Therefore, in the Kerr-(A)dS case an orbit that crosses $\tilde{r}=0$ will not cross $\theta=\frac{\pi}{2}$ and an orbit that crosses $\theta=\frac{\pi}{2}$ will not cross $\tilde{r}=0$.

A further difference to the Kerr case is that terminating orbits do not exist for $q\neq 0$. Only orbits with $\tilde{K}=(E\tilde{a}-\tilde{L}\Xi)^2$ and simultaneously $q=0$ end in the singularity.

\begin{table}[h!]
\begin{center}
\begin{tabular}{|lccll|}\hline
type & zeros & region  & range of $\tilde{r}$ & orbit \\
\hline\hline
A & 0 & Ib &
\begin{pspicture}(-4,-0.2)(3.5,0.2)
\psline[linewidth=0.5pt]{->}(-4,0)(3.5,0)
\psline[linewidth=0.5pt](-2.5,-0.2)(-2.5,0.2)
\psline[linewidth=0.5pt,doubleline=true](-0.5,-0.2)(-0.5,0.2)
\psline[linewidth=0.5pt,doubleline=true](1,-0.2)(1,0.2)
\psline[linewidth=1.2pt]{-}(-4,0)(3.5,0)
\end{pspicture}
  & TrO
\\  \hline
B & 2 & IIb &
\begin{pspicture}(-4,-0.2)(3.5,0.2)
\psline[linewidth=0.5pt]{->}(-4,0)(3.5,0)
\psline[linewidth=0.5pt](-2.5,-0.2)(-2.5,0.2)
\psline[linewidth=0.5pt,doubleline=true](-0.5,-0.2)(-0.5,0.2)
\psline[linewidth=0.5pt,doubleline=true](1,-0.2)(1,0.2)
\psline[linewidth=1.2pt]{-*}(-4,0)(-3.5,0)
\psline[linewidth=1.2pt]{*-}(-3,0)(3.5,0)
\end{pspicture}
  & EO, CTEO
\\ \hline
C  & 2 & IIIa,b &
\begin{pspicture}(-4,-0.2)(3.5,0.2)
\psline[linewidth=0.5pt]{->}(-4,0)(3.5,0)
\psline[linewidth=0.5pt](-2.5,-0.2)(-2.5,0.2)
\psline[linewidth=0.5pt,doubleline=true](-0.5,-0.2)(-0.5,0.2)
\psline[linewidth=0.5pt,doubleline=true](1,-0.2)(1,0.2)
\psline[linewidth=1.2pt]{-*}(-4,0)(-3,0)
\psline[linewidth=1.2pt]{*-}(-1,0)(3.5,0)
\end{pspicture}
& EO, TEO
\\
C$_-$  &  &  &
\begin{pspicture}(-4,-0.2)(3.5,0.2)
\psline[linewidth=0.5pt]{->}(-4,0)(3.5,0)
\psline[linewidth=0.5pt](-2.5,-0.2)(-2.5,0.2)
\psline[linewidth=0.5pt,doubleline=true](-0.5,-0.2)(-0.5,0.2)
\psline[linewidth=0.5pt,doubleline=true](1,-0.2)(1,0.2)
\psline[linewidth=1.2pt]{-*}(-4,0)(-3,0)
\psline[linewidth=1.2pt]{*-}(-0.5,0)(3.5,0)
\end{pspicture}
& EO, TEO
\\
C$_0$ &  & &
\begin{pspicture}(-4,-0.2)(3.5,0.2)
\psline[linewidth=0.5pt]{->}(-4,0)(3.5,0)
\psline[linewidth=0.5pt](-2.5,-0.2)(-2.5,0.2)
\psline[linewidth=0.5pt,doubleline=true](-0.5,-0.2)(-0.5,0.2)
\psline[linewidth=0.5pt,doubleline=true](1,-0.2)(1,0.2)
\psline[linewidth=1.2pt]{-*}(-4,0)(-3.5,0)
\psline[linewidth=1.2pt]{*-}(-2.5,0)(3.5,0)
\end{pspicture}
  & EO, TO/TEO
\\ \hline
D & 4 & IVa &
\begin{pspicture}(-4,-0.2)(3.5,0.2)
\psline[linewidth=0.5pt]{->}(-4,0)(3.5,0)
\psline[linewidth=0.5pt](-2.5,-0.2)(-2.5,0.2)
\psline[linewidth=0.5pt,doubleline=true](-0.5,-0.2)(-0.5,0.2)
\psline[linewidth=0.5pt,doubleline=true](1,-0.2)(1,0.2)
\psline[linewidth=1.2pt]{-*}(-4,0)(-3,0)
\psline[linewidth=1.2pt]{*-*}(-1,0)(1.5,0)
\psline[linewidth=1.2pt]{*-}(2,0)(3.5,0)
\end{pspicture}
  & EO, MBO, EO
\\ 
D$_-$ &  &  &
\begin{pspicture}(-4,-0.2)(3.5,0.2)
\psline[linewidth=0.5pt]{->}(-4,0)(3.5,0)
\psline[linewidth=0.5pt](-2.5,-0.2)(-2.5,0.2)
\psline[linewidth=0.5pt,doubleline=true](-0.5,-0.2)(-0.5,0.2)
\psline[linewidth=0.5pt,doubleline=true](1,-0.2)(1,0.2)
\psline[linewidth=1.2pt]{-*}(-4,0)(-3,0)
\psline[linewidth=1.2pt]{*-*}(-0.5,0)(1.5,0)
\psline[linewidth=1.2pt]{*-}(2,0)(3.5,0)
\end{pspicture}
  & EO, MBO, EO
\\
D$_+$ &  &  &
\begin{pspicture}(-4,-0.2)(3.5,0.2)
\psline[linewidth=0.5pt]{->}(-4,0)(3.5,0)
\psline[linewidth=0.5pt](-2.5,-0.2)(-2.5,0.2)
\psline[linewidth=0.5pt,doubleline=true](-0.5,-0.2)(-0.5,0.2)
\psline[linewidth=0.5pt,doubleline=true](1,-0.2)(1,0.2)
\psline[linewidth=1.2pt]{-*}(-4,0)(-3,0)
\psline[linewidth=1.2pt]{*-*}(-1,0)(1,0)
\psline[linewidth=1.2pt]{*-}(2,0)(3.5,0)
\end{pspicture}
  & EO, MBO, EO
\\
D$_\pm$ &  &  &
\begin{pspicture}(-4,-0.2)(3.5,0.2)
\psline[linewidth=0.5pt]{->}(-4,0)(3.5,0)
\psline[linewidth=0.5pt](-2.5,-0.2)(-2.5,0.2)
\psline[linewidth=0.5pt,doubleline=true](-0.5,-0.2)(-0.5,0.2)
\psline[linewidth=0.5pt,doubleline=true](1,-0.2)(1,0.2)
\psline[linewidth=1.2pt]{-*}(-4,0)(-3,0)
\psline[linewidth=1.2pt]{*-*}(-0.5,0)(1,0)
\psline[linewidth=1.2pt]{*-}(2,0)(3.5,0)
\end{pspicture}
  & EO, MBO, EO
\\
D$_0$ &  &  &
\begin{pspicture}(-4,-0.2)(3.5,0.2)
\psline[linewidth=0.5pt]{->}(-4,0)(3.5,0)
\psline[linewidth=0.5pt](-2.5,-0.2)(-2.5,0.2)
\psline[linewidth=0.5pt,doubleline=true](-0.5,-0.2)(-0.5,0.2)
\psline[linewidth=0.5pt,doubleline=true](1,-0.2)(1,0.2)
\psline[linewidth=1.2pt]{-*}(-4,0)(-3,0)
\psline[linewidth=1.2pt]{*-*}(-2.5,0)(1.5,0)
\psline[linewidth=1.2pt]{*-}(2,0)(3.5,0)
\end{pspicture}
  & EO, TO/MBO, EO
\\
D$_0+$ &  &  &
\begin{pspicture}(-4,-0.2)(3.5,0.2)
\psline[linewidth=0.5pt]{->}(-4,0)(3.5,0)
\psline[linewidth=0.5pt](-2.5,-0.2)(-2.5,0.2)
\psline[linewidth=0.5pt,doubleline=true](-0.5,-0.2)(-0.5,0.2)
\psline[linewidth=0.5pt,doubleline=true](1,-0.2)(1,0.2)
\psline[linewidth=1.2pt]{-*}(-4,0)(-3,0)
\psline[linewidth=1.2pt]{*-*}(-2.5,0)(1.0,0)
\psline[linewidth=1.2pt]{*-}(2,0)(3.5,0)
\end{pspicture}
  & EO, TO/MBO, EO
\\ \hline
E & 4 & IVa &
\begin{pspicture}(-4,-0.2)(3.5,0.2)
\psline[linewidth=0.5pt]{->}(-4,0)(3.5,0)
\psline[linewidth=0.5pt](-2.5,-0.2)(-2.5,0.2)
\psline[linewidth=0.5pt,doubleline=true](-0.5,-0.2)(-0.5,0.2)
\psline[linewidth=0.5pt,doubleline=true](1,-0.2)(1,0.2)
\psline[linewidth=1.2pt]{-*}(-4,0)(-3,0)
\psline[linewidth=1.2pt]{*-*}(-2,0)(-1.5,0)
\psline[linewidth=1.2pt]{*-}(-1,0)(3.5,0)
\end{pspicture}
  & EO, BO, TEO
\\
E$_-$ &  &  &
\begin{pspicture}(-4,-0.2)(3.5,0.2)
\psline[linewidth=0.5pt]{->}(-4,0)(3.5,0)
\psline[linewidth=0.5pt](-2.5,-0.2)(-2.5,0.2)
\psline[linewidth=0.5pt,doubleline=true](-0.5,-0.2)(-0.5,0.2)
\psline[linewidth=0.5pt,doubleline=true](1,-0.2)(1,0.2)
\psline[linewidth=1.2pt]{-*}(-4,0)(-3,0)
\psline[linewidth=1.2pt]{*-*}(-2,0)(-1.5,0)
\psline[linewidth=1.2pt]{*-}(-0.5,0)(3.5,0)
\end{pspicture}
  & EO, BO, TEO
\\
E$_0$ &  &  &
\begin{pspicture}(-4,-0.2)(3.5,0.2)
\psline[linewidth=0.5pt]{->}(-4,0)(3.5,0)
\psline[linewidth=0.5pt](-2.5,-0.2)(-2.5,0.2)
\psline[linewidth=0.5pt,doubleline=true](-0.5,-0.2)(-0.5,0.2)
\psline[linewidth=0.5pt,doubleline=true](1,-0.2)(1,0.2)
\psline[linewidth=1.2pt]{-*}(-4,0)(-3,0)
\psline[linewidth=1.2pt]{*-*}(-2.5,0)(-1.5,0)
\psline[linewidth=1.2pt]{*-}(-1,0)(3.5,0)
\end{pspicture}
  & EO, TO/BO, TEO
\\
E$_0-$ &  &  &
\begin{pspicture}(-4,-0.2)(3.5,0.2)
\psline[linewidth=0.5pt]{->}(-4,0)(3.5,0)
\psline[linewidth=0.5pt](-2.5,-0.2)(-2.5,0.2)
\psline[linewidth=0.5pt,doubleline=true](-0.5,-0.2)(-0.5,0.2)
\psline[linewidth=0.5pt,doubleline=true](1,-0.2)(1,0.2)
\psline[linewidth=1.2pt]{-*}(-4,0)(-3,0)
\psline[linewidth=1.2pt]{*-*}(-2.5,0)(-1.5,0)
\psline[linewidth=1.2pt]{*-}(-0.5,0)(3.5,0)
\end{pspicture}
  & EO, TO/BO, TEO
\\ \hline
F & 6 & Va &
\begin{pspicture}(-4,-0.2)(3.5,0.2)
\psline[linewidth=0.5pt]{->}(-4,0)(3.5,0)
\psline[linewidth=0.5pt](-2.5,-0.2)(-2.5,0.2)
\psline[linewidth=0.5pt,doubleline=true](-0.5,-0.2)(-0.5,0.2)
\psline[linewidth=0.5pt,doubleline=true](1,-0.2)(1,0.2)
\psline[linewidth=1.2pt]{-*}(-4,0)(-3,0)
\psline[linewidth=1.2pt]{*-*}(-1,0)(1.5,0)
\psline[linewidth=1.2pt]{*-*}(2,0)(2.5,0)
\psline[linewidth=1.2pt]{*-}(3,0)(3.5,0)
\end{pspicture}
& EO, MBO, BO, EO
\\
F$_-$ &  &  &
\begin{pspicture}(-4,-0.2)(3.5,0.2)
\psline[linewidth=0.5pt]{->}(-4,0)(3.5,0)
\psline[linewidth=0.5pt](-2.5,-0.2)(-2.5,0.2)
\psline[linewidth=0.5pt,doubleline=true](-0.5,-0.2)(-0.5,0.2)
\psline[linewidth=0.5pt,doubleline=true](1,-0.2)(1,0.2)
\psline[linewidth=1.2pt]{-*}(-4,0)(-3,0)
\psline[linewidth=1.2pt]{*-*}(-0.5,0)(1.5,0)
\psline[linewidth=1.2pt]{*-*}(2,0)(2.5,0)
\psline[linewidth=1.2pt]{*-}(3,0)(3.5,0)
\end{pspicture}
& EO, MBO, BO, EO
\\
F$_+$ &  &  &
\begin{pspicture}(-4,-0.2)(3.5,0.2)
\psline[linewidth=0.5pt]{->}(-4,0)(3.5,0)
\psline[linewidth=0.5pt](-2.5,-0.2)(-2.5,0.2)
\psline[linewidth=0.5pt,doubleline=true](-0.5,-0.2)(-0.5,0.2)
\psline[linewidth=0.5pt,doubleline=true](1,-0.2)(1,0.2)
\psline[linewidth=1.2pt]{-*}(-4,0)(-3,0)
\psline[linewidth=1.2pt]{*-*}(-1,0)(1.0,0)
\psline[linewidth=1.2pt]{*-*}(2,0)(2.5,0)
\psline[linewidth=1.2pt]{*-}(3,0)(3.5,0)
\end{pspicture}
& EO, MBO, BO, EO
\\ \hline
G & 6 & Va &
\begin{pspicture}(-4,-0.2)(3.5,0.2)
\psline[linewidth=0.5pt]{->}(-4,0)(3.5,0)
\psline[linewidth=0.5pt](-2.5,-0.2)(-2.5,0.2)
\psline[linewidth=0.5pt,doubleline=true](-0.5,-0.2)(-0.5,0.2)
\psline[linewidth=0.5pt,doubleline=true](1,-0.2)(1,0.2)
\psline[linewidth=1.2pt]{-*}(-4,0)(-3,0)
\psline[linewidth=1.2pt]{*-*}(-2,0)(-1.5,0)
\psline[linewidth=1.2pt]{*-*}(-1,0)(1.5,0)
\psline[linewidth=1.2pt]{*-}(2,0)(3.5,0)
\end{pspicture}
& EO, BO, MBO, EO
\\
G$_-$ &  &  &
\begin{pspicture}(-4,-0.2)(3.5,0.2)
\psline[linewidth=0.5pt]{->}(-4,0)(3.5,0)
\psline[linewidth=0.5pt](-2.5,-0.2)(-2.5,0.2)
\psline[linewidth=0.5pt,doubleline=true](-0.5,-0.2)(-0.5,0.2)
\psline[linewidth=0.5pt,doubleline=true](1,-0.2)(1,0.2)
\psline[linewidth=1.2pt]{-*}(-4,0)(-3,0)
\psline[linewidth=1.2pt]{*-*}(-2,0)(-1.5,0)
\psline[linewidth=1.2pt]{*-*}(-0.5,0)(1.5,0)
\psline[linewidth=1.2pt]{*-}(2,0)(3.5,0)
\end{pspicture}
& EO, BO, MBO, EO
\\
G$_0$ &  &  &
\begin{pspicture}(-4,-0.2)(3.5,0.2)
\psline[linewidth=0.5pt]{->}(-4,0)(3.5,0)
\psline[linewidth=0.5pt](-2.5,-0.2)(-2.5,0.2)
\psline[linewidth=0.5pt,doubleline=true](-0.5,-0.2)(-0.5,0.2)
\psline[linewidth=0.5pt,doubleline=true](1,-0.2)(1,0.2)
\psline[linewidth=1.2pt]{-*}(-4,0)(-3,0)
\psline[linewidth=1.2pt]{*-*}(-2.5,0)(-1.5,0)
\psline[linewidth=1.2pt]{*-*}(-1,0)(1.5,0)
\psline[linewidth=1.2pt]{*-}(2,0)(3.5,0)
\end{pspicture}
& EO, TO/BO, MBO, EO
\\
G$_0-$ &  &  &
\begin{pspicture}(-4,-0.2)(3.5,0.2)
\psline[linewidth=0.5pt]{->}(-4,0)(3.5,0)
\psline[linewidth=0.5pt](-2.5,-0.2)(-2.5,0.2)
\psline[linewidth=0.5pt,doubleline=true](-0.5,-0.2)(-0.5,0.2)
\psline[linewidth=0.5pt,doubleline=true](1,-0.2)(1,0.2)
\psline[linewidth=1.2pt]{-*}(-4,0)(-3,0)
\psline[linewidth=1.2pt]{*-*}(-2.5,0)(-1.5,0)
\psline[linewidth=1.2pt]{*-*}(-0.5,0)(1.5,0)
\psline[linewidth=1.2pt]{*-}(2,0)(3.5,0)
\end{pspicture}
& EO, TO/BO, MBO, EO
\\ \hline\hline
\end{tabular}
\caption{\label{tab:orbit-types} Types of orbits in the Kerr-Newman-dS-spacetime ($\tilde{R}_0>0$). The range of the orbits is represented by thick lines. The dots show the turning points of the orbits. The positions of the event horizon and the Cauchy horizon are marked by a vertical double line. The cosmological horizon is not displayed here since it is not relevant for the orbits. The single vertical line indicates $\tilde{r}=0$.}
\end{center}
\end{table}


\section{ANALYTICAL SOLUTION OF THE GEODESIC EQUATIONS}\label{analytical solutions}

In this section, we will present the analytical solutions of the geodesic equations (\ref{drd})--(\ref{dtd}) in the Kerr-Newman-(A)dS spacetime. We will treat each equation separately and  give the solutions in terms of the Weierstrass $\wp$, $\zeta$ and $\sigma$ functions as well as the Kleinian $\sigma$ function.

\subsection{$\theta$ motion}\label{thetam}
We start with the differential equation (\ref{dthetad}) describing the $\theta$ motion
\begin{align}
\left(\dfrac{d\theta}{d\gamma} \right)^{2}=
\tilde{\Theta}(\theta)=\Delta_{\theta}(\tilde{K}-\varepsilon
\tilde{a}^{2}\cos^{2}\theta)-\dfrac{1}{\sin^{2}\theta} \left(
\tilde{a}E \sin^{2}\theta -\tilde{L}\Xi \right)^{2},
\end{align}
and substitute $\upsilon=cos^{2}\theta$ to simplify the equation
\begin{align}\label{doo}
\left(\dfrac{d\upsilon}{d\gamma}\right)^{2}=4\upsilon \tilde{\Theta}^{'}({\upsilon})=4\upsilon(1-\upsilon)\left(1+\dfrac{\tilde{R}_{0}}{12}\tilde{a}^{2}\upsilon\right)(\tilde{K}-\varepsilon \tilde{a}^{2}\upsilon)-4\upsilon\big( \tilde{a}E(1-\upsilon)-\tilde{L}\Xi \big)^{2}.
\end{align}

\subsubsection{Timelike geodesics}

The differential equation (\ref{doo}) is of elliptic type, since $4\upsilon \tilde{\Theta}^{'}({\upsilon})$ is in general a polynomial of order four. Here we consider the case $ \varepsilon = 1 $. Assuming that $\tilde{\Theta}^{'}({\upsilon})$
has only simple zeros, equation (\ref{doo}) can be solved in terms of the
Weierstrass elliptic $\wp$ function. To get the solution we
transform $4\upsilon \tilde{\Theta}^{'}({\upsilon})$ into the Weierstrass
form $(4y^{3}-g_{2}y-g_{3})$ with the constants $g_{2}$ and $g_{3}$.
First, we apply the substitution $\upsilon=\xi^{-1}$ yielding
\begin{align}\label{dkhi}
\left(\dfrac{d\xi}{d\gamma}\right)^{2}=\tilde{\Theta}_{\xi},
\end{align}
where
\begin{align}\label{khiii}
\tilde{\Theta}_{\xi}=:4\xi^{3} \left[ \tilde{K}-(E\tilde{a}
-\tilde{L}\Xi)^{2} \right] +4\xi^{2} \left[
\tilde{a}^{2}(-\varepsilon+\frac{1}{12}\tilde{K}\tilde{R}_{0}
+2E^{2})-(\tilde{K}+2E\tilde{L}\Xi \tilde{a}) \right] \nonumber\\
+4\xi \left[ \tilde{a}^{2}(-\frac{1}{12}\tilde{K}\tilde{R}_{0}
+\varepsilon -E^{2}-\frac{1}{12}\tilde{R}_{0}\tilde{a}^{2}
\varepsilon) \right] +\frac{1}{3}\tilde{R}_{0}\tilde{a}^{4}\varepsilon
=:\sum_{i=1}^3 a_{i}\xi^{i}
\end{align}
is a now a polynomial of order three. Second, we substitute $\xi=\frac{1}{a_{3}}(4y-\frac{a_{2}}{3})$, which gives
\begin{align}\label{dy}
(\dfrac{dy}{d\gamma})^{2}=4y^{3}-g_{2}y-g_{3},
\end{align}
where the Weierstrass invariants are
\begin{align}
g_{2}=\frac{1}{16} \left( \frac{4}{3} a_{2}^{2}-4a_{1}a_{3} \right),
\end{align}
\begin{align}\label{g3}
g_{3}=\frac{1}{16} \left( \frac{1}{3}a_{1}a_{2}a_{3}
-\frac{2}{27}a_{2}^{3}-a_{0}a_{3}^{2} \right).
\end{align}
The differential equation (\ref{dy}) represents an elliptic intregal of the first kind, which can be solved by \cite{Hackmann:2008zza,M.Abramowitz, E. T. Whittaker}
\begin{align}\label{PW}
y(\gamma)=\wp \left(\gamma -\gamma_{\theta , in};g_{2},g_{3} \right).
\end{align}
Finally, the solution of Eq.(\ref{dthetad}) is given by
\begin{align}
\theta \big(\gamma \big)=\arccos \big( \pm \sqrt{\dfrac{a_{3}}{4\wp
(\gamma -\gamma_{\theta ,in}; g_{2},g_{3})-\frac{a_{2}}{3}}} \big),
\end{align}
where $\gamma_{\theta ,in}=\gamma_{0}+
\int_{y_{0}}^{\infty}\dfrac{dy'} {\sqrt{4y'^{3}-g_{2}y'-g_{3}}}$
and  $y_{0}=\dfrac{a_{3}}{4 cos^{2}(\theta_{0})}+\dfrac{a_{2}}{12}$
depends only on the initial values $\gamma_{0}$ and $\theta_{0}$.
Since the $\theta$ motion is symmetric with respect to the equatorial plane
$\theta=\frac{\pi}{2}$, the sign of the square root can be chosen so that $\theta(\gamma)$
is either in $(0,\frac{\pi}{2})$ (positive sign) or in
$(\frac{\pi}{2},\pi)$ (negative sign).

\subsubsection{Null geodesics}

The differential 
equation (\ref{doo}) for $ \varepsilon = 0 $ is already a
polynomial of degree three and, thus, with the standard substitution
$\upsilon =\frac{1}{b_{3}} \big(4y-\frac{b_{2}}{3} \big)$ where
$4\upsilon \tilde{\Theta}^{'}({\upsilon})=\sum_{i=1}^3 b_{i}\upsilon^{i}$
transforms the problem to the form Eq.~(\ref{dy}). The solution is
then given by
\begin{align}
\theta \left( \gamma \right)=\arccos \left( \pm \sqrt{\frac{4}{b_{3}}\wp
(\gamma - \gamma_{\theta ,in}; g_{2},g_{3})-\frac{b_{2}}{3 b_{3}}}
\right) 
\end{align}
where $\gamma_{\theta ,in }$,$g_{2}$, and $g_{3}$ are as above with
$a_{i}$ replaced by $b_{i}$.

\subsection{r motion}\label{rmotion}

The dynamics of $r$ are described by the differential equation (\ref{drd})
\begin{align}
\left(\dfrac{d\tilde{r}}{d\gamma}\right)^{2}
=\tilde{R}(\tilde{r})=\Delta_{\tilde{r}}(-\varepsilon
\tilde{r}^{2}-\tilde{K})+ \big[ (\tilde{a}^{2}+\tilde{r}^{2})E
-\tilde{a}\tilde{L}\Xi \big]^{2} \, .
\end{align}
Here the solution procedure is more complicated because $\tilde{R}$ is in general a polynomial of order six. However, for null geodesics the order of the polynomial is reduced to four. In the following we will consider timelike and null geodesics separately.

\subsubsection{Null geodesics} 

Considering light, i.e. $\varepsilon=0$,
$\tilde{R}$ is simplified to a polynomial of degree four and therefore the differential
equation (\ref{drd}) is of elliptic type. Then we can solve it using the method of section \ref{thetam}.
By substituting first $\tilde{r}=\xi^{-1}+\tilde{r}_{\tilde{R}}$,
where $\tilde{r}_{\tilde{R}}$ is a zero of $\tilde{R}$, and then
$\xi=\frac{1}{b_{3}}(4y- \frac{b_{2}}{3})$, where
$b_{i}=\frac{1}{(4-i)!}\frac{d^{(4-i)}\tilde{R}}{d\tilde{r}^{(4-i)}}(\tilde{r}_{\tilde{R}})$,
Eq.~(\ref{drd}) aquires the form  of Eq.~(\ref{dy}). Again, this can be solved with the help of the Weierstrass elliptic $\wp$ function, so that the result is
\begin{align}\label{rr}
\tilde{r}(\gamma)=\dfrac{b_{3}}{4\wp(\gamma
-\gamma_{\tilde{r},in};g_{2},g_{3})-\frac{b_{2}}{3}}+\tilde{r}_{\tilde{R}},
\end{align}
where
$\gamma_{\tilde{r},in}=\gamma_{0}+\int_{y_{0}}^{\infty}\dfrac{dy'}
{\sqrt{4y'^{3}-g_{2,r}y'-g_{3,r}}}$ and
$y_{0}=\frac{b_{3}}{4(\tilde{r}_{0}-r_{R})}+\frac{b_{2}}{12}$
depends only on the initial values $\gamma_{0}$ and $\tilde{r}_{0}$
and $g_{2}$, $g_{3}$ are defined as in Eq.(\ref{g3}) with
$a_{i}=b_{i}$. 

\subsubsection{Timelike geodesics} 

Considering particles, i.e.
$\varepsilon=1$, and assuming that $\tilde{R}$ has only simple zeros,
the differential equation (\ref{drd}) is of hyperelliptic type. As presented in \cite{Hackmann:2010zz}, this equation can be solved in terms of derivatives of the Kleinian $\sigma$ function. To begin with, Eq.~(\ref{drd}) is transformed into the standard form with the substitution
$\tilde{r}=\pm \frac{1}{u}+\tilde{r}_{\tilde{R}}$ where $\tilde{r}_{\tilde{R}}$ is a zero of $\tilde{R}$.
Then we get
\begin{align}\label{udu}
\left( u\dfrac{du}{d\gamma} \right)^{2}=\tilde{R}_{u},
\end{align}
where
\begin{align}\label{Ru}
\tilde{R}_{u}=\sum_{i=0}^{5} c_{i} u^{i}, \qquad
c_{i}=\dfrac{\big( \pm 1
\big)^{i}}{(6-i)!}\dfrac{d^{(6-i)}\tilde{R}}{d
u^{(6-i)}}(\tilde{r}_{\tilde{R}}).
\end{align}
A separation of variables leads to
\begin{equation}\label{Phi}
\gamma - \gamma_{0}
=\int_{u_{0}}^{u}\frac{udu}{\sqrt{\tilde{R}_{u}}},
\end{equation}
where $u_{0}=u(\gamma_{0})$. Considering the solution of the
integral (\ref{Phi}), we have to address two points. First, due to the two branches of the square root the integrand is not well defined in the complex
plane. Second, the solution $ u(\gamma) $ should not depend on the chosen path of
integration \cite{Hackmann:2008zza}. Let $ \zeta $ be a closed
integration path and
\begin{equation}\label{Omega}
\omega=\oint_{\zeta}\frac{udu}{\sqrt{\tilde{R}_{u}}},
\end{equation}
then also
\begin{equation}\label{PhiOmega}
\gamma - \gamma_{0}-\omega
=\int_{u_{0}}^{u}\frac{udu}{\sqrt{\tilde{R}_{u}}},
\end{equation}
should be true. Hence, the solution
$ u(\gamma) $
of our problem has to fulfill
\begin{equation}\label{PO}
u(\gamma)=u(\gamma - \omega)
\end{equation}
for every $ \omega\neq{0} $ obtained from Eq.~(\ref{Omega}). These
two issues can be solved if we consider Eq.~(\ref{Phi}) to be
defined on the Riemann surface $ y^{2}=\tilde{R}_{u}(x) $ of genus $
g=2 $ and introduce a basis of canonical holomorphic and meromorphic
differentials $ dz_{i} $ and $ dr_{i} $, respectively,
\begin{equation}\label{dz}
dz_{1}=\frac{dx}{\sqrt{\tilde{R}_{x}}}, \qquad\qquad\qquad
dz_{2}=\frac{xdx}{\sqrt{\tilde{R}_{x}}},
\end{equation}
\begin{equation}\label{dr}
dr_{1}=\frac{a_{3}x+2a_{4}x^{2}+3a_{5}x^{3}}
{4\sqrt{\tilde{R}_{x}}}dx, \qquad\qquad\qquad
dr_{2}=\frac{x^{2}dx}{4\sqrt{\tilde{R}_{x}}},
\end{equation}
and real $2\omega_{ij}$, $2\eta_{ij}$ and imaginary
$2\omega^{\prime}_{ij}$, $2\eta^{\prime}_{ij}$ period matrices
\begin{equation}\label{2w}
2\omega_{ij}=\oint_{a_{j}} {dz_{i}}, \qquad\qquad\qquad
2\omega^{\prime}_{ij}=\oint_{b_{j}} {dz_{i}},
\end{equation}
\begin{equation}\label{2n}
2\eta_{ij}=\oint_{a_{j}} {dr_{i}}, \qquad\qquad\qquad
2\eta^{\prime}_{ij}=\oint_{b_{j}} {dr_{i}}.
\end{equation}
The equation (\ref{Phi}) is a hyperelliptic integral of the first kind and can be
solved by \cite{Hackmann:2010zz,Enolski:2010if}
\begin{align}\label{ugamma}
u(\gamma)=-\frac{\sigma_{1}}{\sigma_{2}}(\gamma_{\sigma}),
\end{align}
where $ \sigma_{i} $ is the $i$-th derivative of the Kleinian $\sigma$
function
\begin{equation}
\sigma(z)=Ce^{z^{t}kz} \theta[K_{\infty}](2\omega^{-1}z;\tau),
\end{equation}
which is given by the Riemann $\theta $-function with characteristic
$K_{\infty} $,
\begin{equation}
\theta(z;\tau)=\sum_{m\in \mathbb{Z}^{2}}e^{i{\pi}m^{t}({\tau}m+2z)}.
\end{equation}
A number of parameters enter here: the symmetric Riemann matrix
$\tau=(\omega^{-1}\omega^{\prime}) $, the period-matrix $(2\omega,
2\omega^{\prime}) $, the period-matrix of the second kind $ (2\eta,
2\eta^{\prime}) $, the matrix $ k=\eta(2\omega)^{-1} $, and the
vector of Riemann constants with base point at infinity $
2K_{\infty} = (0, 1)^{t} + (1, 1)^{t}\tau$. The constant $C$ can be
given explicitly, see e.g.~\cite{V. M. Buchstaber}, but is not
important here. In Eq.(\ref{ugamma}) the argument $\gamma_{\sigma}$  is
an element of the one-dimensional $\sigma$ divisor: $ \gamma_{\sigma} =
(f(\gamma -\gamma_{\tilde{r},in}),\gamma
-\gamma_{\tilde{r},in})^{t} $ with
$\gamma_{\tilde{r},in}=\sqrt{c_{5}}\gamma_{0}+\int_{u_{0}}^{\infty}\dfrac{u'
du'}{\sqrt{\tilde{R}_{u'}}}$ and $u_{0}=\pm
(\tilde{r}_{0}-\tilde{r}_{\tilde{R}})^{-1}$ depends only on the
initial values $\gamma_{0}$ and $\tilde{r}_{0}$, and the function $
f$ can be found from the vanishing condition
$\sigma((f(x),x)^{t})=0$  \cite{Hackmann:2010zz}, so it describes the $\theta$-divisor. Then the solution of the
$\tilde{r}$ equation is given by
\begin{align}
\tilde{r}(\gamma)=\mp \frac{\sigma_{2}}{\sigma_{1}}(\gamma_{\sigma}).
\end{align}
Here the sign depends on the sign that was chosen in the substitution
$\tilde{r}=\pm \frac{1}{u}+\tilde{r}_{\tilde{R}}$. The functions $ \sigma_{1}
$ and $ \sigma_{2} $ depend on $\gamma_{\sigma},\omega, \eta, \tau $ and also on the polynomial $ \tilde{R}_{u}$
according to Eqs.~($ \ref{Phi}-\ref{2n} $), which contains all the
parameter-dependence of the modified gravity solution. The solution of $
\tilde{r} $ is the analytic solution of the equation of motion of a
test particle in the Kerr-Newman-(A)dS spacetime. This solution is valid in all regions of
this spacetime.

\subsection{$\varphi$ motion}\label{fii}

The $\varphi$-equation (\ref{dphi}) can be rewritten using the
$\tilde{r}$- and $\theta$-equations, (\ref{drd}) and (\ref{dthetad})
\begin{align}
d\varphi = \dfrac{\tilde{a}E \Xi
(\tilde{a}^{2}+\tilde{r}^{2})-\tilde{a}^{2}\Xi^{2}\tilde{L}}{\Delta_{\tilde{r}}\sqrt{\tilde{R}}}d\tilde{r}
- \dfrac{\tilde{a}\Xi E
\sin^{2}\theta-\Xi^{2}\tilde{L}}{\Delta_{\theta}\sin^{2}\theta
\sqrt{\tilde{\Theta}(\theta)}}d\theta .
\end{align}
Intergrating this equation gives an $\tilde{r}$-dependent integral
$I_{r}$ and a $\theta$-dependent integral $I_{\theta}$ which can be
solved separately
\begin{align}\label{phi}
\varphi - \varphi_{0}= \int_{\tilde{r}_{0}}^{\tilde{r}}
\dfrac{\tilde{a}E \Xi
(\tilde{a}^{2}+\tilde{r}^{2})-\tilde{a}^{2}\Xi^{2}\tilde{L}}{\Delta_{\tilde{r}}\sqrt{\tilde{R}}}d\tilde{r}
- \int_{\theta_{0}}^{\theta}\dfrac{\tilde{a}\Xi E
\sin^{2}\theta-\Xi^{2}\tilde{L}}{\Delta_{\theta}\sin^{2}\theta
\sqrt{\tilde{\Theta}(\theta)}}d\theta = I_{r}-I_{\theta} .
\end{align}

\subsubsection{The $\theta$-dependent integral}

Let us first consider the $\theta$-dependent integral
\begin{align}
I_{\theta}=\int_{\theta_{0}}^{\theta}\dfrac{(\tilde{a}E\Xi
\sin^{2}\theta - \Xi^{2}\tilde{L})d\theta}
{\Delta_{\theta}\sin^{2}\theta \sqrt{\tilde{\Theta}(\theta)}},
\end{align}
which can be simplified by the substitution $\upsilon=\cos^{2}\theta$
\begin{align}
I_{\theta}=\mp\int_{\upsilon_{0}}^{\upsilon}\dfrac{\tilde{a}E \Xi
(1-\upsilon)-\Xi^{2}\tilde{L}}{\Delta_{\upsilon}(1-\upsilon)\sqrt{4\upsilon
\tilde{\Theta}^{'}({\upsilon})}}d\upsilon '\, ,
\end{align}
where the polynomial $\Theta^{'}({\upsilon})$ is given in Eq.~(\ref{thetanoo}) and
$\Delta_{\upsilon}=1+\frac{\tilde{R}_{0}}{12}\tilde{a}^{2}\upsilon$.
Assuming $4\upsilon\Theta^{'}({\upsilon})$ has only simple zeros and is a polynomial of order four, then $I_{\theta}$ is an elliptic integral of the third kind. In this case the solution to $I_{\theta}$ is given by~\cite{Hackmann:2010zz}
\begin{align}\label{thetafi}
I_{\theta}=\frac{|a_{3}|}{a_{3}} \bigg\lbrace(\tilde{a}\Xi
E-\Xi^{2}\tilde{L})(\upsilon - \upsilon_{0})-
\sum_{i=1}^{4}\dfrac{a_{3}}{4\wp'(\upsilon_{i})}
\bigg(\zeta(\upsilon_{i})(\upsilon -\upsilon_{0})+\log
\dfrac{\sigma(\upsilon
-\upsilon_{i})}{\sigma(\upsilon_{0}-\upsilon_{i})}+2\pi i k_{i}
\bigg) \nonumber\\ \left(\tilde{a}^{3}\frac{\tilde{R_{0}}}{12}(E\Xi
-\tilde{a}\frac{\tilde{R_{0}}}{12}\Xi^{2}
\tilde{L})(\delta_{i1}+\delta_{i2})+\Xi^{2}
\tilde{L}(\delta_{i3}+\delta_{i4}) \right) \bigg\rbrace ,
\end{align}
where the coefficients $a_{i}$ of the polynomial $4\upsilon\Theta^{'}({\upsilon})$ are given in subsection \ref{thetam} and
\begin{align}
\wp(\upsilon_{1})=\frac{a_{2}}{12}-\frac{1}{48}\tilde{a}^{2}
\tilde{R_{0}}a_{3}=\wp(\upsilon_{2}),
\nonumber\\
\wp(\upsilon_{3})=\frac{a_{2}}{12}+\frac{1}{4}a_{3}=\wp(\upsilon_{4}) \, .
\end{align}
Also we have $\upsilon=\upsilon(\gamma)=\gamma-\gamma_{\theta,in}$, where
$\gamma_{\theta,in}$ is defined in Eq.~(\ref{PW}) and
$\upsilon_{0}=\upsilon(\gamma_{0})$. The different branches of the logarithm are represented by the integers $k_{i}$. The details of the computation can be
found in ref.~\cite{Hackmann:2010zz}.

\subsubsection{The $r$-dependent integral}

Next we will solve the
$\tilde{r}$-dependent integral 
\begin{align}
I_{r}=\int_{\tilde{r}_{0}}^{\tilde{r}}\dfrac{\tilde{a}E
\Xi (\tilde{a}^{2}+\tilde{r}^{2})-\tilde{a}^{2}\Xi^{2}\tilde{L}}
{\Delta_{\tilde{r}}\sqrt{\tilde{R}}}d\tilde{r}.
\end{align}
Here we will distinguish between timelike and null geodesics as the equation simplifies in the latter case.\\

\paragraph{Null geodesics}
Considering light, i.e.~$\varepsilon=0$, 
the polynomial $\tilde{R}$ is of order four, and therefore $I_{r}$ is an elliptic integral of the third kind and can be solved
analogously to $I_{\theta}$. We apply the same substitutions
$\tilde{r}=\frac{1}{\xi}+\tilde{r}_{\tilde{R}}$ and
$\xi=\frac{1}{b_{3}}(4y-\frac{b_{2}}{3})$, as in subsection
(\ref{rmotion}) for the case $\varepsilon=0$, then perform a partial
fraction decomposition, and finally substitute $y=\wp
(\upsilon)$. Then we get
\begin{align}
\frac{b_{3}}{\mid
b_{3}\mid}I_{r}=\sum_{i=1}^{4}C_{i}\int_{\upsilon_{0}}^{\upsilon}\dfrac{d\upsilon}{\wp
(\upsilon)-y_{i}}-\dfrac{\tilde{a}E\Xi
(\tilde{a}^{2}+\tilde{r}^{2}_{\tilde{R}})-\tilde{a}^{2}\Xi^{2}\tilde{L}}{\Delta_{\tilde{r}=
\tilde{r}_{\tilde{R}}}}\int_{\upsilon_{0}}^{\upsilon}d\upsilon \, ,
\end{align}
where $b_{3}$ is given in Eq.~(\ref{g3}), and the $y_{i}$ are the four zeros of 
$\Delta_{y(\tilde{r})}$. The constants $C_{i}$ arise from the partial
fraction decomposition and depend on the parameters of the test particle and the metric.
The integrand $(\wp
(\upsilon)-y_{i})^{-1}$ has simple poles $\upsilon_{i1}$,
$\upsilon_{i2}$, where $\wp (\upsilon_{i1})=y_{i}=\wp (\upsilon_{i2})$.

$I_{r}$ can be integrated according to ref.~\cite{Hackmann:2010zz}, and the solution is
\begin{align}
I_{r}=\frac{|b_{3}|}{b_{3}} \bigg\lbrace\sum_{i=1}^{4}
\sum_{j=1}^{2}\dfrac{C_{i}}{\wp ' (\upsilon_{ij})} \big[ \xi
(\upsilon_{ij})(\upsilon-\upsilon_{0})+  \log \sigma
(\upsilon-\upsilon_{ij}) \nonumber\\ - \log
\sigma(\upsilon_{0}-\upsilon_{ij}) \big] -
\dfrac{\tilde{a}E\Xi
(\tilde{a}^{2}+\tilde{r}^{2}_{\tilde{R}})-\tilde{a}^{2}\Xi^{2}\tilde{L}}{\Delta_{\tilde{r}
=\tilde{r}_{\tilde{R}}}}(\upsilon-\upsilon_{0})\bigg\rbrace,
\end{align}
with $\upsilon=\upsilon(\gamma)=\gamma-\gamma_{\tilde{r},in}$ and $
\upsilon_{0}=\upsilon (\gamma_{0})$, where $\gamma_{\tilde{r},in}$ is given 
in Eq.~(\ref{rr}).\\

\paragraph{Timelike geodesics} 
Considering particles, i.e.~$\varepsilon=1$, 
and assuming that the polynomial $\tilde{R}$ has only simple zeros,
$I_{r}$ is a hyperelliptic integral of the third kind. 

The first step in the solution procedure is to transform $\tilde{R}$ to the standard form by the substitution $\tilde{r}=\pm 1/u+\tilde{r}_{\tilde{R}}$, where  $\tilde{r}_{\tilde{R}}$ is a zero of $\tilde{R}$ (see section \ref{rmotion}). Next we apply a partial fraction decomposition to the integrand, so that the solution method of ref.~\cite{Hackmann:2010zz} can be used. The solution of $I_{r}$ is 
\begin{align}\label{akh}
I_{r}=\mp\dfrac{\tilde{a}u_{0}}{\mid u_{0}  \mid} \big
\lbrace C_{1}(\omega -\omega_{0})+ C_{0}(f(\omega)-f(\omega_{0}))
\nonumber\\ +\sum_{i=1}^{4}\dfrac{C_{2,i}}{\sqrt{\tilde{R}_{u_{i}}}}
[\frac{1}{2} \log
\frac{\sigma(W^{+}(\omega))}{\sigma(W^{-}(\omega))}-\frac{1}{2}\log
\frac{\sigma(W^{+}(\omega_{0}))}{\sigma(W^{-}(\omega_{0}))}
\nonumber\\ -(f(\omega)-f(\omega_{0}), \omega - \omega_{0}) \big(
\int_{u_{i}^{-}}^{u_{i}^{+}} d\vec{r} \big) ] \big \rbrace .
\end{align}
with $\omega =\omega (\gamma)=\gamma
-\gamma_{\tilde{r},in}$ and $\omega_{0}=\omega (\gamma_{0})$. Again the constants $C_{i}$ arise from the partial fraction decomposition. 
$\tilde{R}_{u}$ is defined in Eq.~(\ref{udu}), and the $u_{i}$ are the four zeros of
$\Delta_{\tilde{r}=\pm1/u+\tilde{r}_{\tilde{R}}}, u_{0}=\pm
(\tilde{r}-\tilde{r}_{\tilde{R}})^{-1}$.
The functions $W^{\pm}$ are
given by $W^{\pm}(\omega):=
(f(\omega),\omega)^{t}-2\int_{\infty}^{u_{i}^{\pm}} d\vec{z}$ with $u_{i}^{\pm}=(u_{i}\pm \sqrt{\tilde{R}_{u_{i}}})$ (compare~\cite{Hackmann:2010zz}).

\subsection{t motion}

The $\tilde{t}$-equation (\ref{dtd}) can be rewritten using the
$\tilde{r}$- and $\theta$-equations, (\ref{drd}) and (\ref{dphi})
\begin{align}
d\tilde{t}=\dfrac{E(\tilde{r}^{2}+\tilde{a}^{2})^{2}-\tilde{a}\tilde{L}\Xi(\tilde{r}^{2}
+\tilde{a}^{2})}
{\Delta_{\tilde{r}}}\frac{dr}{\sqrt{\tilde{R}}}
-\frac{\sin^{2}\theta}{\Delta_{\theta}}(E
\tilde{a}^{2}-\frac{\tilde{L}\Xi
\tilde{a}}{\sin^{2}\theta})\dfrac{d\theta}{\sqrt{\tilde{\Theta}(\theta)}},
\end{align}
and has the same structure as the equation for the $\varphi$ motion. Integrating the $\tilde{t}$-equation yields
\begin{align}
\tilde{t}-\tilde{t}_{0}=\big[ \int_{r_{0}}^{r}\dfrac{E(\tilde{r}^{2}
+\tilde{a}^{2})^{2}-\tilde{a}
\tilde{L}\Xi(\tilde{r}^{2}+\tilde{a}^{2})}{\Delta_{\tilde{r}}} \frac{dr}{\sqrt{\tilde{R}}}
- \int_{\theta_{0}}^{\theta} \frac{\sin^{2}\theta}{\Delta_{\theta}}(E
\tilde{a}^{2}-\frac{\tilde{L}\Xi
\tilde{a}}{\sin^{2}\theta})\dfrac{d\theta}{\sqrt{\tilde{\Theta}(\theta)}}
\big] = \tilde{I}_{r}-\tilde{I}_{\theta}.
\end{align}
The solutions can be found in the same way as in section \ref{fii}. For the $\theta$-dependent part we have \cite{Hackmann:2010zz}
\begin{align}
\tilde{I}_{\theta}= a_{3}(\upsilon -\upsilon_{0}) -
\sum_{i=1}^{2}\dfrac{a_{3}\tilde{a}^{2}R_{0}}{4\wp' (\upsilon_{i})}
\bigg[\zeta(\upsilon_{i})(\upsilon - \upsilon_{0}) +\log \sigma
(\upsilon - \upsilon_{i})-\log \sigma (\upsilon_{0}-\upsilon_{i}) \bigg],
\end{align}
where $a_{2}$ and $a_{3}$ are given in
Eq.~(\ref{khiii}), $\wp(\upsilon_{1})=\frac{a_{2}}{12}+\frac{1}{4}\tilde{a}^{2}R_{0}a_{3}=\wp(\upsilon_{2})$,
and $\upsilon=\upsilon(\gamma)=2\gamma -\gamma_{\theta,in}$ with the initial value
$\upsilon_{0}=\upsilon(\gamma_{0})$. 

Considering light, i.e.~$\varepsilon=0$, 
the solution for the $\tilde{r}$-dependent part is very simple and
given by \cite{Hackmann:2010zz}
\begin{align}
\tilde{I}_{r}=\frac{|b_{3}|}{b_{3}} \bigg\lbrace\sum_{i=1}^{4}
\sum_{j=1}^{2}\dfrac{\tilde{C}_{i}}{\wp ' (\upsilon_{ij})} \big[ \xi
(\upsilon_{ij})(\upsilon-\upsilon_{0})+  \log \sigma
(\upsilon-\upsilon_{ij}) \nonumber\\ - \log
\sigma(\upsilon_{0}-\upsilon_{ij}) \big] -
\dfrac{\tilde{a}\tilde{L}\Xi (\tilde{r}_{\tilde{R}}^{2}
+\tilde{a}^{2})-E(\tilde{r}_{\tilde{R}}^{2}+\tilde{a}^{2})^{2}}{\Delta_{\tilde{r}=\tilde{r}_{\tilde{R}}}}
(\upsilon-\upsilon_{0})\bigg\rbrace ,
\end{align}
where $b_{3}$ is given in Eq.~(\ref{rr}), the $\tilde{C}_{i}$ arise from the partial fraction decomposition, and
$\wp(\upsilon_{i1})=y_{i}=\wp(\upsilon_{i2})$, where 
$y_{i}$ are the four zeros of $\Delta_{y(\tilde{r})}$. The variable $\upsilon=\upsilon(\gamma)=\gamma -\gamma_{\tilde{r},in}$ has the initial value
$\upsilon_{0}=\upsilon(\gamma_{0})$.\\ 

In the case of timelike geodesics, i.e.~$\varepsilon=1$, the solution of the now hyperelliptic $\tilde{r}$-dependent part is
given by \cite{Hackmann:2010zz}
\begin{align}
\tilde{I}_{r}=\dfrac{u_{0}}{\sqrt{c_{5}} \mid u_{0}  \mid} \big
\lbrace \tilde{C}_{1}(\omega -\omega_{0})+
\tilde{C}_{0}(f(\omega)-f(\omega_{0})) \nonumber\\
+\sum_{i=1}^{4}\dfrac{\tilde{C}_{2,i}}{\sqrt{\tilde{\tilde{R}}_{u_{i}}}}
[\frac{1}{2} \log
\frac{\sigma(W^{+}(\omega))}{\sigma(W^{-}(\omega))}-\frac{1}{2}\log
\frac{\sigma(W^{+}(\omega_{0}))}{\sigma(W^{-}(\omega_{0}))}
\nonumber\\ -(f(\omega)-f(\omega_{0}), \omega - \omega_{0}) \big(
\int_{u_{i}^{-}}^{u_{i}^{+}} d\vec{r} \big) ] \big \rbrace .
\end{align}
For the notation see Eq.~(\ref{akh}). The constants
$\tilde{C}_{0},\tilde{C}_{1},\tilde{C}_{2,i}$ result from the partial fraction decomposition.

\section{THE ORBITS}\label{sec:orbits}

The analytical solutions can be used to plot the orbits of test particles and light rays. We present example of the orbits around the static charged (A)dS black hole (Reissner-Nordstr\"om-(A)dS) and the rotating charged (A)dS black hole (Kerr-Newman-(A)dS).

\subsection{The static case}

\begin{figure}[h!]
	\centering
	\subfigure[Bound orbit with parameters $\varepsilon=1$, $\tilde{R}_0=\frac{1}{3}\cdot 10^{-5}$, $\tilde{q}=0.75$, $\mathcal{L}=0.076$, $E=\sqrt{0.918}$.]{
		\includegraphics[width=0.45\textwidth]{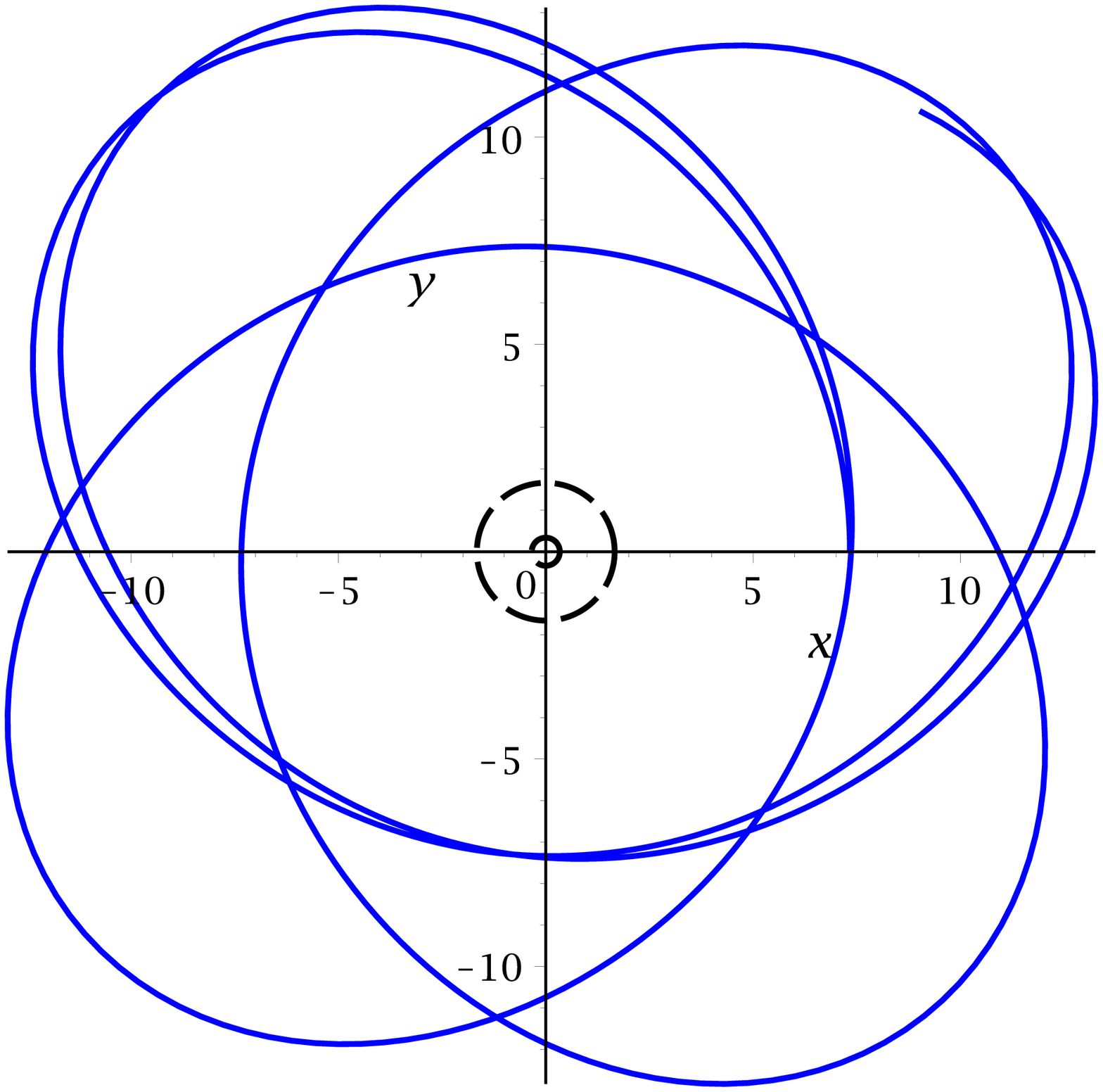}
	}
	\subfigure[Many-world bound orbit with parameters $\varepsilon=1$, $\tilde{R}_0=\frac{1}{3}\cdot 10^{-5}$, $\tilde{q}=0.995$, $\mathcal{L}=0.8$, $E=\sqrt{0.2}$. The particle is reflected at the potential barrier arising from the charge.]{
		\includegraphics[width=0.45\textwidth]{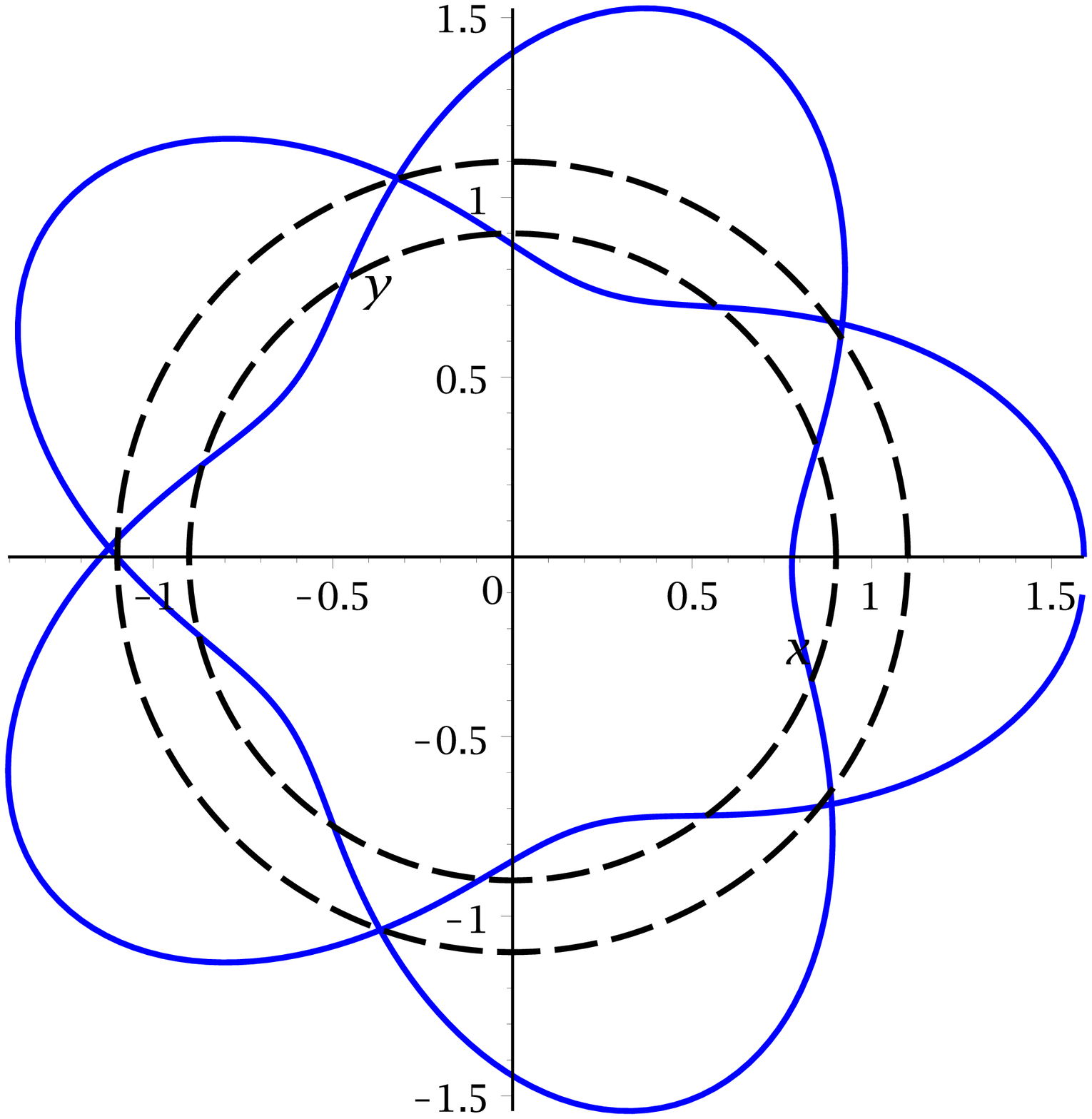}
	}
	\caption{Two examples of particle orbits in the Reissner-Nordstr\"om-(A)dS spacetime. 
The blue curves depict the orbits and the black dashed circle indicte the positions of the horizons.}
 \label{pic:orbits1}
\end{figure}

\begin{figure}[h!]
	\centering
	\subfigure[Escape orbit with parameters $\varepsilon=0$, $\tilde{R}_0=\frac{1}{3}\cdot 10^{-5}$, $\tilde{q}=0.75$, $\mathcal{L}=0.1$, $E=\sqrt{0.46}$.]{
		\includegraphics[width=0.45\textwidth]{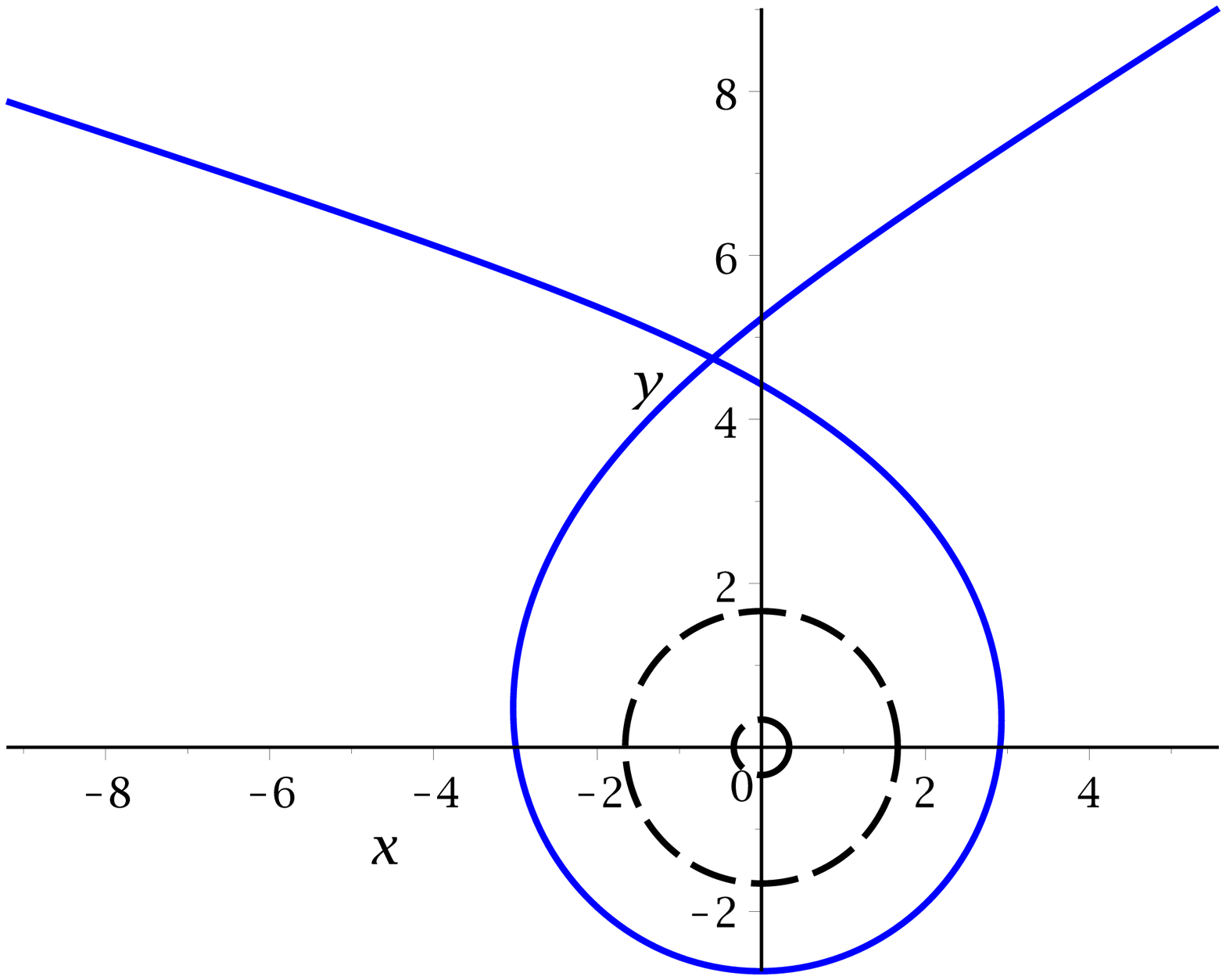}
	}
	\subfigure[Two-world escape orbit with parameters $\varepsilon=0$, $\tilde{R}_0=\frac{1}{3}\cdot 10^{-5}$, $\tilde{q}=0.95$, $\mathcal{L}=5$, $E=\sqrt{0.8}$. Light is reflected at the potential barrier arising from the charge.]{
		\includegraphics[width=0.45\textwidth]{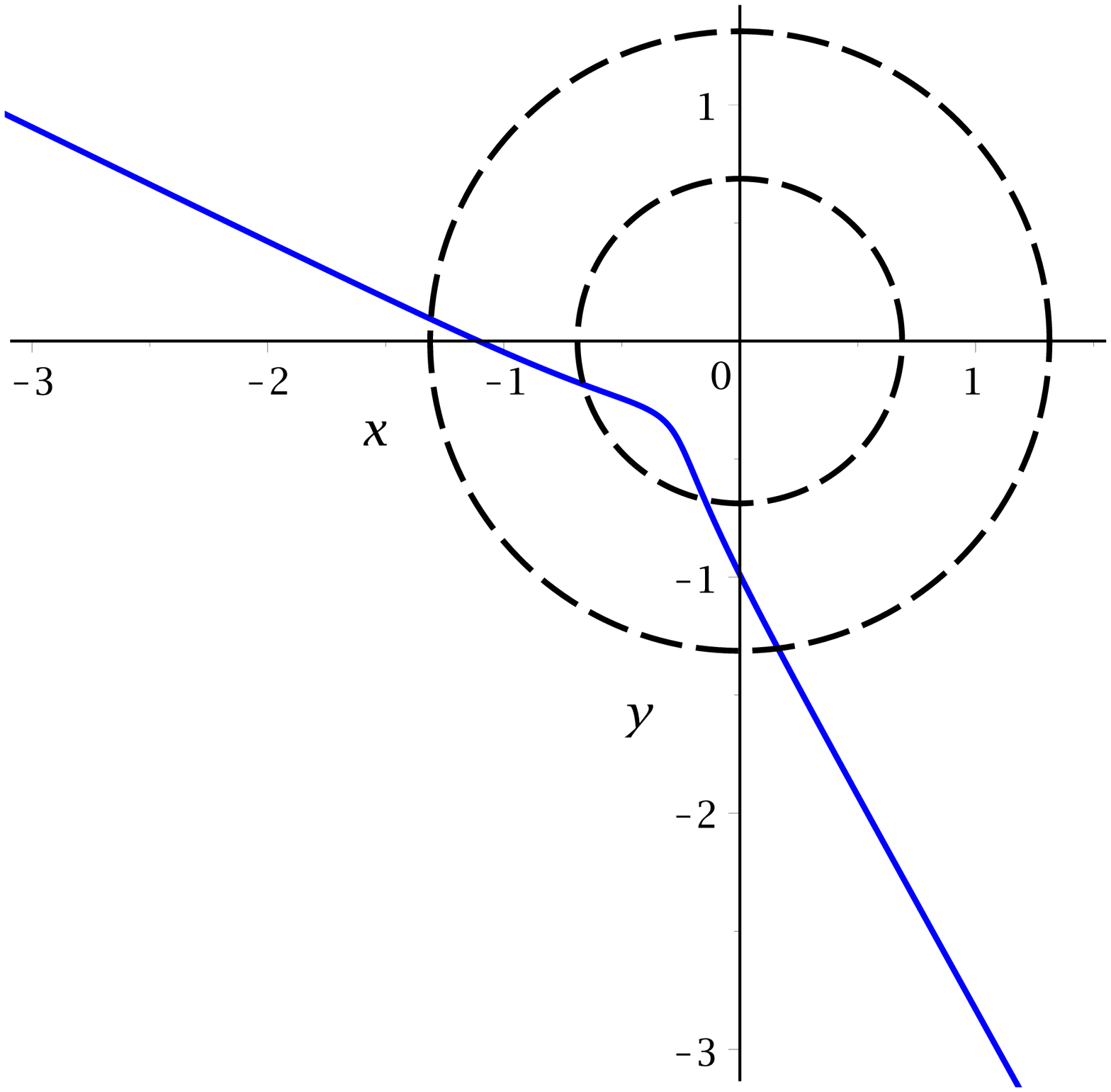}
	}
	\caption{Two examples of light orbits in the Reissner-Nordstr\"om-(A)dS spacetime. 
The blue curves depict the orbits and the black dashed circle indicte the positions of the horizons.}
 \label{pic:orbits2}
\end{figure}

Some examples of timelike and null geodesics in the static case can be found in Figs.~\ref{pic:orbits1} and \ref{pic:orbits2}.
In Fig.~\ref{pic:orbits1}, two bound orbits of test particles are shown: a bound orbit outside the horizons (Fig.~\ref{pic:orbits1}(a)) and a many-world bound orbit (Fig.~\ref{pic:orbits1}(b)). On the many-world bound orbit both horizons are crossed several times and each time the test particles emerge into
another universe. Note that the test particle is reflected at the potential barrier behind the horizons arising from the charge.

An escape orbit and a two-world escape orbit are depicted in Figs.~\ref{pic:orbits2}(a) and~\ref{pic:orbits2}(b) respectively.
The two-world escape orbit crosses both horizons twice and escapes to another universe. Also the reflection at the potential barrier is visible.

\clearpage

\subsection{The rotating case}

\begin{figure}[h!]
 \centering
 \subfigure[Bound orbit with parameters $\varepsilon=1$, $a=0.7$, $K=2$, $q=0.7$, $R_0=4\cdot 10^{-5}$, $L=1.9$, $E=0.84$.]{
  \includegraphics[width=0.45\textwidth]{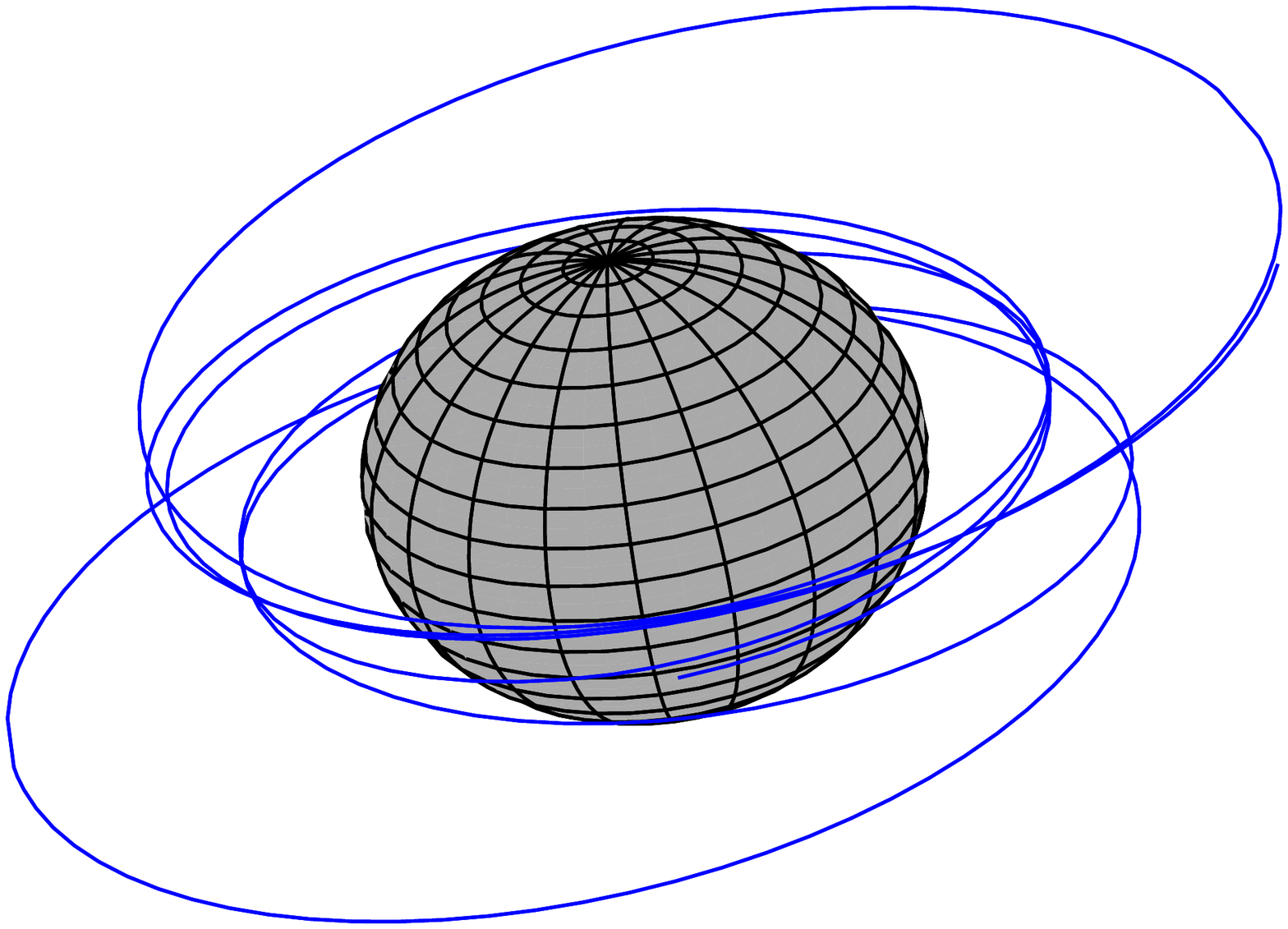}
 }
 \subfigure[Escape orbit with parameters $\varepsilon=0$, $a=0.7$, $K=2$, $q=0.7$, $R_0=4\cdot 10^{-5}$, $L=1.9$, $E=0.75$.]{
  \includegraphics[width=0.45\textwidth]{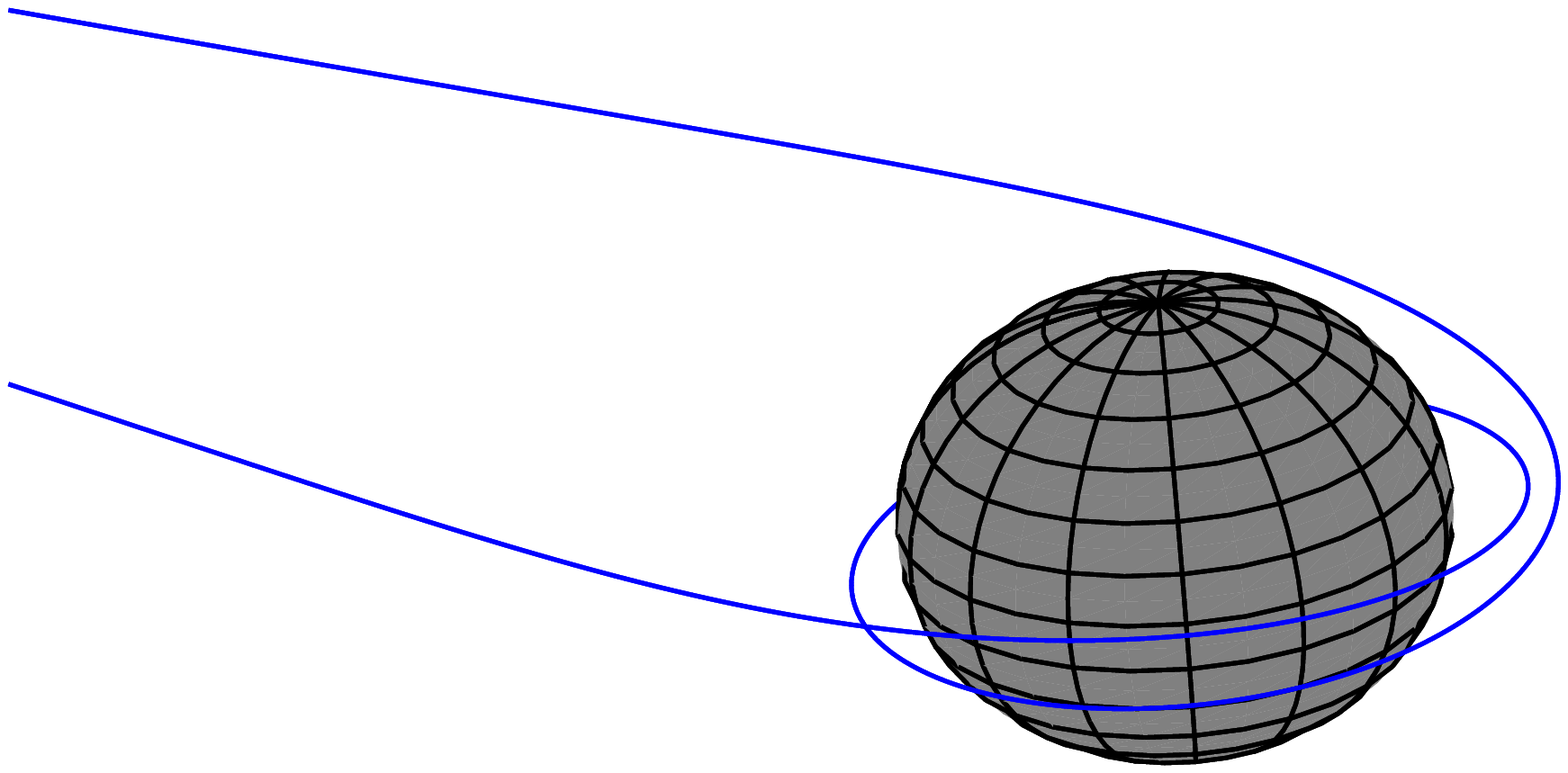}
 }
 \caption{Two examples of possible orbits in the Kerr-Newman-(A)dS spacetime. The blue lines show the path of the orbits and the sphere represents the event horizon.}
\label{pic:bo-eo}
\end{figure}

\begin{figure}[h!]
 \centering
 \subfigure[Transit orbit with parameters $\varepsilon=0$, $a=0.7$, $K=2$, $q=0.5$, $R_0=4\cdot 10^{-5}$, $L=0.5$, $E=3.75$.]{
  \includegraphics[width=0.45\textwidth]{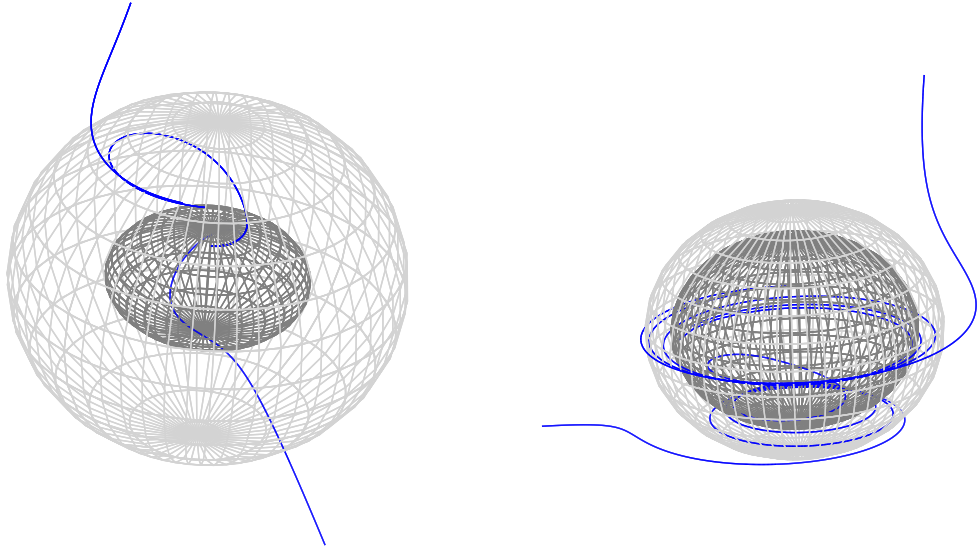}
 }
 \subfigure[Two-world escape orbit with parameters $\varepsilon=0$, $a=0.7$, $K=1$, $q=0.7$, $R_0=4\cdot 10^{-5}$, $L=0.5$, $E=1.5$.]{
  \includegraphics[width=0.45\textwidth]{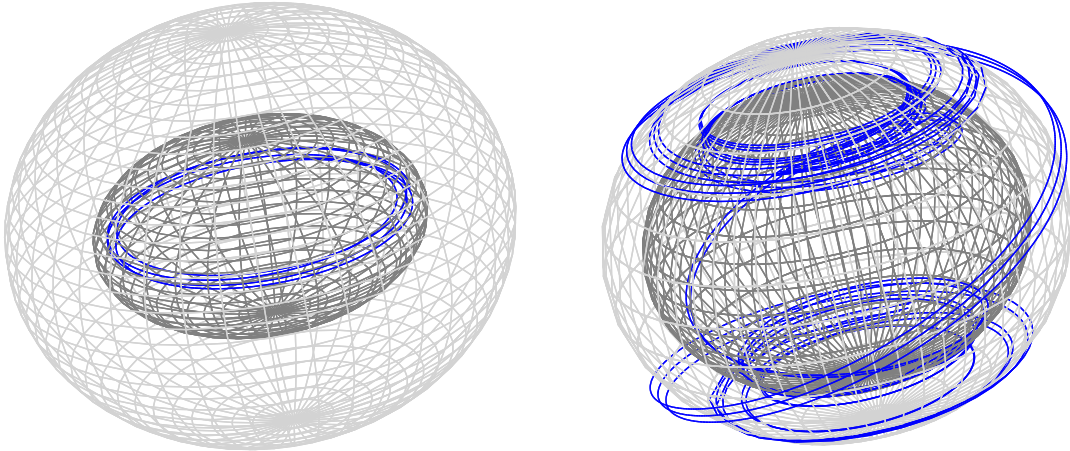}
 }
 \caption{Two examples of possible orbits in the Kerr-Newman-(A)dS spacetime. The blue lines show the path of the orbits and the spheres represent the inner and outer horizon.}
\label{pic:tro-teo}
\end{figure}

\begin{figure}[h!]
 \centering
 \subfigure[Bound orbit behind the inner horizon with parameters $\varepsilon=1$, $a=0.9$, $K=0.3$, $q=0.2$, $R_0=4\cdot 10^{-5}$, $L=1.45$, $E=1.02$.]{
  \includegraphics[width=0.45\textwidth]{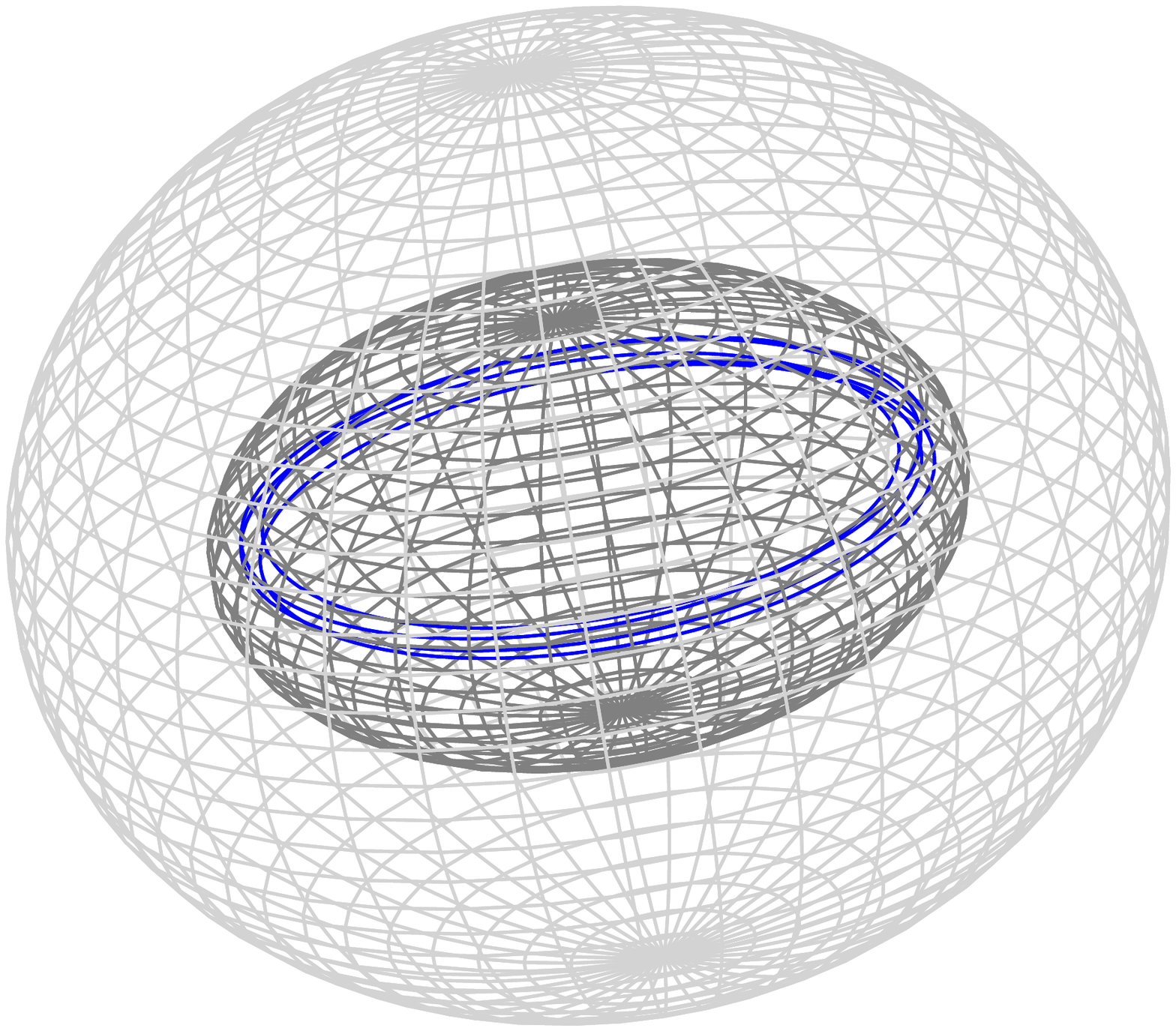}
 }
 \subfigure[Many-world bound orbit with parameters $\varepsilon=0$, $a=0.7$, $K=1$, $q=0.7$, $R_0=4\cdot 10^{-5}$, $L=0.5$, $E=0.05$.]{
  \includegraphics[width=0.45\textwidth]{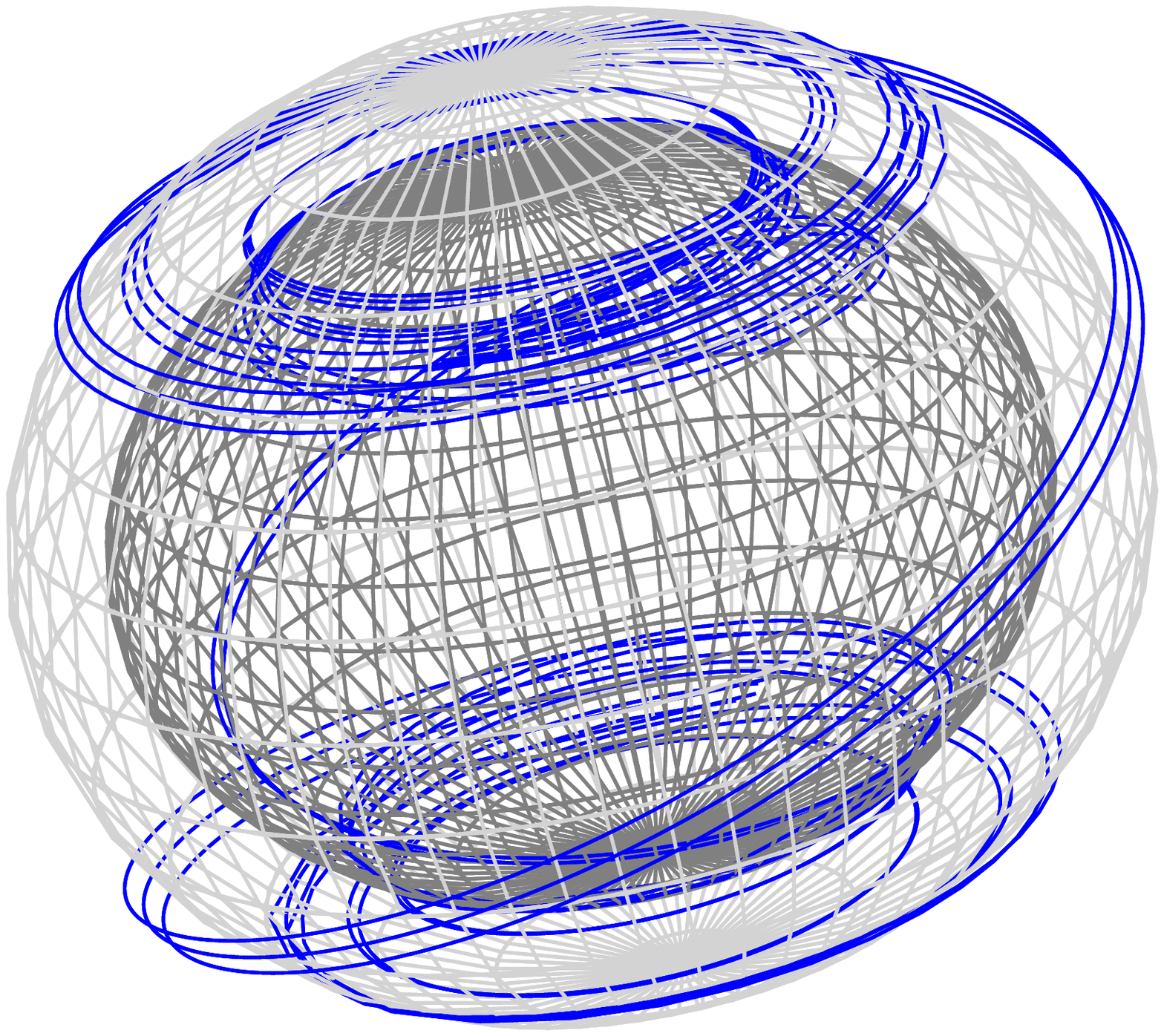}
 }
 \caption{Two examples of possible orbits in the Kerr-Newman-(A)dS spacetime. The blue lines show the path of the orbits and the spheres represent the inner and outer horizon.}
\label{pic:innerbo-mbo}
\end{figure}

Here we show some orbits in the Kerr-Newman-(A)dS spacetime. Figure \ref{pic:bo-eo}, shows two example plots of a bound orbit for particles and an escape orbit for light. A transit orbit crossing $r=0$ can be seen in Figure \ref{pic:tro-teo}(a). A two-world escape orbit which crosses both horizons twice and escapes to another universe is depicted in \ref{pic:tro-teo}(b). In Figure \ref{pic:innerbo-mbo}(a), a bound orbit hidden behind the inner horizon is shown. Figure \ref{pic:innerbo-mbo}(b), shows a many-world bound orbit, where both horizons are crossed several times.

\section{conclusions}\label{conclusions}

In this paper we discussed the motion of test particles and light
rays in the spacetime of the static and rotating charged black hole (Kerr-Newman-(A)dS spacetime). 
After reviewing the spacetime and presenting the corresponding
equations of motion, we classified the possible types of geodesic motion
by an analysis of the zeros of the polynomials underlying the
$\theta$- and  $r$-motion. The geodesic equations were solved in
terms of Weierstrass elliptic functions and derivatives of Kleinian
$\sigma$ functions.
Using effective potential techniques and parametric diagrams, the
possible types of orbits were derived.
Finally, a number of orbits were illustrated.

The techniques employed in this
paper, present a useful tool to calculate the exact orbits,
and the results obtained should prove valuable in order to analyze their properties, 
including observables like the periastron shift of bound
orbits, the light deflection of flyby orbits, the deflection angle
and the Lense-Thirring effect. For the calculation of these
observables analogous formulas to those given in~\cite{Hackmann:2008zza,Rindler:2007zz,
Bhattacharya:2009rv,Bhattacharya:2010xh,Kraniotis:2003ig,Kraniotis:2004cz} may be used.

The analytical solutions of the equations of motion are also useful in the context of AdS/CFT, since geodesics in an AdS spacetime can be related to CFT propagators (see e.g. \cite{Balasubramanian:1999zv}).

\begin{acknowledgements}
S.G.~and J.K.~would like to acknowledge support by the DFG
Research Training Group \emph{Models of Gravity}.
\end{acknowledgements}


\bibliographystyle{amsplain}

\begin{thebibliography}{99}
\bibitem{Riess:1998cb}
A.~G.~Riess {\it et al.}  [Supernova Search Team Collaboration],
Astron.\ J.\  {\bf 116}, 1009 (1998) [astro-ph/9805201].
S.~Perlmutter {\it et al.}  [Supernova Cosmology Project
Collaboration],
Astrophys.\ J.\  {\bf 517}, 565 (1999) [astro-ph/9812133].
J.~L.~Tonry {\it et al.}  [Supernova Search Team Collaboration],
Astrophys.\ J.\  {\bf 594}, 1 (2003) [astro-ph/0305008].
A. G.~Riess et al., Astrophys. J. \textbf{607}, 665 (2004).


\bibitem{Spergel:2006hy} 
  D.~N.~Spergel {\it et al.} [WMAP Collaboration],
  Astrophys.\ J.\ Suppl.\  {\bf 170}, 377 (2007)
  [astro-ph/0603449].
  E.~Komatsu {\it et al.} [WMAP Collaboration],
  Astrophys.\ J.\ Suppl.\  {\bf 180}, 330 (2009)
  [arXiv:0803.0547 [astro-ph]].
  E.~Komatsu {\it et al.} [WMAP Collaboration],
  Astrophys.\ J.\ Suppl.\  {\bf 192}, 18 (2011)
  [arXiv:1001.4538 [astro-ph.CO]].

\bibitem{Ade:2013sjv} 
  P.~A.~R.~Ade {\it et al.} [Planck Collaboration],
  Astron.\ Astrophys.\  {\bf 571}, A1 (2014)
  [arXiv:1303.5062 [astro-ph.CO]].
  P.~A.~R.~Ade {\it et al.} [Planck Collaboration],
  Astron.\ Astrophys.\  {\bf 571}, A16 (2014)
  [arXiv:1303.5076 [astro-ph.CO]].
  P.~A.~R.~Ade {\it et al.} [Planck Collaboration],
  arXiv:1502.01589 [astro-ph.CO].

\bibitem{Jain:2003tba} 
  B.~Jain and A.~Taylor,
  Phys.\ Rev.\ Lett.\  {\bf 91}, 141302 (2003)
  [astro-ph/0306046].
  H.~Hoekstra and B.~Jain,
  Ann.\ Rev.\ Nucl.\ Part.\ Sci.\  {\bf 58}, 99 (2008)
  [arXiv:0805.0139 [astro-ph]].



\bibitem{Eisenstein:2005su} 
  D.~J.~Eisenstein {\it et al.} [SDSS Collaboration],
  Astrophys.\ J.\  {\bf 633}, 560 (2005)
  [astro-ph/0501171].
  S.~Cole {\it et al.} [2dFGRS Collaboration],
  Mon.\ Not.\ Roy.\ Astron.\ Soc.\  {\bf 362}, 505 (2005)
  [astro-ph/0501174].
  W.~J.~Percival {\it et al.} [SDSS Collaboration],
  Mon.\ Not.\ Roy.\ Astron.\ Soc.\  {\bf 401}, 2148 (2010)
  [arXiv:0907.1660 [astro-ph.CO]].

\bibitem{Tegmark:2003ud} 
  M.~Tegmark {\it et al.} [SDSS Collaboration],
  Phys.\ Rev.\ D {\bf 69}, 103501 (2004)
  [astro-ph/0310723].
  U.~Seljak {\it et al.} [SDSS Collaboration],
  Phys.\ Rev.\ D {\bf 71}, 103515 (2005)
  [astro-ph/0407372].
  M.~Betoule {\it et al.} [SDSS Collaboration],
  Astron.\ Astrophys.\  {\bf 568}, A22 (2014)
  [arXiv:1401.4064 [astro-ph.CO]].




\bibitem{Y. Hagihara}
Y. Hagihara, 
Japan. J. Astron. Geophys. {\bf 8}, 67 (1931).

\bibitem{Chandrasekhar:1985kt} 
S. Chandrasekhar, \textit{The Mathematical Theory of Black Holes}, 
(Oxford University Press, Oxford, 1983).

\bibitem{Hackmann:2008zza} 
  E.~Hackmann and C.~L\"ammerzahl,
  Phys.\ Rev.\ Lett.\  {\bf 100}, 171101 (2008)
  [arXiv:1505.07955 [gr-qc]].
  E.~Hackmann and C.~L\"ammerzahl,
  Phys.\ Rev.\ D {\bf 78}, 024035 (2008)
  [arXiv:1505.07973 [gr-qc]].

\bibitem{Hackmann:2008tu} 
  E.~Hackmann, V.~Kagramanova, J.~Kunz and C.~L\"ammerzahl,
  Phys.\ Rev.\ D {\bf 78}, 124018 (2008)
  Erratum: [Phys.\ Rev.\  {\bf 79}, 029901 (2009)]
  [arXiv:0812.2428 [gr-qc]].

\bibitem{Hackmann:2009nh} 
  E.~Hackmann, V.~Kagramanova, J.~Kunz and C.~L\"ammerzahl,
  Europhys.\ Lett.\  {\bf 88}, 30008 (2009)
  [arXiv:0911.1634 [gr-qc]].

\bibitem{Hackmann:2010zz} 
  E.~Hackmann, C.~L\"ammerzahl, V.~Kagramanova and J.~Kunz,
  Phys.\ Rev.\ D {\bf 81}, 044020 (2010)
  [arXiv:1009.6117 [gr-qc]].

\bibitem{Grunau:2010gd} 
  S.~Grunau and V.~Kagramanova,
  Phys.\ Rev.\ D {\bf 83}, 044009 (2011)
  doi:10.1103/PhysRevD.83.044009
  [arXiv:1011.5399 [gr-qc]].

\bibitem{Enolski:2010if} 
  V.~Z.~Enolski, E.~Hackmann, V.~Kagramanova, J.~Kunz and C.~L\"ammerzahl,
  J.\ Geom.\ Phys.\  {\bf 61}, 899 (2011)
  [arXiv:1011.6459 [gr-qc]].

\bibitem{Kagramanova:2010bk} 
  V.~Kagramanova, J.~Kunz, E.~Hackmann and C.~L\"ammerzahl,
  Phys.\ Rev.\ D {\bf 81}, 124044 (2010)
  [arXiv:1002.4342 [gr-qc]].

\bibitem{Kagramanova:2013mwv} 
  V.~Diemer and E.~Smolarek,
  Class.\ Quant.\ Grav.\  {\bf 30}, 175014 (2013)
  [arXiv:1302.1705 [gr-qc]].


\bibitem{Kagramanova:2012hw} 
  V.~Kagramanova and S.~Reimers,
  Phys.\ Rev.\ D {\bf 86}, 084029 (2012)
  [arXiv:1208.3686 [gr-qc]].

\bibitem{Diemer:2014lba} 
  V.~Diemer, J.~Kunz, C.~L\"ammerzahl and S.~Reimers,
  Phys.\ Rev.\ D {\bf 89}, no. 12, 124026 (2014)
  [arXiv:1404.3865 [gr-qc]].

\bibitem{Diemer:2013fza} 
  V.~Diemer and J.~Kunz,
  Phys.\ Rev.\ D {\bf 89}, no. 8, 084001 (2014)
  [arXiv:1312.6540 [gr-qc]].

\bibitem{Grunau:2012ai} 
  S.~Grunau, V.~Kagramanova, J.~Kunz and C.~L\"ammerzahl,
  Phys.\ Rev.\ D {\bf 86}, 104002 (2012)
  [arXiv:1208.2548 [gr-qc]].

\bibitem{Grunau:2012ri} 
  S.~Grunau, V.~Kagramanova and J.~Kunz,
  Phys.\ Rev.\ D {\bf 87}, no. 4, 044054 (2013)
  [arXiv:1212.0416 [gr-qc]].


\bibitem{Aliev:1988wv} 
  A.~N.~Aliev and D.~V.~Galtsov,
  Sov.\ Astron.\ Lett.\  {\bf 14}, 48 (1988).

\bibitem{Galtsov:1989ct} 
  D.~V.~Galtsov and E.~Masar,
  Class.\ Quant.\ Grav.\  {\bf 6}, 1313 (1989).

\bibitem{Chakraborty:1991mb} 
  S.~Chakraborty and L.~Biswas,
  Class.\ Quant.\ Grav.\  {\bf 13}, 2153 (1996).

\bibitem{Ozdemir:2003km} 
  N.~Ozdemir,
  Class.\ Quant.\ Grav.\  {\bf 20}, 4409 (2003).

\bibitem{Ozdemir:2004ne} 
  F.~Ozdemir, N.~Ozdemir and B.~T.~Kaynak,
  Int.\ J.\ Mod.\ Phys.\ A {\bf 19}, 1549 (2004).

\bibitem{Grunau:2013oca} 
  S.~Grunau and B.~Khamesra,
  Phys.\ Rev.\ D {\bf 87}, no. 12, 124019 (2013)
  [arXiv:1303.6863 [gr-qc]].

\bibitem{Hackmann:2009rp} 
  E.~Hackmann, B.~Hartmann, C.~Laemmerzahl and P.~Sirimachan,
  Phys.\ Rev.\ D {\bf 81}, 064016 (2010)
  [arXiv:0912.2327 [gr-qc]].

\bibitem{Hackmann:2010ir} 
  E.~Hackmann, B.~Hartmann, C.~L\"ammerzahl and P.~Sirimachan,
  Phys.\ Rev.\ D {\bf 82}, 044024 (2010)
  [arXiv:1006.1761 [gr-qc]].

 \bibitem{Soroushfar:2015wqa} 
  S.~Soroushfar, R.~Saffari, J.~Kunz and C.~L\"ammerzahl,
  Phys.\ Rev.\ D {\bf 92}, 044010 (2015)  
  [arXiv:1504.07854 [gr-qc]].  

\bibitem{Enolski:2011id} 
  V.~Enolski, B.~Hartmann, V.~Kagramanova, J.~Kunz, C.~L\"ammerzahl and P.~Sirimachan,
Journal of mathematical physics {\bf 53}, 012504 (2012) 
[arXiv:1106.2408 [gr-qc]].

  
\bibitem{Soroushfar:2015dfz} 
  S.~Soroushfar, R.~Saffari and A.~Jafari,
  Phys.\ Rev.\ D {\bf 93}, 
  [arXiv:1512.08449 [gr-qc]].
  
\bibitem{Soroushfar:2016yea} 
  S.~Soroushfar, R.~Saffari and E.~Sahami,
  [arXiv:1601.03143 [gr-qc]].



\bibitem{Bertone:2004pz} 
  G.~Bertone, D.~Hooper and J.~Silk,
  Phys.\ Rept.\  {\bf 405}, 279 (2005)
  [hep-ph/0404175].




\bibitem{Capozziello:2011et} 
  S.~Capozziello and M.~De Laurentis,
  Phys.\ Rept.\  {\bf 509}, 167 (2011)
  [arXiv:1108.6266 [gr-qc]].
\bibitem{Clifton:2011jh} 
  T.~Clifton, P.~G.~Ferreira, A.~Padilla and C.~Skordis,
  Phys.\ Rept.\  {\bf 513}, 1 (2012)
  [arXiv:1106.2476 [astro-ph.CO]].
\bibitem{Joyce:2014kja} 
  A.~Joyce, B.~Jain, J.~Khoury and M.~Trodden,
  Phys.\ Rept.\  {\bf 568}, 1 (2015)
  [arXiv:1407.0059 [astro-ph.CO]].
\bibitem{Jain:2010ka} 
  B.~Jain and J.~Khoury,
  Annals Phys.\  {\bf 325}, 1479 (2010)
  [arXiv:1004.3294 [astro-ph.CO]].
\bibitem{Koyama:2015vza} 
  K.~Koyama,
  Rept.\ Prog.\ Phys.\  {\bf 79}, no. 4, 046902 (2016)
  [arXiv:1504.04623 [astro-ph.CO]].

\bibitem{Lovelock:1971yv} 
  D.~Lovelock,
  J.\ Math.\ Phys.\  {\bf 12}, 498 (1971)

\bibitem{Sotiriou:2008rp} 
  T.~P.~Sotiriou and V.~Faraoni,
  Rev.\ Mod.\ Phys.\  {\bf 82}, 451 (2010)
  [arXiv:0805.1726 [gr-qc]].

\bibitem{DeFelice:2010aj} 
  A.~De Felice and S.~Tsujikawa,
  Living Rev.\ Rel.\  {\bf 13}, 3 (2010)
  [arXiv:1002.4928 [gr-qc]].

\bibitem{Berti:2015itd} 
  E.~Berti {\it et al.},
  Class.\ Quant.\ Grav.\  {\bf 32}, 243001 (2015)
  [arXiv:1501.07274 [gr-qc]].

\bibitem{Brevik:2004sd} 
  I.~H.~Brevik, S.~Nojiri, S.~D.~Odintsov and L.~Vanzo,
  Phys.\ Rev.\ D {\bf 70}, 043520 (2004)
  [hep-th/0401073].

\bibitem{Cognola:2005de} 
  G.~Cognola, E.~Elizalde, S.~Nojiri, S.~D.~Odintsov and S.~Zerbini,
  JCAP {\bf 0502}, 010 (2005)
  [hep-th/0501096].

\bibitem{Saffari:2007zt} 
  R.~Saffari and S.~Rahvar,
  Phys.\ Rev.\ D {\bf 77}, 104028 (2008)
  [arXiv:0708.1482 [astro-ph]].

\bibitem{delaCruzDombriz:2009et} 
  A.~de la Cruz-Dombriz, A.~Dobado and A.~L.~Maroto,
  Phys.\ Rev.\ D {\bf 80}, 124011 (2009)
  Erratum: [Phys.\ Rev.\ D {\bf 83}, 029903 (2011)]
  [arXiv:0907.3872 [gr-qc]].

\bibitem{Capozziello:2009jg} 
  S.~Capozziello, M.~de Laurentis and A.~Stabile,
  Class.\ Quant.\ Grav.\  {\bf 27}, 165008 (2010)
  [arXiv:0912.5286 [gr-qc]].

\bibitem{Sebastiani:2010kv} 
  L.~Sebastiani and S.~Zerbini,
  Eur.\ Phys.\ J.\ C {\bf 71}, 1591 (2011)
  [arXiv:1012.5230 [gr-qc]].


\bibitem{Larranaga:2011fv} 
  A.~Larranaga,
  Pramana {\bf 78}, 697 (2012)
  [arXiv:1108.6325 [gr-qc]].

\bibitem{Cembranos:2011sr} 
  J.~A.~R.~Cembranos, A.~de la Cruz-Dombriz and P.~Jimeno Romero,
  Int.\ J.\ Geom.\ Meth.\ Mod.\ Phys.\  {\bf 11}, 1450001 (2014)
  [arXiv:1109.4519 [gr-qc]].

\bibitem{delaCruzDombriz:2012xy} 
  A.~de la Cruz-Dombriz and D.~Saez-Gomez,
  Entropy {\bf 14}, 1717 (2012)
  [arXiv:1207.2663 [gr-qc]].

\bibitem{Moon:2011hq} 
  T.~Moon, Y.~S.~Myung and E.~J.~Son,
  Gen.\ Rel.\ Grav.\  {\bf 43}, 3079 (2011)
  [arXiv:1101.1153 [gr-qc]].

\bibitem{Hendi:2014mba} 
  S.~H.~Hendi, B.~Eslam Panah and R.~Saffari,
  Int.\ J.\ Mod.\ Phys.\ D {\bf 23}, no. 11, 1450088 (2014)
  [arXiv:1408.5570 [hep-th]].


\bibitem{Hackmann:2013pva} 
  E.~Hackmann and H.~Xu,
  Phys.\ Rev.\ D {\bf 87}, no. 12, 124030 (2013)
  doi:10.1103/PhysRevD.87.124030
  [arXiv:1304.2142 [gr-qc]].

\bibitem{Stuchlik:1997gk} 
  Z.~Stuchlik, G.~Bao, E.~Ostgaard and S.~Hledik,
  Phys.\ Rev.\ D {\bf 58}, 084003 (1998).
  doi:10.1103/PhysRevD.58.084003

\bibitem{Heisnam:2014} 
  S. Heisnam, I. Meitei and K. Singh,
  International Journal of Astronomy and Astrophysics {\bf 4}, 365-373 (2014). 
  doi: 10.4236/ijaa.2014.42031

\bibitem{Kraniotis:2014paa} 
  G.~V.~Kraniotis,
  Gen.\ Rel.\ Grav.\  {\bf 46}, no. 11, 1818 (2014)
  doi:10.1007/s10714-014-1818-8
  [arXiv:1401.7118 [gr-qc]].

\bibitem{Carter:1968rr} 
  B.~Carter,
  Phys.\ Rev.\  {\bf 174}, 1559 (1968).

\bibitem{Mino:2003yg} 
  Y.~Mino,
  Phys.\ Rev.\ D {\bf 67}, 084027 (2003)
  [gr-qc/0302075].


\bibitem{M.Abramowitz}
M. Abramowitz and I. E. Stegun, 
\textit{Handbook of Mathematical Functions},  
(Dover Publications,New York,1968).

\bibitem{E. T. Whittaker}
E. T. Whittaker and G. N. Watson, \textit{A course of Modern Analysis}, 
(Cambrige University Press, Cambrige, 1950).


\bibitem{V. M. Buchstaber}
V. M. Buchstaber, V. Z. Enolskii, and D. V. Leykin,
\textit{Hyperelliptic Kleinian Functions and Applications}, 
(Gordon and Breach, New York, 1997).


\bibitem{Rindler:2007zz} 
  W.~Rindler and M.~Ishak,
  Phys.\ Rev.\ D {\bf 76}, 043006 (2007)
  [arXiv:0709.2948 [astro-ph]].
 
\bibitem{Bhattacharya:2009rv} 
  A.~Bhattacharya, A.~Panchenko, M.~Scalia, C.~Cattani and K.~K.~Nandi,
  JCAP {\bf 1009}, 004 (2010)
  [arXiv:0910.1112 [gr-qc]].

\bibitem{Bhattacharya:2010xh} 
  A.~Bhattacharya, G.~M.~Garipova, E.~Laserra, A.~Bhadra and K.~K.~Nandi,
  JCAP {\bf 1102}, 028 (2011)
  [arXiv:1002.2601 [gr-qc]].
  
\bibitem{Kraniotis:2003ig} 
  G.~V.~Kraniotis and S.~B.~Whitehouse,
  Class.\ Quant.\ Grav.\  {\bf 20}, 4817 (2003)
  [astro-ph/0305181].
  
\bibitem{Kraniotis:2004cz} 
  G.~V.~Kraniotis,
  Class.\ Quant.\ Grav.\  {\bf 21}, 4743 (2004)
  [gr-qc/0405095].

\bibitem{Balasubramanian:1999zv} 
  V.~Balasubramanian and S.~F.~Ross,
  Phys.\ Rev.\ D {\bf 61}, 044007 (2000)
  doi:10.1103/PhysRevD.61.044007
  [hep-th/9906226].

\end{thebibliography}

\end{document}